\documentclass[final,english]{elsarticle}
\usepackage[T1]{fontenc}
\usepackage[latin9]{inputenc}
\usepackage{varioref}
\usepackage{units}
\usepackage{amsmath}
\usepackage{amssymb}
\usepackage{graphicx}
\usepackage{esint}
\usepackage{caption}
\usepackage{subcaption}
\usepackage{appendix}

\pdfpageheight\paperheight
\pdfpagewidth\paperwidth

\journal{Elsevier}
\usepackage{upgreek}
\usepackage[bookmarks,bookmarksopen,bookmarksdepth=6]{hyperref}
\usepackage{cases}
\usepackage{url}
\usepackage{lineno}

\usepackage{a4wide}

\usepackage{newtxtext,newtxmath,amsmath}
\makeatother

\usepackage{babel}
\begin{document}

\title{Position reconstruction for segmented detectors}

\author[]{A.~Ebrahimi$^{a,}$\footnote{Now at Paul Scherrer Institut, 5232 Villingen, Switzerland.}}
\author[]{F.~Feindt$^a$}
\author[]{E.~Garutti$^a$}
\author[]{M.~Hajheidari$^a$}
\author[]{R.~Klanner$^{a,}$\corref{cor1}}
\author[]{D.~Pitzl$^b$}
\author[]{J.~Schwandt$^a$}
\author[]{G.~Steinbr\"uck$^a$}
\author[]{I.~Zoi$^a$}

 \cortext[cor1]{Corresponding author, Email address: Robert.Klanner@desy.de,
 Tel.: +49 40 8998 2558.}
 \address{$^a$Institute for Experimental Physics, University of Hamburg, Luruper Chaussee 149, 22761 Hamburg, Germany}
 \address{$^b$ DESY, Notkestr. 85, 22607, Hamburg, Germany}


\begin{abstract}

 The topic of the paper is the position reconstruction from signals of segmented detectors.
 With the help of a simple simulation, it is shown that the position reconstruction using the centre-of-gravity method is strongly biased, if the width of the charge (or e.g. light) distribution at the electrodes (or photo detectors) is less than the read-out pitch.
 A method is proposed which removes this bias for  events with signals in two or more read-out channels and thereby improves the position resolution.
 The method also provides an estimate of the position-response function for every event.
 Examples are given for which its width as a function of the reconstructed position varies by as much as an order of magnitude.

  A fast Monte Carlo program is described which simulates the signals from a silicon pixel detector traversed by charged particles under different angles, and the results obtained with the proposed reconstruction method and with the centre-of-gravity method are compared.
  The simulation includes the local energy-loss fluctuations, the position-dependent electric field, the diffusion of the charge carriers, the electronics noise and charge thresholds for clustering,
 A comparison to test-beam-data is used to validate the simulation.

\end{abstract}


\maketitle
 \tableofcontents
 \pagenumbering{arabic}

\section{Introduction}
 \label{sect:Introduction}

 The position reconstruction using analogue signals from segmented detectors is a standard analysis problem.
 Frequently the "centre-of-gravity" method for clusters of signals exceeding a threshold is used, which is simple and robust, but results in most cases in a biased position reconstruction\,\cite{Belau:1983, Landi:2020}.
 "Bias" means that $x_\mathit{true}$, the position at which particles hit the detector, and $\langle x_\mathit{rec} \rangle$, the mean reconstructed position for a given $x_\mathit{true}$ differ.
 This paper describes a general method which can be used to correct such a bias for a given position-reconstruction algorithm, and also determine the position-response function, i.\,e. the probability-density distribution of
$\Delta x$, where $\Delta x$ is the difference between true and reconstructed position.
 The method does not require additional data beyond those to be reconstructed.

 Compared to methods reported in the literature and discussed in more detail in the conclusions of Sect.\,\ref{sect:Model}, the proposed method is simpler to implement:
 It does not require the selection of a fixed cluster size, which has to be adjusted as a function of the track angle, nor an externally predicted position.
 The position resolution achieved by the proposed method is similar to reconstruction methods described in the literature.

 To demonstrate the method, Monte Carlo data are analysed, which simulate the signals from  silicon pixel sensors for minimum ionizing particles at different incident angles, and the resolutions using the "centre-of-gravity" and the proposed method are compared.
 The validity of the simulated data is verified by a comparison to test-beam data taken with three parallel pixel detectors with a pixel pitch of $25\,\upmu$m~$\times 100\,\upmu$m.
 In addition, the expected improvement of the position resolution for the test-beam data as a function of track angle is presented.
 It is noted that the method is not limited to the centre-of-gravity algorithm, but can be used for other reconstruction methods.

 The paper is structured in the following way.
 In the next section a simple model for the signals from strip detectors is used to explain the reason of the bias of the centre-of-gravity method,
 and a reconstruction method which avoids this bias is developed.
 This is followed by a discussion of the effects of charge-carrier diffusion on the charge distribution arriving at the electrodes for charged particles traversing a silicon strip detector at different angles.
 A short description of the Monte Carlo program used to simulate the signals in silicon pixel detectors follows.
 Next, the fluctuations of the mean position of the charge distribution at the electrodes due to the fluctuations of the energy loss along the particle track is presented.
 In Sect.\,\ref{sect:Resolution}, simulations of the signals from a pixel silicon sensors for track angles of $0^{\,\circ}$, $10^{\,\circ}$, $20^{\,\circ}$ and $32^{\,\circ}$ are used to study the position resolution as a function of the threshold used to assign signals to a cluster.
 Finally, the results of the simulations are compared to test-beam data for track angles between $0^{\,\circ}$ and $30^{\,\circ}$, and the expected improvement in position resolution of the proposed method compared to the centre-of-gravity algorithm  presented.
 Sect.\,\ref{sect:Conclusions} summarizes the results.
 The Appendices give details of the Monte Carlo program used, and presents a method for determining the average distribution of tracks in the pixels from the measured distribution of particles over the entire sensor.



 \section{Method}
  \label{sect:Model}

\begin{figure}[!ht]
  \centering
    \includegraphics[width=0.4\textwidth]{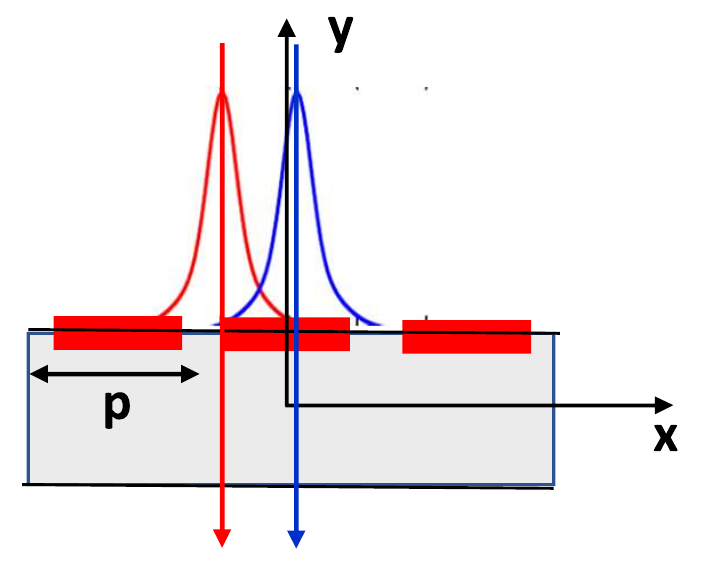}
    \caption{Sketch of a detector with three readout strips with pitch $p = 25\,\upmu$m.
    The charge distributions, $\mathrm{d}Q / \mathrm{d}x$,  induced by two particles with normal incidence, one at $x = - 10\,\upmu$m and the other at $1\,\upmu$m, are shown;
  $x = 0$ is at the centre of the central strip.
    Independent of the $x$\,position of the particle the charge distribution induced at the electrode plane, $\mathrm{d}Q / \mathrm{d}x$, is simulated by two Gauss\,functions with equal weights, one with $\sigma = 2\,\upmu$m and the other with $\sigma = 5\,\upmu$m.
    This simple model is only used to demonstrate the method.
    For a more realistic model we refer to Sect.~\ref{sect:Simulation} and Appendix\,\ref{sect:Appendix_Signal}.}
  \label{fig:Fig_Sketch}
 \end{figure}

 The method of correcting the bias of a given reconstruction algorithm, is explained with the help of Fig.\,\ref{fig:Fig_Sketch}, which shows a strip detector with 3 readout channels of pitch $p$.
 The signal distributions, $\mathrm{d}Q/ \mathrm{d}x$, at the electrode plane by two particles at normal incidence are shown as solid lines.
 The signal, $Q_i$, collected by strip $i$ is given by the integral of $\mathrm{d}Q/ \mathrm{d}x$ over the corresponding strip pitch.
 A reconstruction algorithm is used to obtain the reconstructed position, $x_\mathit{rec}$, from the $Q_i$.
 Frequently, the centre-of-gravity is used for which $x_\mathit{rec} = \sum (c_i \cdot Q_i) / \sum Q_i $, where the $c_i$ are the positions of the strip centres.

 To illustrate the proposed method, $10^6$ Monte Carlo events are generated in the following way:
 The strip pitch is $p = 25 \, \upmu $m, and the signal distribution, $\mathrm{d}Q / \mathrm{d}x$, is simulated by the sum of two normalized Gauss-distributed random numbers with weights 1/2, mean values $x$, and $\sigma = 2$ and $5\,\upmu$m, respectively.
 For the position $x$, a uniformly distributed random number in the range $ -p/2 \leq x < p/2$ is generated, and the signal $Q_i$ of strip $i$ is obtained by integrating $\mathrm{d}Q / \mathrm{d}x$ over the strip pitch .
 To account for  noise, Gauss-distributed random numbers with $\sigma _\mathit{el} = 0.025$ are added to the $Q_i$.
 This very simple simulation approximates the response of a silicon strip sensor with $25 \, \upmu $m pitch in the $x$\,direction, to particles with normal incidence uniformly distributed in $x$ over the pitch of the central strip.
 It is used to explain the method and allows for analytical calculations for checking the analysis code.
 The widths of the Gaussians, $\sigma = 2\,\upmu$m and $5\,\upmu$m, correspond to charge collection times for electrons of 0.6 and 3.6\,ns, which are typical for different regions in thin silicon detectors.
 For the detailed studies and the comparison to experimental data, the more realistic simulation of pixel sensors presented in Sect.~\ref{sect:Simulation} is used.

\begin{figure}[!ht]
   \centering
   \begin{subfigure}[a]{0.5\textwidth}
    \includegraphics[width=\textwidth]{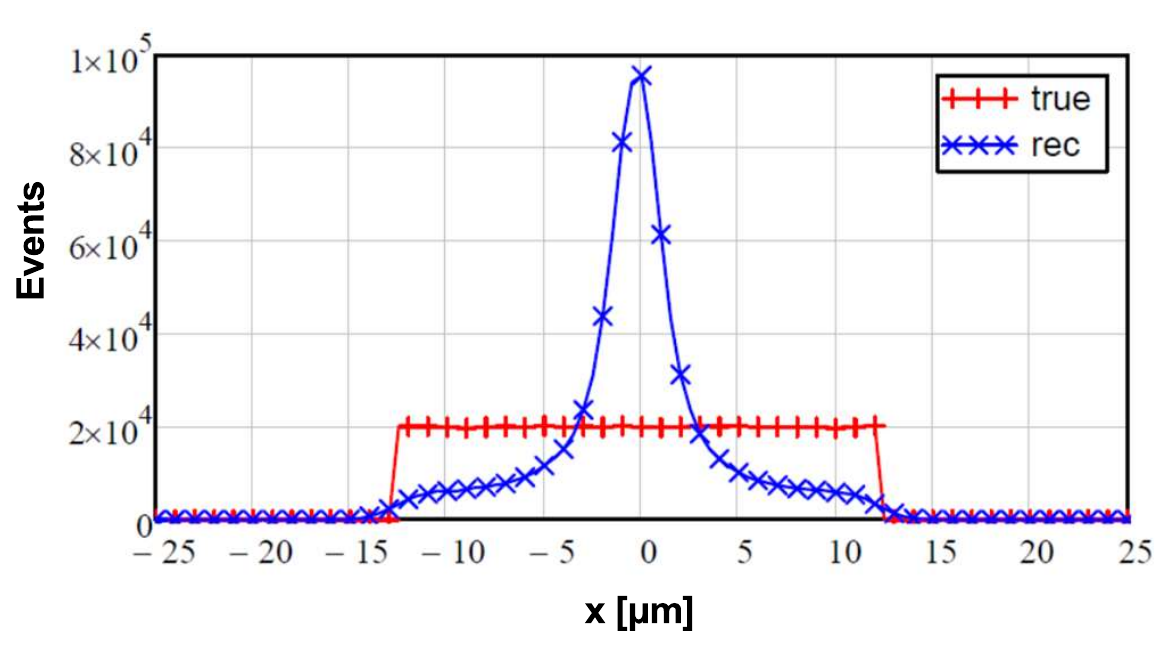}
    \caption{ }
    \label{fig:Fig_dNdx-true}
   \end{subfigure}%
    ~
   \begin{subfigure}[a]{0.46\textwidth}
    \includegraphics[width=\textwidth]{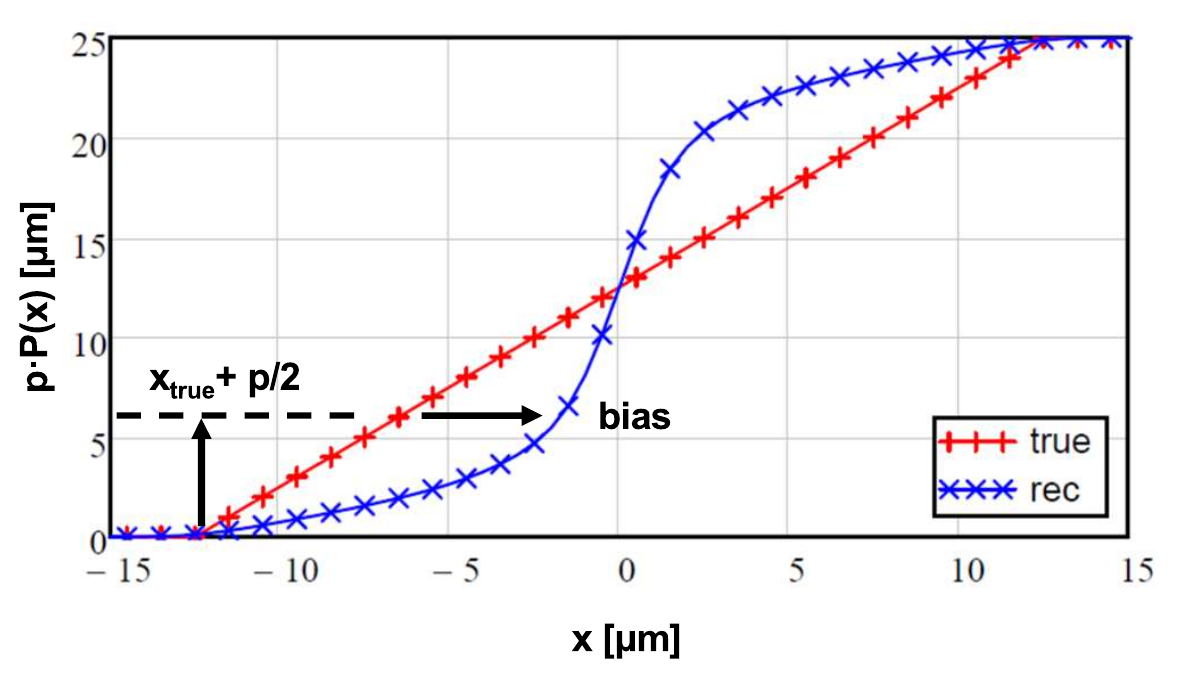}
    \caption{ }
    \label{fig:Fig_Nx-true}
   \end{subfigure}%
   \caption{ (a) Distributions $\mathrm{d}N / \mathrm{d}x$ within an electrode of width $p = 25\,\upmu$m for $x_\mathit{true}$ and the reconstructed position $x_\mathit{rec}$.
   (b) Normalised cumulative distributions of the curves shown in (a) multiplied with $p$.
   The horizontal arrow shows the mean bias used to correct $x_\mathit{rec}$. }
  \label{fig:dNdx}
 \end{figure}

 In Fig.\,\ref{fig:Fig_dNdx-true} the event distributions of $\mathrm{d}N / \mathrm{d}x$ for $x_\mathit{true}$ and $x_\mathit{rec}$ are shown.
 The generated $x$\,position is denoted $x_\mathit{true}$, and the position reconstructed with the centre-of-gravity method without threshold cuts $x_\mathit{rec}$.
 The generated distribution $\mathrm{d}N / \mathrm{d}x_\mathit{true}$ is flat,
 whereas $\mathrm{d}N / \mathrm{d}x_\mathit{rec}$ peaks at $x_\mathit{rec} = 0 $.
 The reason is that the signal distribution at the strips, $\mathrm{d}Q / \mathrm{d}x$, peaks at the track position $x_\mathit{true}$ which biases the $x_\mathit{rec}$ values towards $x_\mathit{rec} = 0 $.
 It is noted that only for flat signal distributions with widths, which are multiples of $p$, the centre-of-gravity has no bias.
 Removing the bias means finding the function $x _\mathit{corr} (x _\mathit{rec})$ so that $\mathrm{d}N / \mathrm{d}x_\mathit{corr}$ is a flat distribution of width $p$, i.e. has the same shape as $\mathrm{d}N / \mathrm{d}x_\mathit{true}$.

 This can be achieved by using the normalized cumulative distributions
 \begin{equation}\label{equ:CumDistr}
 P_\mathit{true}(x) = \frac{1}{N_\mathit{tot}} \int_{-p/2}^{x} (\mathrm{d}N / \mathrm{d} x _\mathit{true})\,\mathrm{d} x _\mathit{true} \hspace{3mm} \mathrm{and} \hspace{3mm}
 P_\mathit{rec} (x) = \frac{1}{N_\mathit{tot}} \int_{-p/2}^{x} (\mathrm{d}N / \mathrm{d} x _\mathit{rec})\,\mathrm{d}x _\mathit{rec}.
 \end{equation}
 Fig.\,\ref{fig:Fig_Nx-true} shows $p \cdot P_\mathit{rec} (x)$ and $p \cdot P_\mathit{true} (x)$.
 The mean bias, $\langle x_\mathit{rec} - x_\mathit{true} \rangle $ is obtained from $P_\mathit{rec}(x_\mathit{rec}) = P_\mathit{true}(x_\mathit{true} - x_\mathit{rec})$,
  which follows from the requirement of equal number of events between $-p/2$ and $x_\mathit{true}$ for $\mathrm{d}N_\mathit{true}/\mathrm{d}x$, and between $-p/2$ and $x_\mathit{rec}$ for $\mathrm{d}N_\mathit{rec}/\mathrm{d}x$.
 For a uniform $\mathrm{d}N/\mathrm{d}x _\mathit{true}$, $P_\mathit{true}(x)$ is the diagonal between $x = - p/2$ and $x =  p/2$, and
 \begin{equation}\label{equ:shift}
   x_\mathit{corr} = p \cdot \big(P_\mathit{rec}(x_\mathit{rec}) - 0.5\big).
 \end{equation}
 For a non-uniform $\mathrm{d}N/\mathrm{d}x _\mathit{true}$, the horizontal distance shown in Fig.\,\ref{fig:Fig_Nx-true} has to be used.
  However, as shown in the Appendix~\ref{sect:Appendix_SpatialDistribution}, for most situations the assumption of a flat $x_\mathit{true}$ is valid.
 Using Eq.\,\ref{equ:shift}, $x_\mathit{corr}$ can be calculated on an event-by-event basis, and
 the ($x_\mathit{corr} -  x_\mathit{true} $)\,distributions for events in  $x_\mathit{corr}$\,intervals are estimates of the $x_\mathit{corr}$-dependent response functions of the position determination.


\begin{figure}[!ht]
   \centering
   \begin{subfigure}[a]{0.5\textwidth}
    \includegraphics[width=\textwidth]{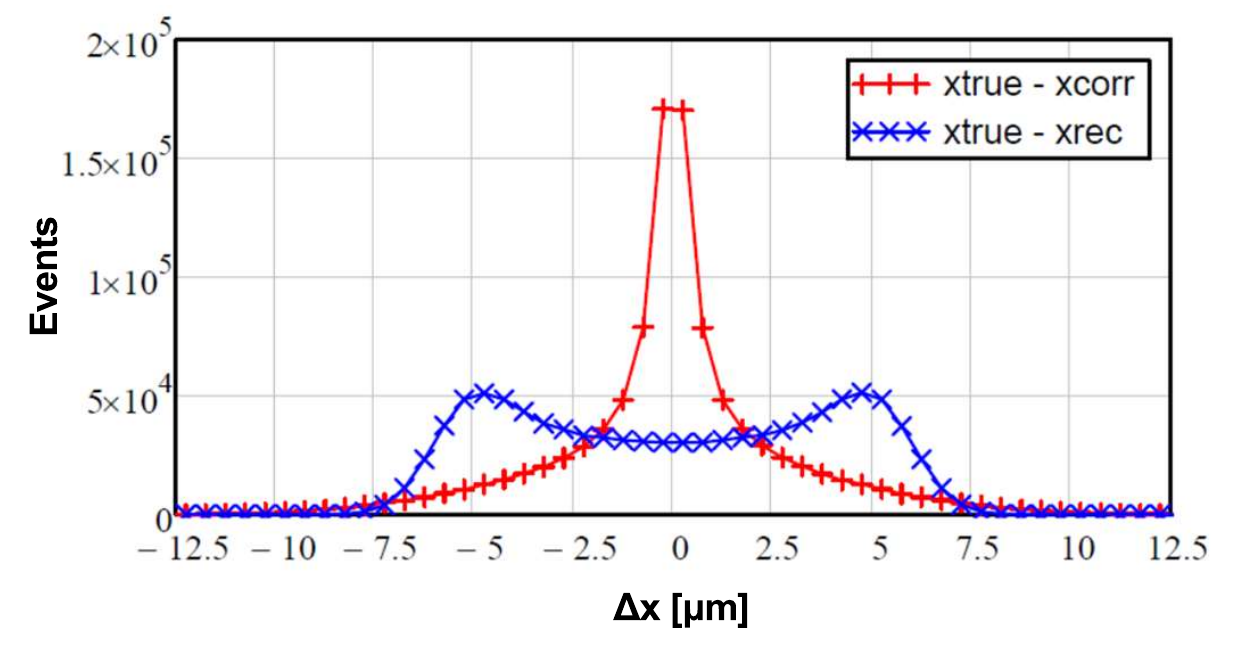}
    \caption{ }
    \label{fig:Fig_Res_nothr}
   \end{subfigure}%
    ~
   \begin{subfigure}[a]{0.5\textwidth}
    \includegraphics[width=\textwidth]{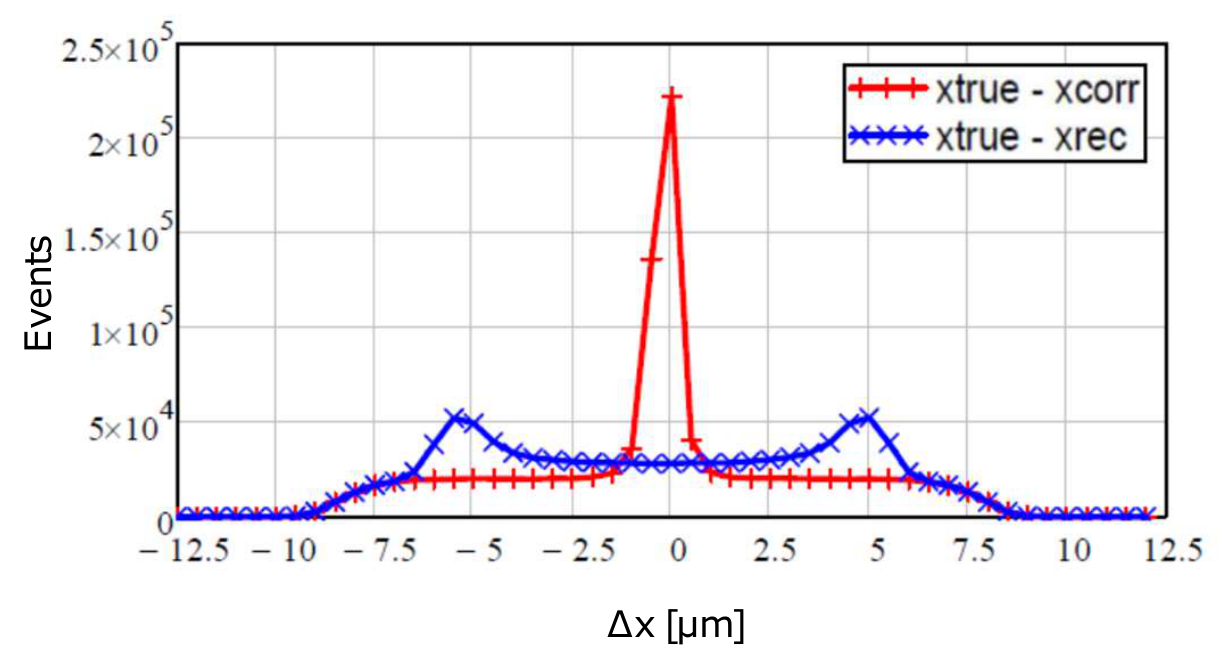}
    \caption{ }
    \label{fig:Fig_Res_thr}
   \end{subfigure}%
   \caption{ Distributions of $\Delta x$ for $x _\mathit{rec}$ and $x _\mathit{corr}$ for the simulated events described in the text.
   (a) Without, and
   (b) with a threshold cut at 10\,\% of the mean signal. }
  \label{fig:Fig_Res}
 \end{figure}

 To illustrate the improvement of the position determination by the correction method, Fig.\,\ref{fig:Fig_Res_nothr} shows the residual distributions $\Delta x _\mathit{rec} = x_\mathit{true} - x_\mathit{rec}$ and $\Delta x _\mathit{corr} = x_\mathit{true} - x_\mathit{corr}$.
 A significant improvement is seen, which is mainly caused by the events in the region of the boundaries between electrodes.
 These events have a very good spatial resolution, however they are reconstructed with a bias of up to $5\,\upmu$m by the centre-of-gravity method and produce the enhancements at $|\Delta x _\mathit{rec}| \approx 5\,\upmu$m.
 It is noted that the position-response function is very different from a Gauss\,function.

 So far, $x_\mathit{rec}$ has been obtained from the $Q_i$ without threshold cuts.
 As a result, for events in which most of the charge is induced in the central strip, the contribution of the signal in the adjacent strips is dominated by random noise and thus worsens the position resolution.
 In the following the effect of threshold cuts will be discussed.
 For the study, the simulated events discussed above with a threshold cut, $\mathit{thr} = 4 \cdot \sigma _\mathit{el} = 0.1$ for a mean total signal of $\sum Q_i =1$ is used.
 Threshold cuts result in events with different  cluster sizes, $\mathit{cls}$.
 The cluster-size distribution $\mathrm{d}N / \mathrm{d}\mathit{cls}$ depends on the track angle $\theta $ and so does the optimal algorithm for $x_\mathit{rec}$\,\cite{Turchetta:1993}.

 Depending on $cls$, the value for $x_\mathit{rec} $ differs:

 \begin{tabular}{l}
  \hspace{2mm} $cls = 0$ \hspace{2mm} $x_\mathit{rec}$ no information available\\
  \hspace{2mm} $cls = 1$ \hspace{2mm} $x_\mathit{rec} = 0$ \\
  \hspace{2mm} $cls > 1$ \hspace{2mm} $x_\mathit{rec}$ from a reconstruction algorithm\\
 \end{tabular}

\begin{figure}[!ht]
   \centering
   \begin{subfigure}[a]{0.5\textwidth}
    \includegraphics[width=\textwidth]{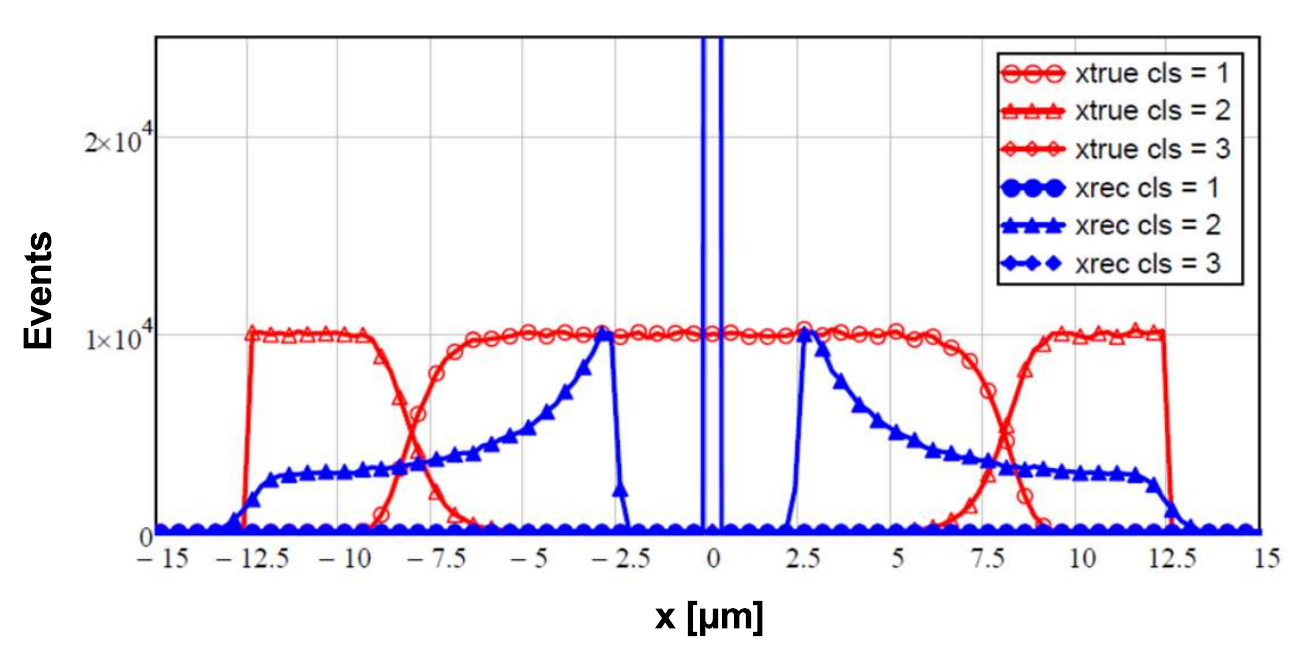}
    \caption{ }
    \label{fig:Fig_dNdx-thr}
   \end{subfigure}%
    ~
   \begin{subfigure}[a]{0.45\textwidth}
    \includegraphics[width=\textwidth]{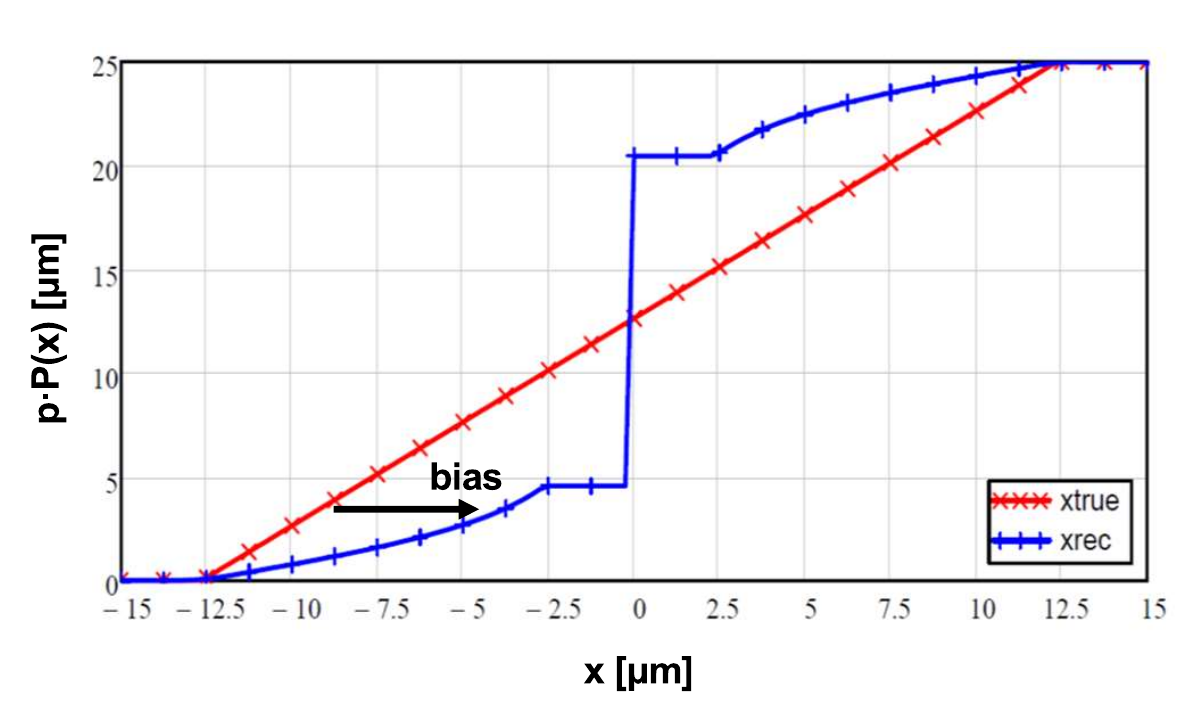}
    \caption{ }
    \label{fig:Fig_Nx-thr}
   \end{subfigure}%
   \caption{ Results of the analysis of the simulated data with a threshold cut of 0.1.
   (a) Event-number distribution $\mathrm{d}N / \mathrm{d}x$ for different  cluster sizes, $\mathit{cls}$, for $x_\mathit{true}$ and for $x_\mathit{rec}$.
       The $\delta $\,function peak at x = 0 comes from the $cls = 1$\,events.
   (b) Normalised cumulative distributions of the distributions shown in (a) multiplied with $p$.
   The horizontal arrow shows the bias used to correct $x_\mathit{rec}$ for $\mathit{cls} > 1$. }
  \label{fig:Fig_NdN-thr}
 \end{figure}

 In order to assure that $\mathrm{d}N / \mathrm{d}x_\mathit{true}$ for the events for which $x_\mathit{rec}$ can be calculated remains flat, the fraction of events with $cls = 0$ (inefficiency) has to be small.
 For the simulated events the fractions of events with $cls = 0, 1,\, 2\, , \mathrm{and}\,3$ are 0, 63.5\,\%, 36.5\,\%, and $10^{-3}\,\%$, respectively.
 Fig.\,\ref{fig:Fig_dNdx-thr} shows the distributions $\mathrm{d}N / \mathrm{d}x_\mathit{true}$ and  $\mathrm{d}N / \mathrm{d}x_\mathit{rec}$ for different $cls$.
 The $6.35 \times 10^5\,\mathit{cls} = 1$ events cover the central region of $x _\mathit{true}$ and are reconstructed at $x _\mathit{rec} = 0 $.
 The $3.65 \times 10^5\,\mathit{cls} = 2$ events cover the outer region of $x _\mathit{true}$. Their $x _\mathit{rec}$ values are shifted towards the centre because of the bias of the centre-of-gravity algorithm.
 Because of the $\mathit{thr} = 0.1$\,cut, no $\mathit{cls} = 2$\,events are reconstructed in the region $| x_\mathit{rec}| \lesssim 2.5\,\upmu$m.
 Fig.\,\ref{fig:Fig_Nx-thr} shows the normalised cumulative distributions, $P _\mathit{true} (x)$ and $P _\mathit{rec} (x)$,  multiplied with $p$, with the bias correction according to Eq.\,\ref{equ:shift} indicated by the arrow.
 This correction can only be applied to the $\mathit{cls} > 1$\,events.
 Fig.\,\ref{fig:Fig_Res_thr} compares the $(\Delta x _\mathit{rec} = x _\mathit{true} - x _\mathit{rec})$- with the $(\Delta x _\mathit{corr} =x _\mathit{true} - x _\mathit{corr})$-distribution.
 Similar to the situation without threshold cut shown in Fig.\,\ref{fig:Fig_Res_nothr}, a significant improvement of the position resolution is observed:
 Although for $\mathit{thr} = 0.1$ only 36.5\,\% of the events have $\mathit{cls} = 2$, the \emph{rms} of the $\Delta x$\,distributions decreases from $4.5\,\upmu$m for the centre-of-gravity\,method to $3.7\,\upmu$m.
 The $\mathit{cls} = 2$\,events, which cause the peaks around $|x _\mathit{true} - x _\mathit{rec}| = 5\,\upmu$m are moved to the narrow peak at $x _\mathit{true} - x _\mathit{corr} = 0$, whereas the $\mathit{cls} = 1$\,events cause the  flat distribution below the peak.
 It should be noted that decreasing $\mathit{thr}$, increases the fraction of $\mathit{cls} = 2$\,events and improves the overall position resolution:
 The \emph{rms} of the $\Delta x _\mathit{corr}$\,distribution is $2.7\,\upmu$m for a threshold $\mathit{thr} = 0.04$.
 It is noted that the position resolution has a strong dependence on the particle position relative to the strip centre.

\begin{figure}[!ht]
   \centering
   \begin{subfigure}[a]{0.5\textwidth}
    \includegraphics[width=\textwidth]{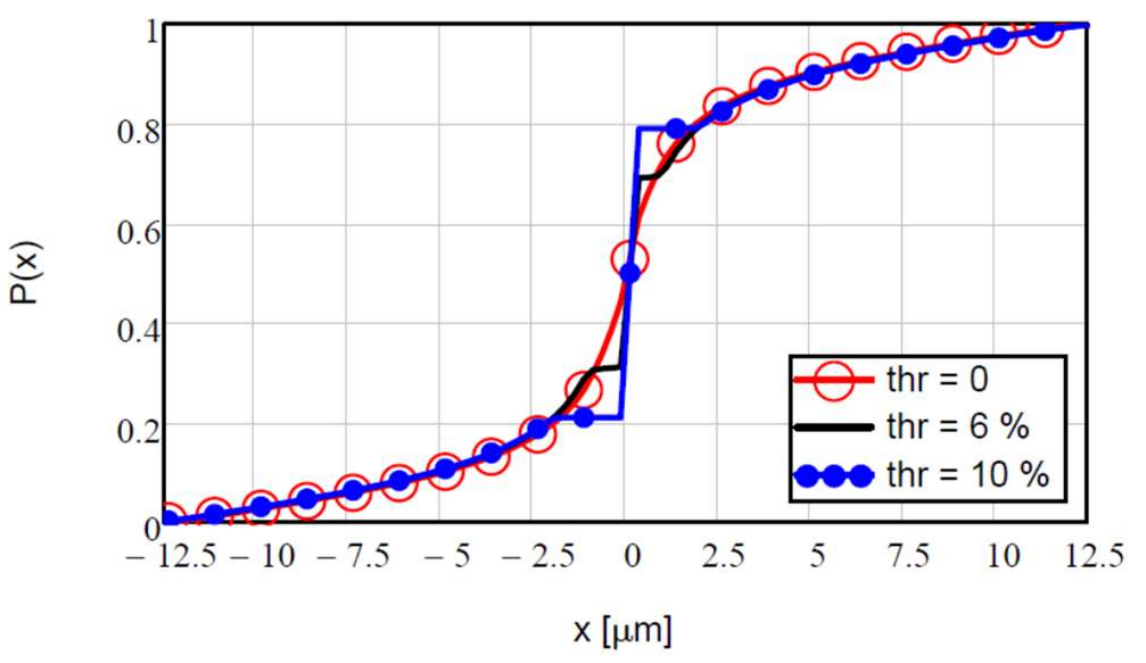}
    \caption{ }
    \label{fig:Fig_Cumthr}
   \end{subfigure}%
    ~
   \begin{subfigure}[a]{0.5\textwidth}
    \includegraphics[width=\textwidth]{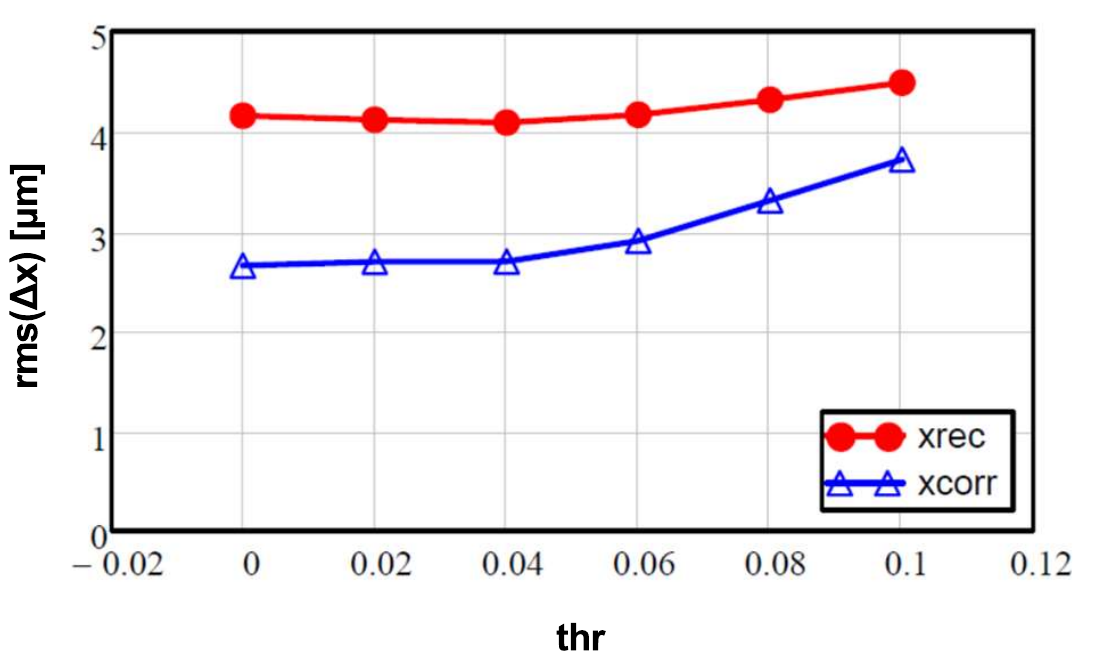}
    \caption{ }
    \label{fig:Fig_RMSthr}
   \end{subfigure}%
   \caption{(a) Normalized cumulative distribution,$P(x)$, for 3 threshold values, $\mathit{thr}$.
   (b) \emph{rms} values of the overall resolution for $x _\mathit{rec}$, the position reconstructed with the centre-of-gravity method, and for $x_\mathit{corr}$, the position corrected for bias for events with  cluster size $ \mathit{cs} \geq 2 $ as a function of $\mathit{thr}$. }
  \label{fig:Fig_Cumrmsvsth}
 \end{figure}

 Next the question is addressed, if the correction depends on the charge, $Q$, generated by the particle.
 This is studied by varying the threshold, $\mathit{thr}$, for a fixed $Q$.
 Fig.\,\ref{fig:Fig_Cumthr} shows the normalised cumulative distributions, $P_\mathit{rec}$, for $\mathit{thr} = 0,\,0.06$ and 0.1.
 As expected for the charge-sharing region $|x_\mathit{rec}| \gtrsim 2.5 \, \upmu$m, $P_\mathit{rec}$ is independent of $\mathit{thr}$.
 A decrease of $\mathit{thr}$ increases the fraction of events with $\mathit{cls} > 1$, deceases the step at $x = 0$, and extends the continuous region of $P_\mathit{rec}$ to smaller $|x_\mathit{rec}|$\,values.
 The study suggests that for energetic charged particles, where $Q$ follows a Landau distribution, a  single correction function can be used as long as no energetic $\delta $-electrons are produced.

 Fig.\,\ref{fig:Fig_RMSthr} compares \emph{rms}$_\mathit{rec}$, the root-mean-square of the resolution for normal incidence for $x_\mathit{rec}$, the position reconstructed by the centre-of-gravity method, with \emph{rms}$_\mathit{corr}$ using the proposed method, as a function of $\mathit{thr}$. A significant improvement is observed, which decreases with increasing $\mathit{thr}$, because of the increase of events with $\mathit{cls} = 1$.

 To summarize this section:
 A method which corrects the bias of position-reconstruction algorithms for segmented detectors with analogue readout is proposed.
 The method is explained and demonstrated with the help of a very simple simulation of charged particles traversing a silicon strip detector at normal incidence using the centre-of-gravity algorithm for the position reconstruction.
 It it shown that the proposed method corrects the bias of the centre-of-gravity algorithm for events with $cls > 1$, and achieves a significant improvement of the spatial resolution for the simulated data.
 The method can also be used for other reconstruction methods.

 The method is similar to the $\eta$~algorithm of Refs.\,\cite{Belau:1983, Turchetta:1993}:
  For $cls \leq 1$ there is insufficient information for any correction,
  for $cls = 2$ both methods are identical, and
  for $cls > 2$ the proposed algorithm automatically performs the bias correction independent of the reconstruction method, whereas for the $\eta$~algorithm it has to be decided how to assign measured charges to what is called the \emph{RIGHT} and the \emph{LEFT} signal in Ref.\,\cite{Belau:1983}.
 An extension of the $\eta$~algorithm, the \emph{multi-pixel $\eta$-correction}~(mp$- \eta$) is presented in Ref.\,\cite{Bugiel:2021}.
 The only difference to the correction method used in this paper is that the method is applied to fixed read-out patterns, and thus a fixed  cluster size.
 Therefore, it can not be used if only signals above a fixed threshold are recorded, which results in different  cluster sizes.
 In Ref.~\cite{Gorelov:2002} an $\eta $~corrections for the individual  cluster sizes is applied.
 The method requires the knowledge of $\Delta _\mathit{cls}$, the widths of the regions in pixel coordinates producing clusters of size $\mathit{cls}$.
 For fully depleted detectors with high charge collection efficiency, the calculation of $\Delta _\mathit{cls}$ is straight forward.
 However, for highly-irradiated detectors with significant charge losses, in particular if they have
depletion regions on both front and rear side with a non-depleted region in between, it is not clear how to estimate $\Delta _\mathit{cls}$.
 The step shown in Fig.\,\ref{fig:Fig_Nx-thr} actually determines $\Delta _{cls = 1}$, however this information is not used by the authors.
 Examples for other bias-correction methods are given in Refs.~\cite{Akiba:2012, Dannheim:2020, Dannheim:2021}, where the corrections are derived from the difference in track position predicted by a beam telescope and the one measured in the detector under study.
 The mean of the difference distribution as a function of the predicted position in the sensor is  fitted by a fifth-order polynomial.
 This method achieves a  similar performance as the $\eta$\,algorithm.

 It is concluded: The method presented does not achieve a higher precision than previously used methods, but it is more easily implemented and does not require additional information beyond the signals recorded in the individual detector elements.

  \section{Angular dependence of the charge distribution at the electrode plane}
  \label{sect:Qsurface}

 In this section the effect of diffusion on the charge distribution arriving at the electrode plane is investigated using a simple calculation.
 Shown are the charge distributions, $\mathrm{d}Q/\mathrm{d}x$, for particles  passing through the centre with angles $\theta = 0^{\,\circ},\, 10^{\,\circ}$ and $20^{\,\circ}$ to the sensor normal.
 For simplicity a uniform charge distribution without fluctuations along the particle track is assumed.

 \begin{figure}[!ht]
  \centering
    \includegraphics[width=0.5\textwidth]{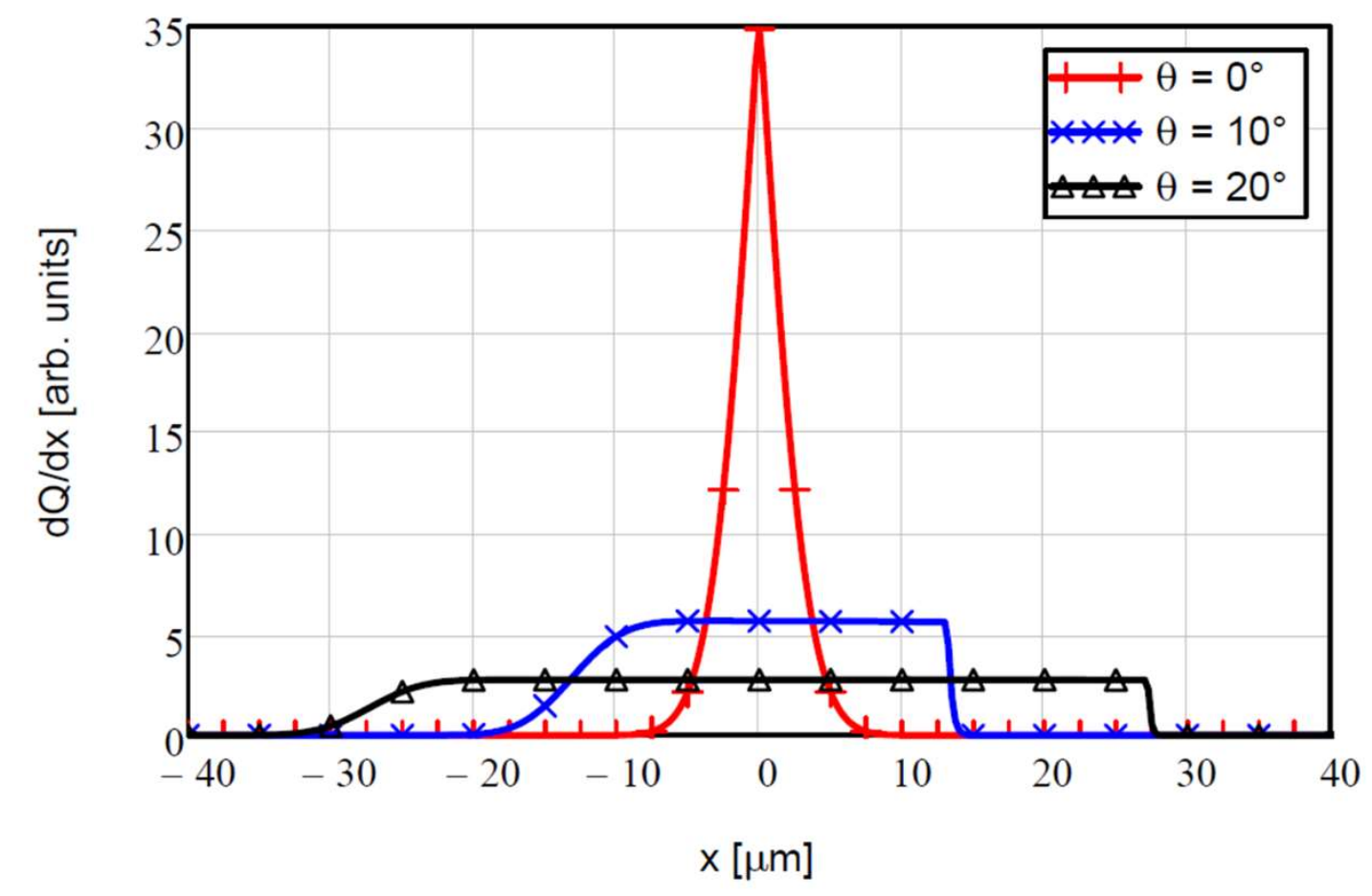}
    \caption{Simulated charge distributions, $\mathrm{d}Q/\mathrm{d}x$ induced at the electrode plane for particles passing through the centre of a $150\,\upmu$m thick sensor at angles of $\theta = 0^{\,\circ},\, 10^{\,\circ}$ and $20^{\,\circ}$ to the sensor normal.
  For finite angles $\theta > 0^{\,\circ}$ the drop of $\mathrm{d}Q/\mathrm{d}x$ to zero for negative $x$ is shallower than for positive $x$, as in the former case the charges which reach the electrodes have to drift through the entire sensor, which results in an increased diffusion.
  The temperature used for the calculation is $20^{\,\circ}$C.
 }
  \label{fig:Fig_Qsurf}
 \end{figure}

 The calculation of the charge distribution induced in the electrode plane for a sensor of thickness $d = 150\,\upmu$m and $p$-doping density $4 \times 10^{12}$\,cm$^{-3}$ at $T = 20^{\,\circ}$C takes into account the electric field in the sensor and the diffusion of the electron cloud.
 For  the coordinate system of Fig.\,\ref{fig:Fig_Sketch}, $\mathrm{d}Q/\mathrm{d}x$ is obtained by integrating Gauss distributions with mean positions $\mu(y) = y \cdot \tan(\theta)$ and \emph{rms} widths $\sigma(y)$ from Eq.\,\ref{equ:sigma} of Appendix\,\ref{sect:Appendix_Signal} between $y = -d/2$ and $y = d/2 - \epsilon$.
 $\epsilon = 0.1\,\upmu$m is introduced to avoid the divergence due to $\sigma (y = d/2) = 0 $.
 The results of the calculation are shown in Fig.\,\ref{fig:Fig_Qsurf}.
 As expected, for angles $\theta$ different from zero, the full width at half maximum of the $\mathrm{d}Q/\mathrm{d}x$ distribution is $d \cdot |\tan(\theta)|$.
 For $\theta > 0^{\,\circ}$, the drop of $\mathrm{d}Q/\mathrm{d}x$ to zero is shallower for negative $x$ than for positive $x$.
 The reason is that in the former case the electrons drift through the entire detector, whereas in the latter one their drift distance is close to zero.
 The increased drift time results in an increased spread of the electron cloud by diffusion.

 As a result a threshold used for the cluster selection will affect the two regions differently and cause a bias for the reconstructed position, $x_\mathit{rec}$.
 Thus even for the optimal angle, $\theta = \mathrm{atan}(p/d)$, where the width of the $\mathrm{d}Q/\mathrm{d}x$ distribution is equal to the electrode pitch $p$, the centre-of-gravity reconstruction is biased.
 However, as shown in Sect.\,\ref{sect:Threshold}, the effect is small, but the proposed correction method corrects for this bias, too.

 \section{Signal simulation of signals in a pixel sensor}
  \label{sect:Simulation}

 In Sect.~\ref{sect:Model} a highly simplified model for a silicon strip sensor has been used to explain the proposed position-reconstruction method and illustrate some of its features.
 The model has the advantage that for most results analytical calculations can be used to verify the analysis.
 In this section a significantly more realistic model for a pixel sensor is introduced, which is used for more detailed investigations and a comparison to experimental results.
 Details of the model and of its implementation are given in the Appendices.

  \begin{figure}[!ht]
   \centering
   \begin{subfigure}[a]{0.4\textwidth}
    \includegraphics[width=\textwidth]{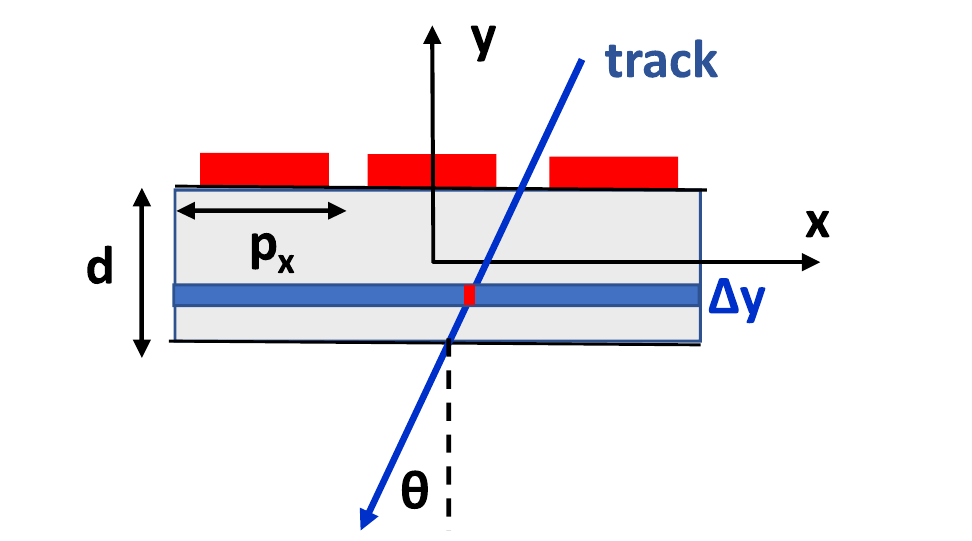}
    \caption{ }
    \label{fig:Fig_Sensor01}
   \end{subfigure}%
    ~
   \begin{subfigure}[a]{0.4\textwidth}
    \includegraphics[width=\textwidth]{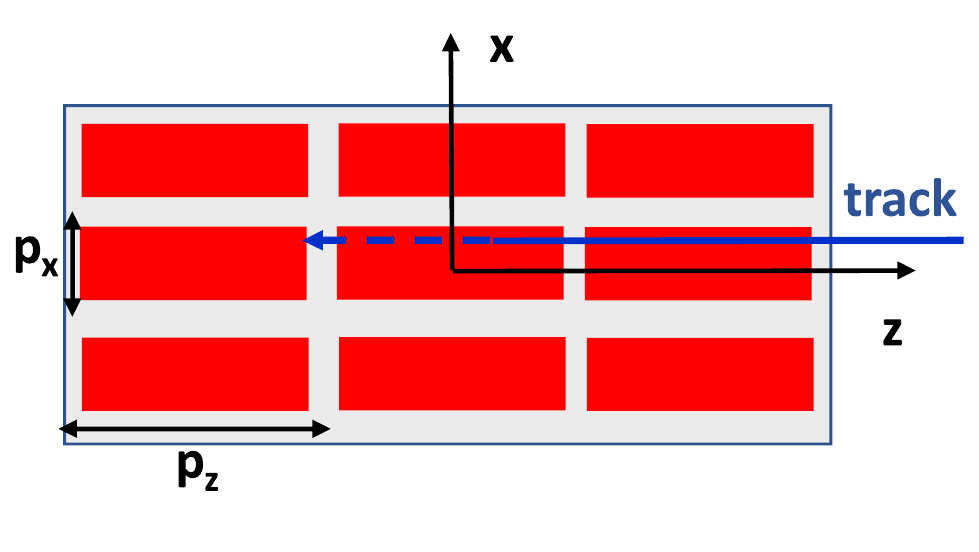}
    \caption{ }
    \label{fig:Fig_Sensor02}
   \end{subfigure}%
   \caption{Sketch of the simulated sensor and illustration of the coordinate system.
   The simulated tracks are uniformly distributed over the central pixel and located in the $x$ - $y$~plane with an angle $\theta $ to the $y$~axis.
   (a) Cross section, and
   (b) top view.   }
  \label{fig:Fig_Sensor}
 \end{figure}

 Fig.~\ref{fig:Fig_Sensor} shows a cross section and a top view of the simulated pixel sensor and a track traversing the central pixel.
 The track is in the $x$ - $y$~plane with an angle $\theta $ relative to the $y$~axis.
 For the electric field the 1D-field of a pad sensor of thickness, $d$, with constant doping, and a bias voltage above full depletion is assumed.
 For the simulation, the sensor is segmented into $n_y$ segments of equal thickness, $\Delta y$.
 For every segment a random number corresponding to the energy loss of a 5~GeV particle for a track length $\Delta y / \cos(\theta)$ is generated and the corresponding charge assigned to the centre of the segment.
 The energy loss model of Bichsel~\cite{Bichsel:1988} is implemented and details are given in Appendix~\ref{sect:Appendix_Eloss}.
 The electron clouds generated in the individual $\Delta y$~segments are drifted to the readout plane, taking into account the effects of diffusion.
 Their spatial distribution at the readout plane is integrated over the $x$ and the $z$~pitch of the pixels for obtaining the charges in the individual pixels.
 Electronics noise, threshold cuts and position reconstruction are performed in the analysis of the simulated events.
 The simulation program generates approximately 500~k~events in one minute on a standard PC.

 \section{Energy-loss fluctuations and position resolution}
  \label{sect:Efluctuations}

\begin{figure}[!ht]
   \centering
   \begin{subfigure}[a]{0.5\textwidth}
    \includegraphics[width=\textwidth]{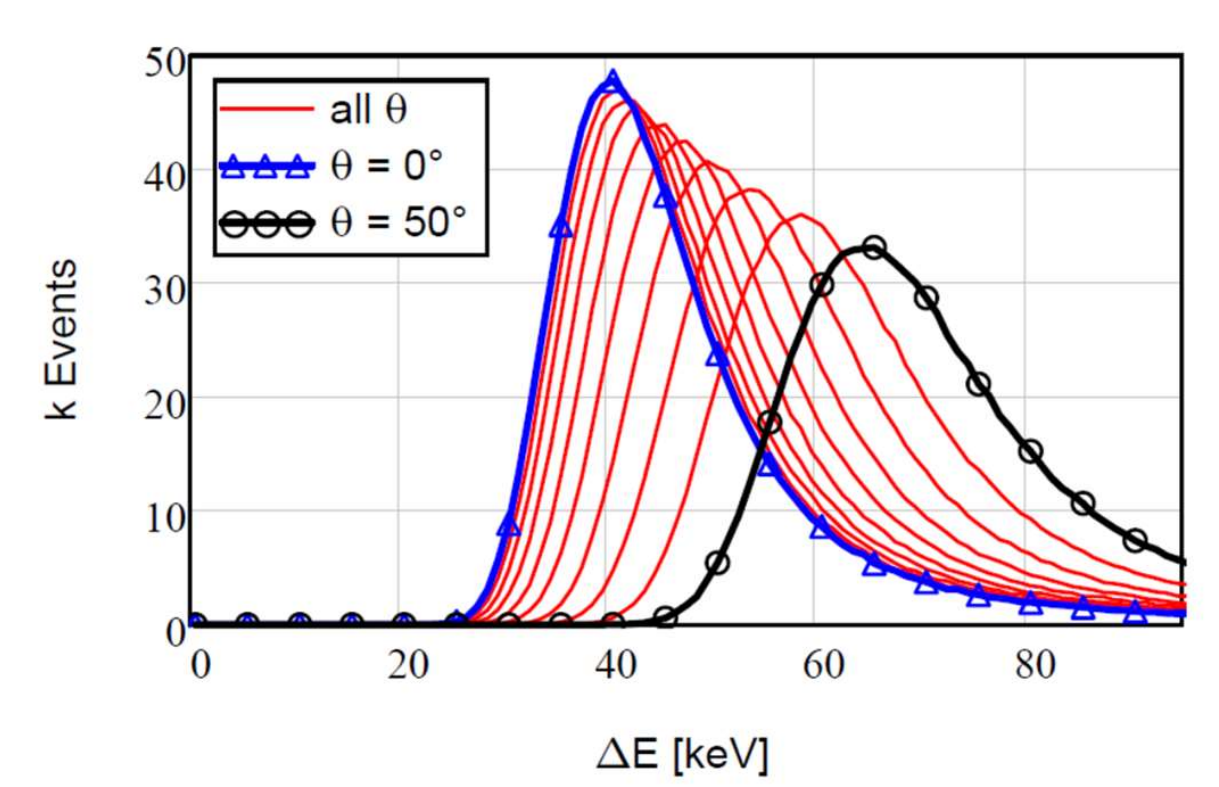}
    \caption{ }
    \label{fig:Fig_dEvsth}
   \end{subfigure}%
    ~
   \begin{subfigure}[a]{0.5\textwidth}
    \includegraphics[width=\textwidth]{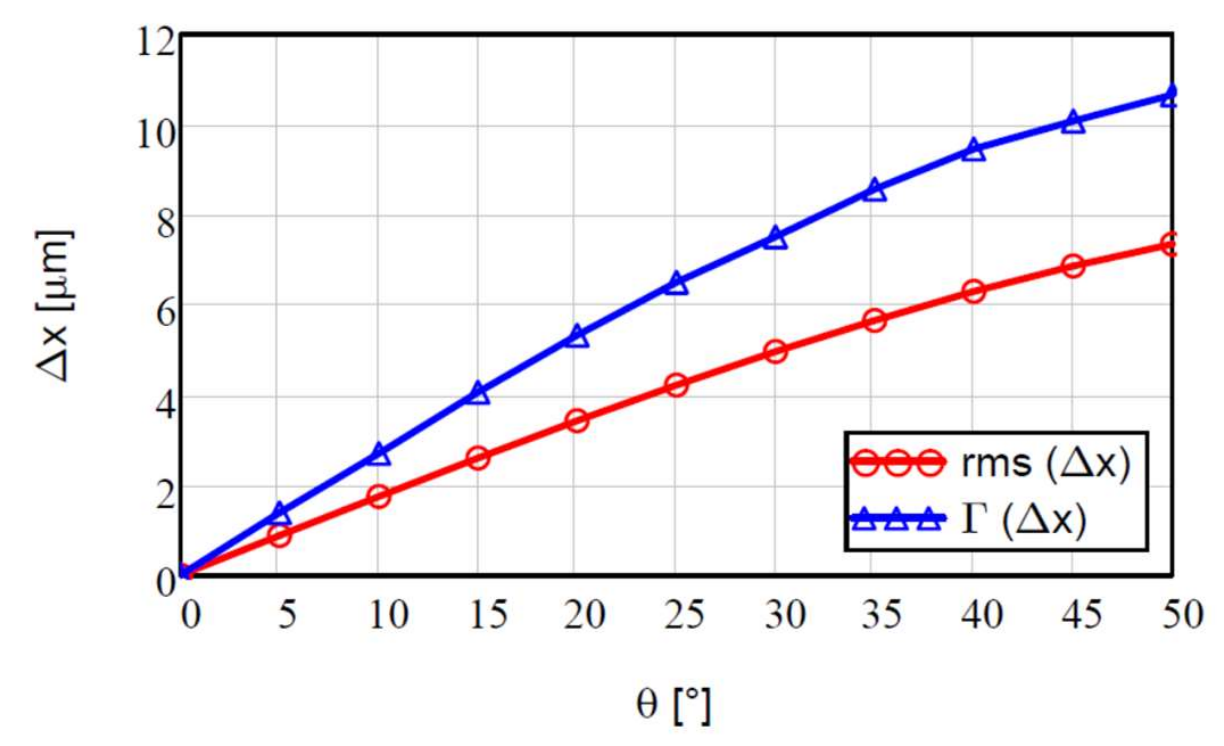}
    \caption{ }
    \label{fig:Fig_dxvsth}
   \end{subfigure}%
   \caption{ (a) Simulated $\mathrm{d}N/\mathrm{d} \Delta E$\, distribution for track angles between $0^{\,\circ}$ and $50^{\,\circ}$ in steps of $5^{\,\circ}$.
  (b) Contribution of the fluctuations of the deposited energy, $\Delta E (y)$, to the position resolution for track angles between $0^{\,\circ}$ and $50^{\,\circ}$.
 As the distribution has non-Gaussian tails, both the $\mathit{rms}$ and the full width at half maximum, $\Gamma $, are shown.
 An example of the effect of the tails on the \emph{rms}: For $\theta = 45^{\,\circ}$, $\Gamma = 10\,\upmu$m.
 The corresponding \emph{rms} for a Gaussian is $\sigma = \Gamma /2.35 = 4.3\,\upmu$m, whereas $\mathit{rms} = 6.8\,\upmu$m for the simulated $\Delta x$\,distribution shown in Fig.\,\ref{fig:Fig_dxth45}.
 The additional broadening due to the finite range of $\delta $-electrons is not taken into account in the simulation.
 }
  \label{fig:Fig_dEdxvsth}
 \end{figure}

\begin{figure}[!ht]
  \centering
    \includegraphics[width=0.5\textwidth]{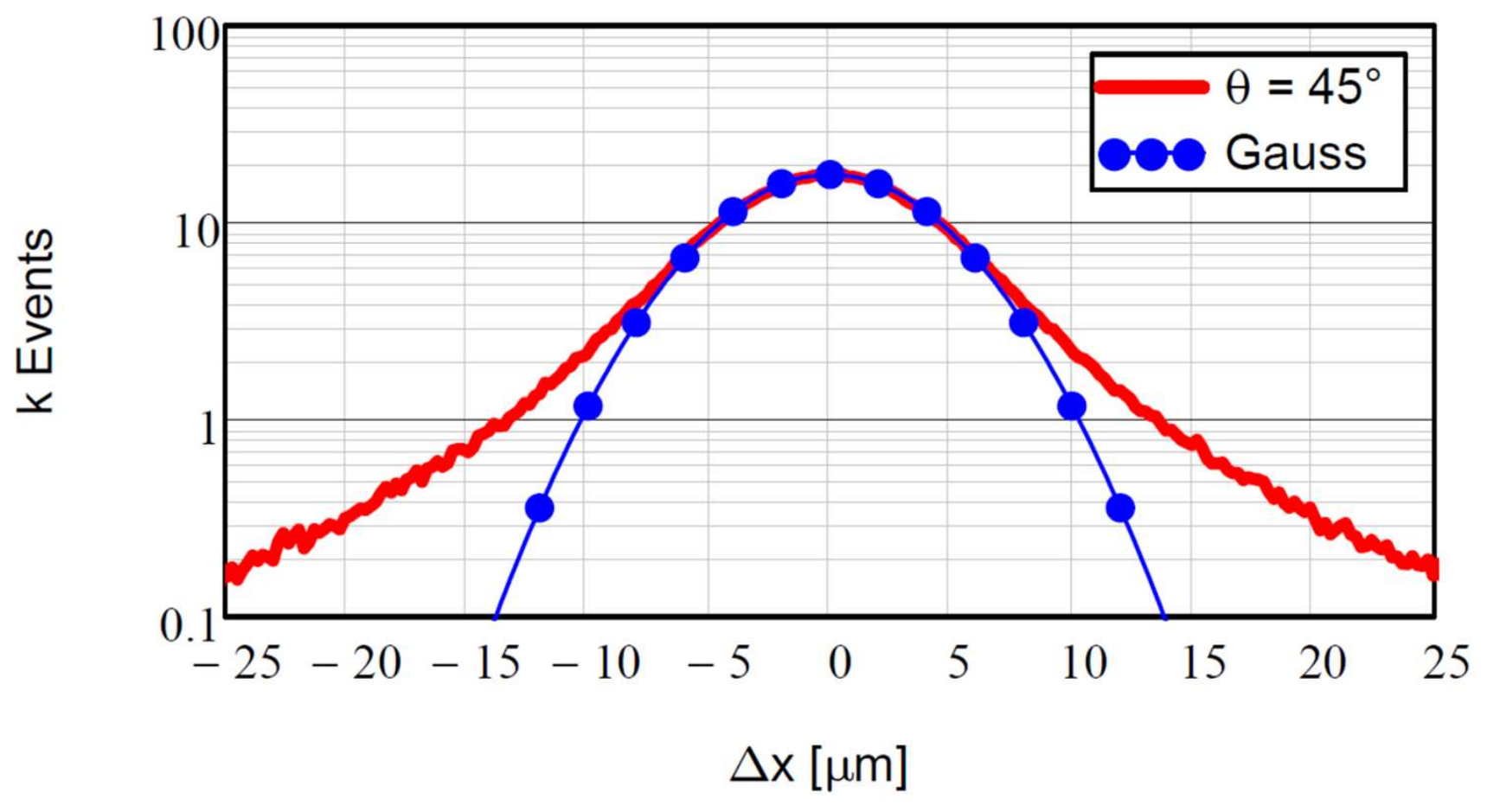}
    \caption{Influence of the local energy-loss fluctuations on the position resolution of segmented Si detectors;
 $\Delta x$, is the difference between the generated particle position and the centre-of-gravity of the charge distribution at the electrodes for a particle incident with an angle $\theta = 45^{\,\circ}$ on a $150\,\upmu$m thick sensor.
 To illustrate the non-Gaussian tails, the result of the simulation (solid line) is compared to a Gauss distribution with the same full width at half maximum (line with markers).}
  \label{fig:Fig_dxth45}
 \end{figure}

 In this section the worsening of the position resolution with track angle due to local energy-loss fluctuations is investigated.
 The Monte Carlo program discussed in the previous section is used to generate for tracks with $x = z = 0$ the energy deposited in the $n_y$ segments, and the energy-weighted mean position, $\langle y \rangle$, is calculated for every simulated event.
 The event-to-event fluctuations of $\langle y \rangle$ multiplied with $\sin(\theta )$
 are an estimate of the contribution of the energy-loss fluctuations to the spatial resolution.
 Results of $10^6$ events for a $150\,\upmu$m thick detector and $n_y = 15$ and $\theta $ between $0^{\,\circ}$ and $50^{\,\circ}$ are shown in Fig.\,\ref{fig:Fig_dEdxvsth}.
 Fig.\,\ref{fig:Fig_dEvsth} shows the  energy lost in the entire sensor which increases with $\theta $ because of the $\propto 1/\cos(\theta )$ increase in path length.
 In Fig.\,\ref{fig:Fig_dxvsth} the width of the distribution of $\Delta x = \langle y \rangle \cdot \sin(\theta ) $, which results from the position-dependent energy losses, is displayed.
 As shown in Fig.\,\ref{fig:Fig_dxth45}, the distribution of the fluctuations has significant non-Gaussian tails due to infrequent large local energy losses.
  Therefore, both the root-mean square, \emph{rms}, and the full width at half maximum, $\Gamma $, are presented.
 It has been verified that for $n_y \gtrsim 10$ the results do not depend on $n_y$.
 It is noted that the finite range of $\delta $-electrons is not taken into account in the simulation and their energy is deposited locally;
 $\delta $-electrons cause an additional broadening of the $\Delta x$~distribution.

 It is concluded that for a $150\,\upmu$m thick detector at angles above $ 10^{\,\circ}$ the contribution of the local fluctuations of energy loss to the spatial resolution can be important, and algorithms, like the head-tail algorithm\,\cite{Turchetta:1993}, which ignore the charge measured in the central electrode for large clusters, typically give superior results than the centre-of-gravity method.

 \section{Position resolution of silicon pixel sensors}
  \label{sect:Resolution}

  \subsection{Position resolution as function of threshold for track angles between $0^{\,\circ}$ and $32^{\,\circ}$}
   \label{sect:Threshold}

 In this section the spatial resolution for the centre-of-gravity algorithm and the method described in Sect.\,\ref{sect:Model} is compared for four angles in the range $\theta = 0 ^{\,\circ}$ to $32^{\,\circ}$, as a function of the threshold used for the cluster reconstruction.
 The sensor parameters for the simulation, defined in table\,\ref{tab:InputSim} of Appendix\,\ref{sect:Appendix_Signal}, are:
 $d = 150\,\upmu$m,
 $p_x = 25\,\upmu$m,
 $p_z = 100\,\upmu$m,
 $n_{pz} = 3$,
 $n_y = 15$,
 $\Delta x_{min} = 2\,\upmu$m,
 $N_d = 4.5 \times 10^{12}$\,cm$^{-3}$,
 $U = 150$\,V,
 $\sigma _\mathit{el} = 275$\,e (elementary charges),
 $\Delta E _\mathit{max} = 2$, and
 $T = 20^{\,\circ}$C.
 The parameters correspond to prototype sensors developed for the luminosity upgrade of the CMS Inner Tracker\,\cite{CMS:2017}.
 For $\theta = 0^{\,\circ}$ and $10^{\,\circ}$, $n_{px}$ is set to 3, and for the larger angles to 5, to account for the increase of the width of the charge distribution collected by the pixels with $\theta$.
 The threshold, \emph{thr}, for assigning signals to a cluster is varied between 0 and 1100~e (elementary charges) in steps of 220~e, corresponding to 0, 2\,\%, 4\,\% up to 10\,\% of the most probable value of the charge distribution from the entire sensor at normal incidence ($\mathit{MPV} = 11\,000$).

 \begin{figure}[!ht]
   \centering
   \begin{subfigure}[a]{0.5\textwidth}
    \includegraphics[width=\textwidth]{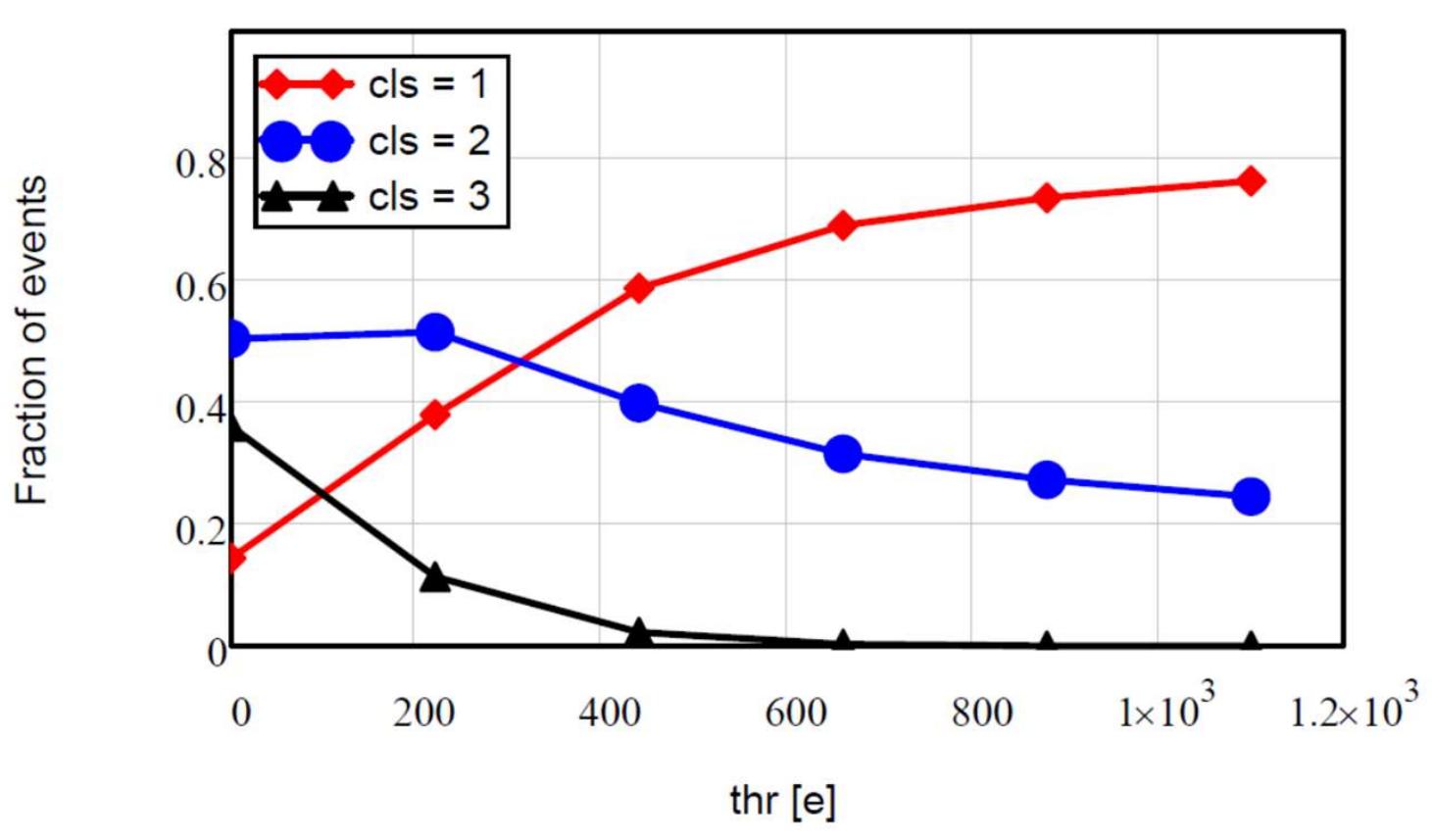}
    \caption{ }
    \label{fig:Fig_clsth0}
   \end{subfigure}%
    ~
   \begin{subfigure}[a]{0.5\textwidth}
    \includegraphics[width=\textwidth]{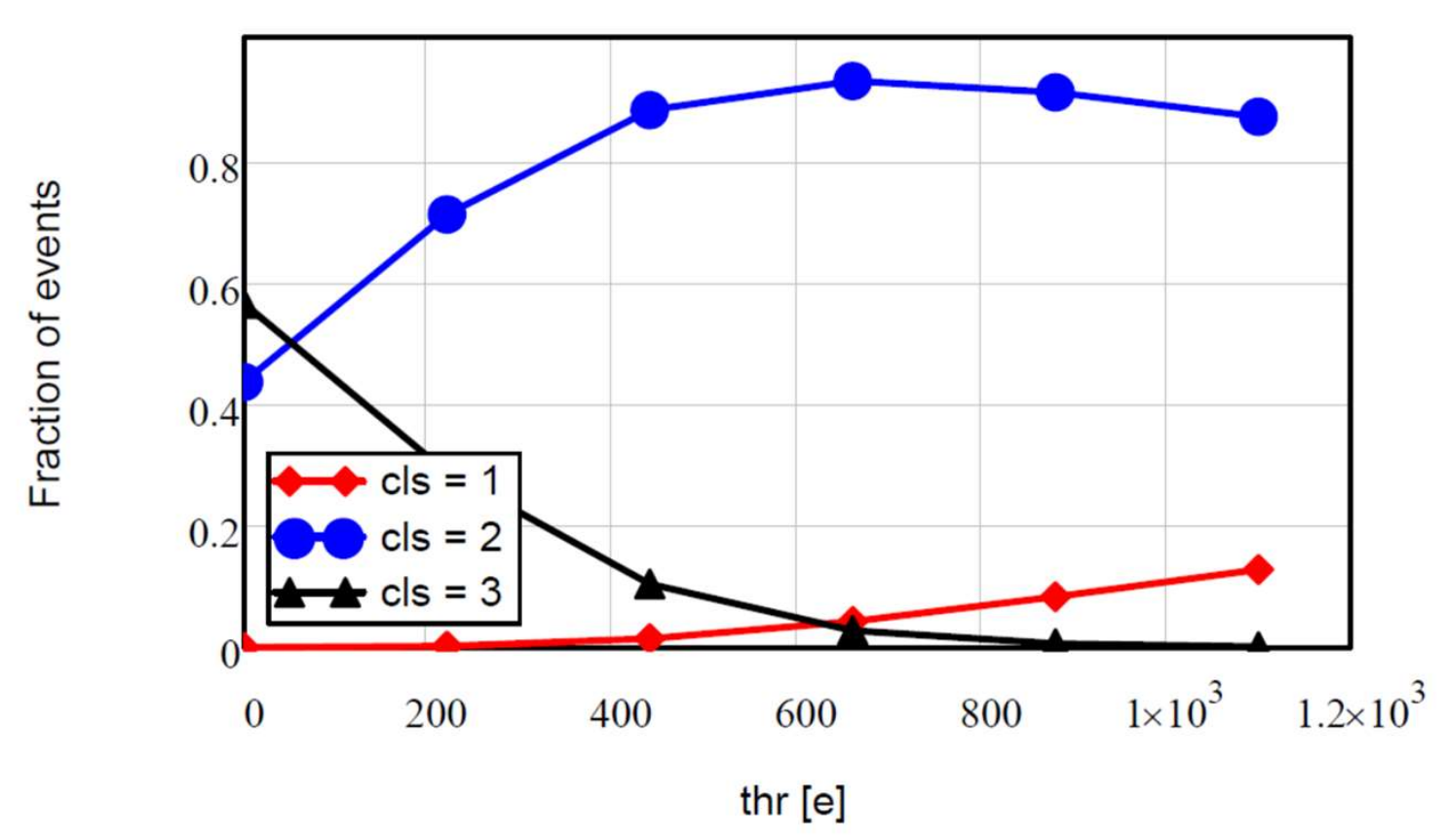}
    \caption{ }
    \label{fig:Fig_clsth10}
   \end{subfigure}%
  \newline
  \centering
   \begin{subfigure}[a]{0.5\textwidth}
    \includegraphics[width=\textwidth]{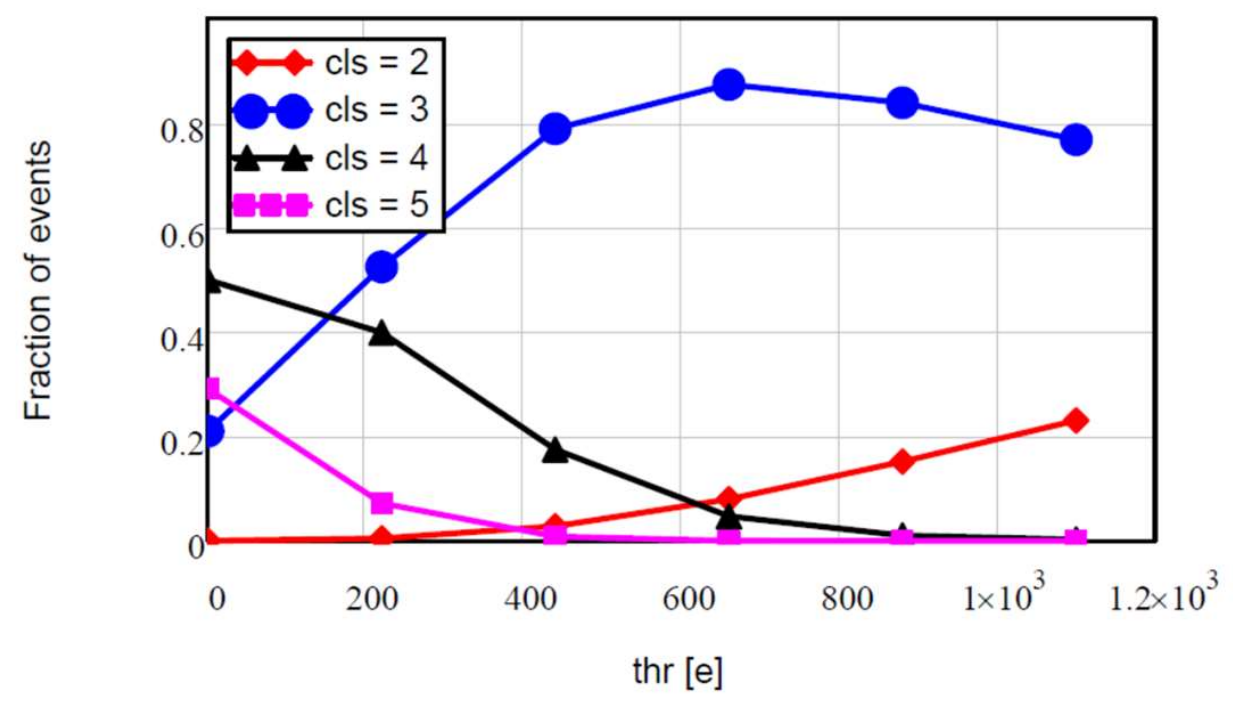}
    \caption{ }
    \label{fig:Fig_clsth20}
   \end{subfigure}%
    ~
   \begin{subfigure}[a]{0.5\textwidth}
    \includegraphics[width=\textwidth]{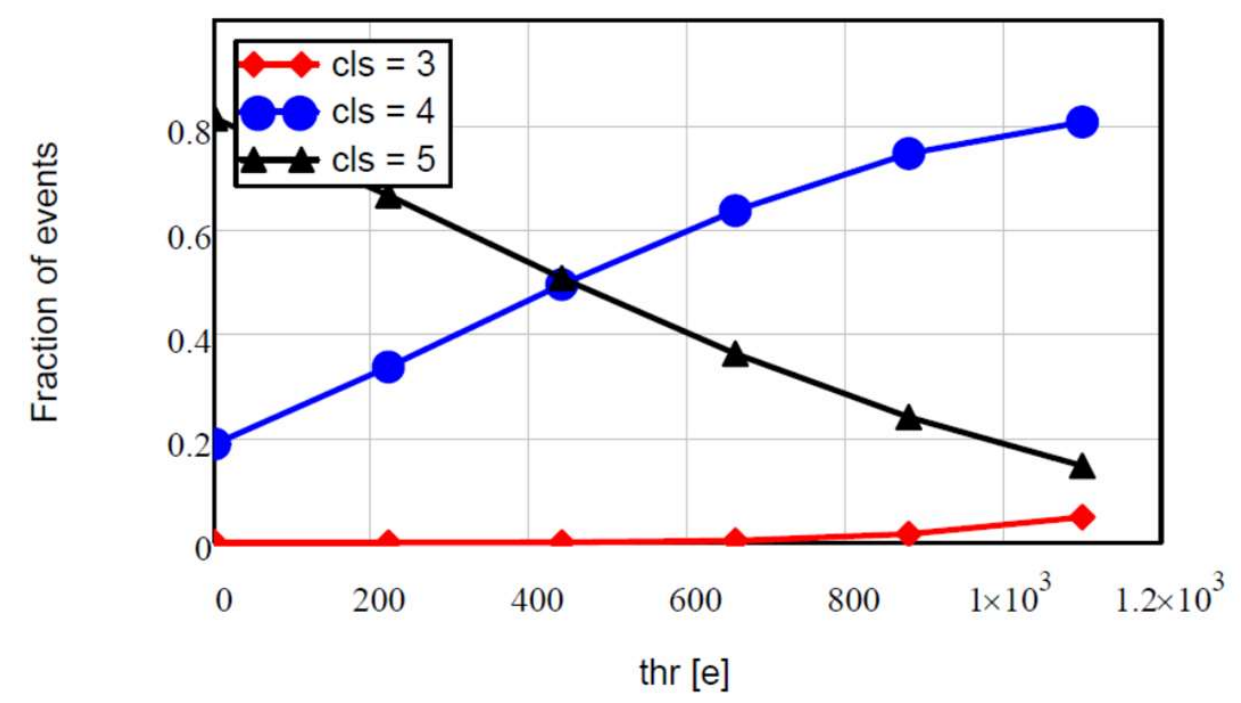}
    \caption{ }
    \label{fig:Fig_clsth32}
   \end{subfigure}%
   \caption{Fraction of events with  cluster size $\mathit{cls} = 1$ to 5 as a function of threshold, $\mathit{thr}$, for tracks with angles of
    (a) $0^{\,\circ}$,
    (b) $10^{\,\circ}$,
    (c) $20^{\,\circ}$, and
    (d) $32^{\,\circ}$ to the sensor normal.
  The threshold is given in units of elementary charges, e, and the most probable signal charge at $0^{\,\circ}$ is $\approx 11\,000$\,e. }
  \label{fig:Fig_clsth}
 \end{figure}

 Fig.\,\ref{fig:Fig_clsth} shows the fraction of events with different projected cluster sizes, \emph{cls}, as a function of threshold, $\mathit{thr}$ for the four $\theta$~values.
 The cluster size \emph{cls} is obtained by adding the signals with the same $z$~values and counting only those which are above \emph{thr}.
 For $\theta = 0^{\,\circ}$ and $\mathit{thr} \geq 400$~e the $\mathit{cls}= 1$ events,
 for $\theta = 10^{\,\circ}$ the $\mathit{cls}= 2$ events, and
 for $\theta = 20^{\,\circ}$ the $\mathit{cls} = 3$ events dominate.
 For $\theta = 32^{\,\circ}$, most events have $cls = 4$ for small and $\mathit{cls} = 3$ for large $\emph{thr}$.

 \begin{figure}[!ht]
   \centering
   \begin{subfigure}[a]{0.5\textwidth}
    \includegraphics[width=\textwidth]{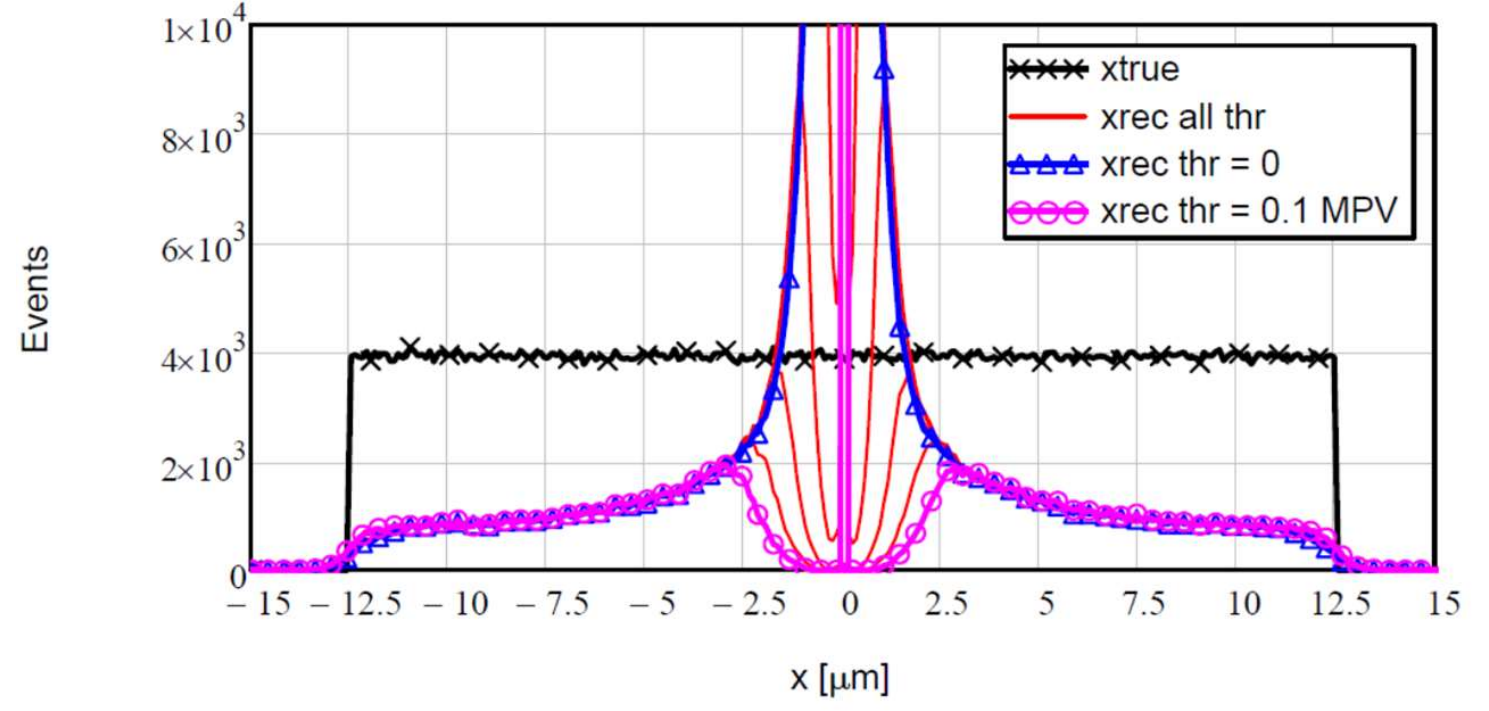}
    \caption{ }
    \label{fig:Fig_xrecth0}
   \end{subfigure}%
    ~
   \begin{subfigure}[a]{0.5\textwidth}
    \includegraphics[width=\textwidth]{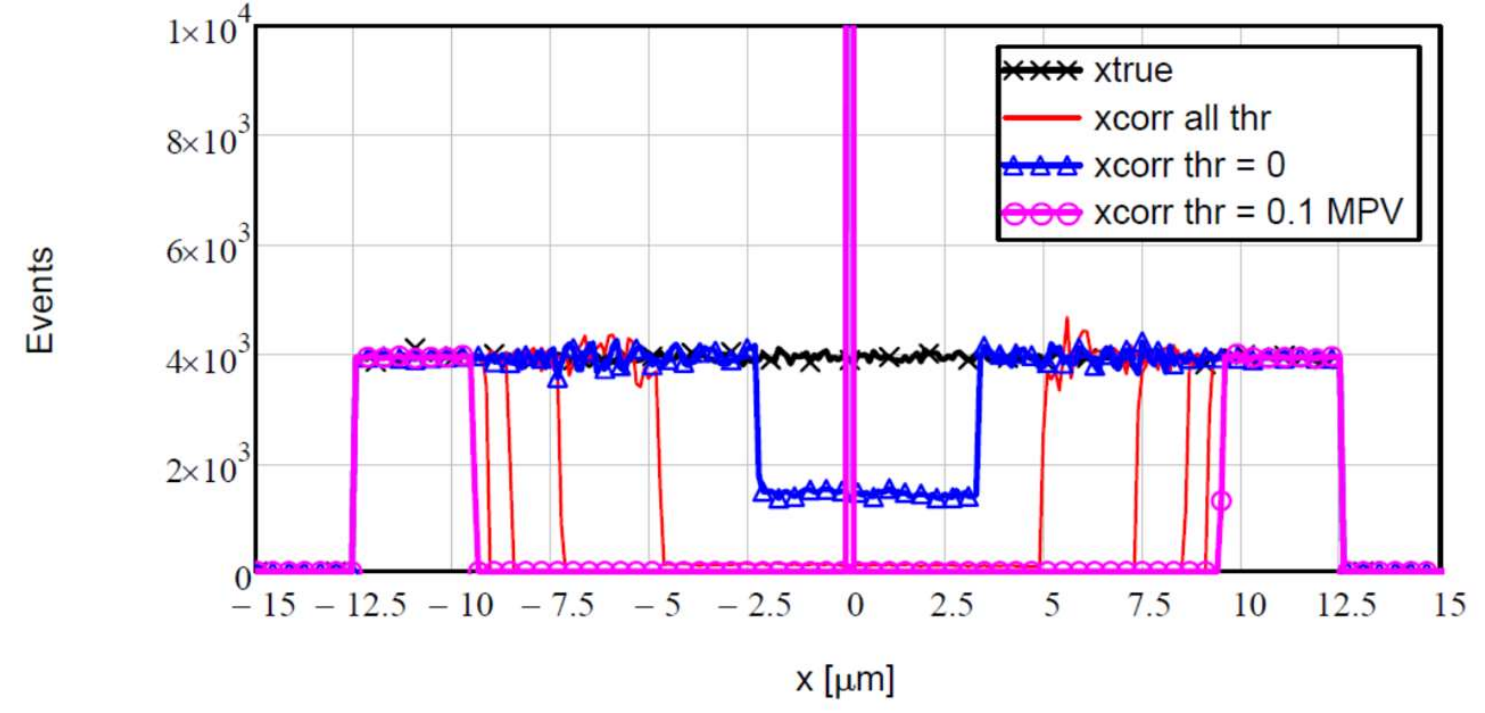}
    \caption{ }
    \label{fig:Fig_xcorth0}
   \end{subfigure}%
  \newline
  \centering
   \begin{subfigure}[a]{0.5\textwidth}
    \includegraphics[width=\textwidth]{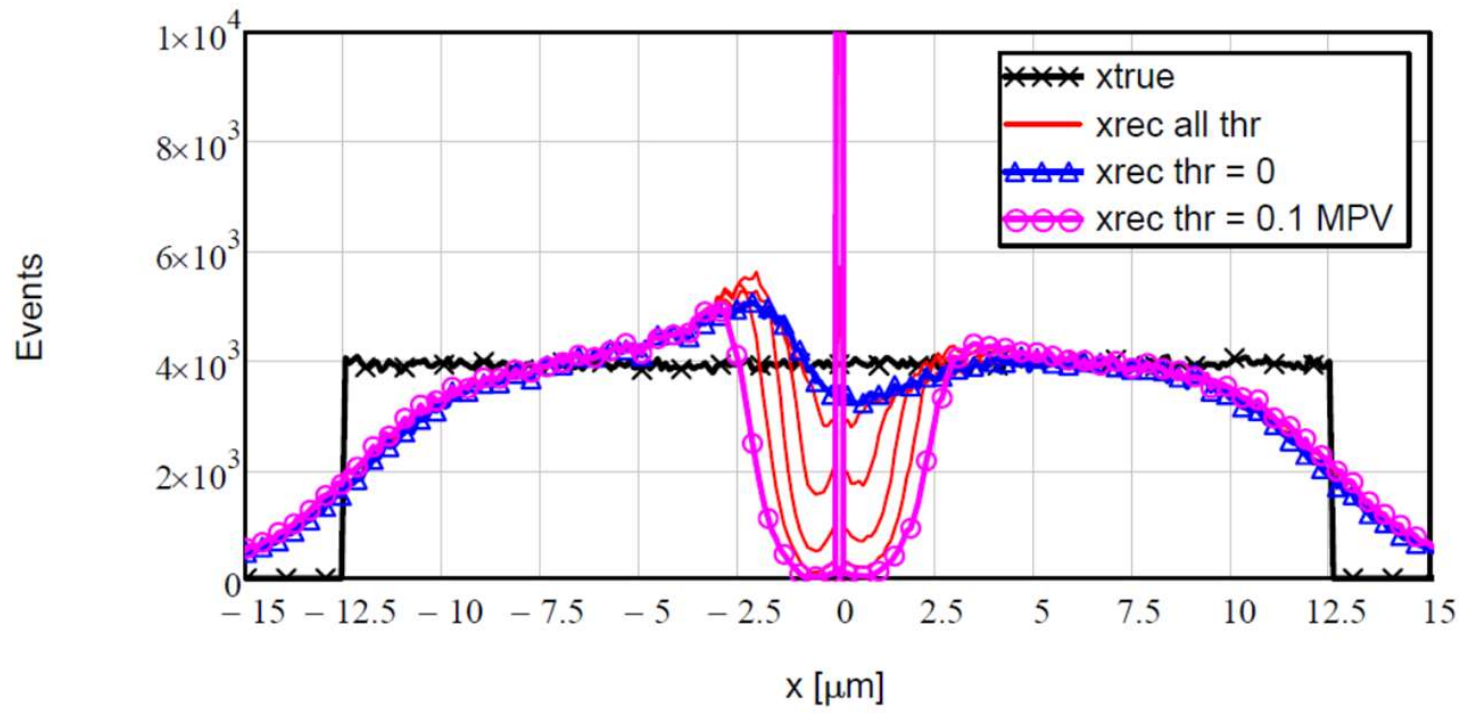}
    \caption{ }
    \label{fig:Fig_xrecth10}
   \end{subfigure}%
    ~
   \begin{subfigure}[a]{0.5\textwidth}
    \includegraphics[width=\textwidth]{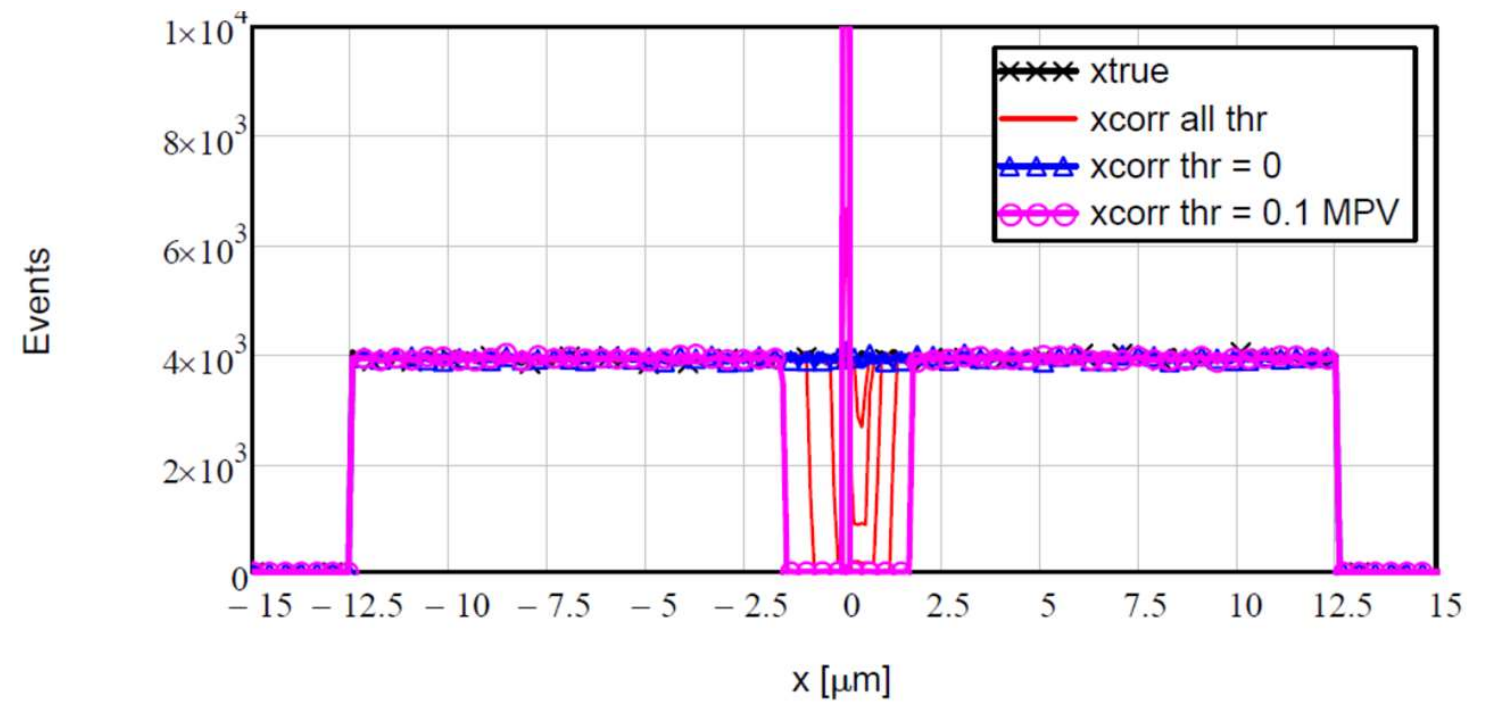}
    \caption{ }
    \label{fig:Fig_xcorth10}
   \end{subfigure}%
  \newline
  \centering
   \begin{subfigure}[a]{0.5\textwidth}
    \includegraphics[width=\textwidth]{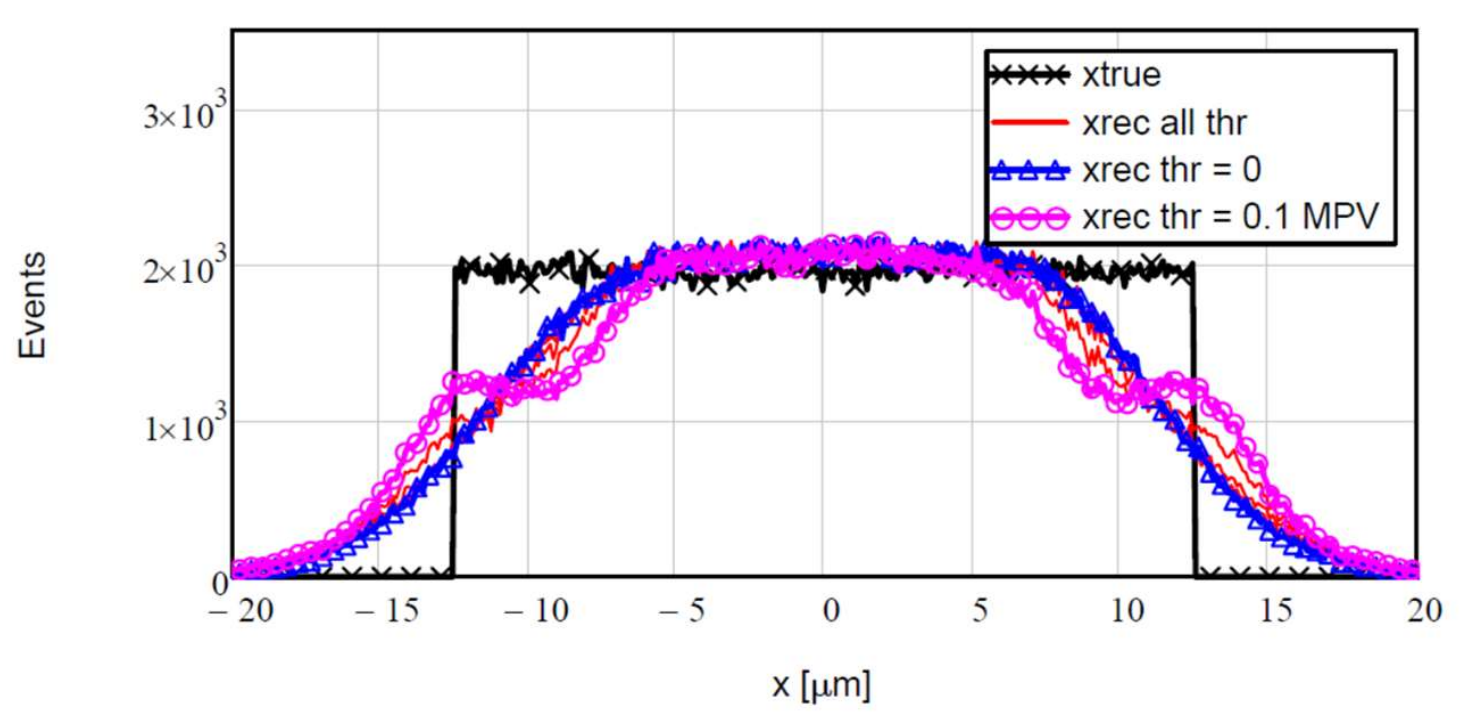}
    \caption{ }
    \label{fig:Fig_xrecth20}
   \end{subfigure}%
    ~
   \begin{subfigure}[a]{0.5\textwidth}
    \includegraphics[width=\textwidth]{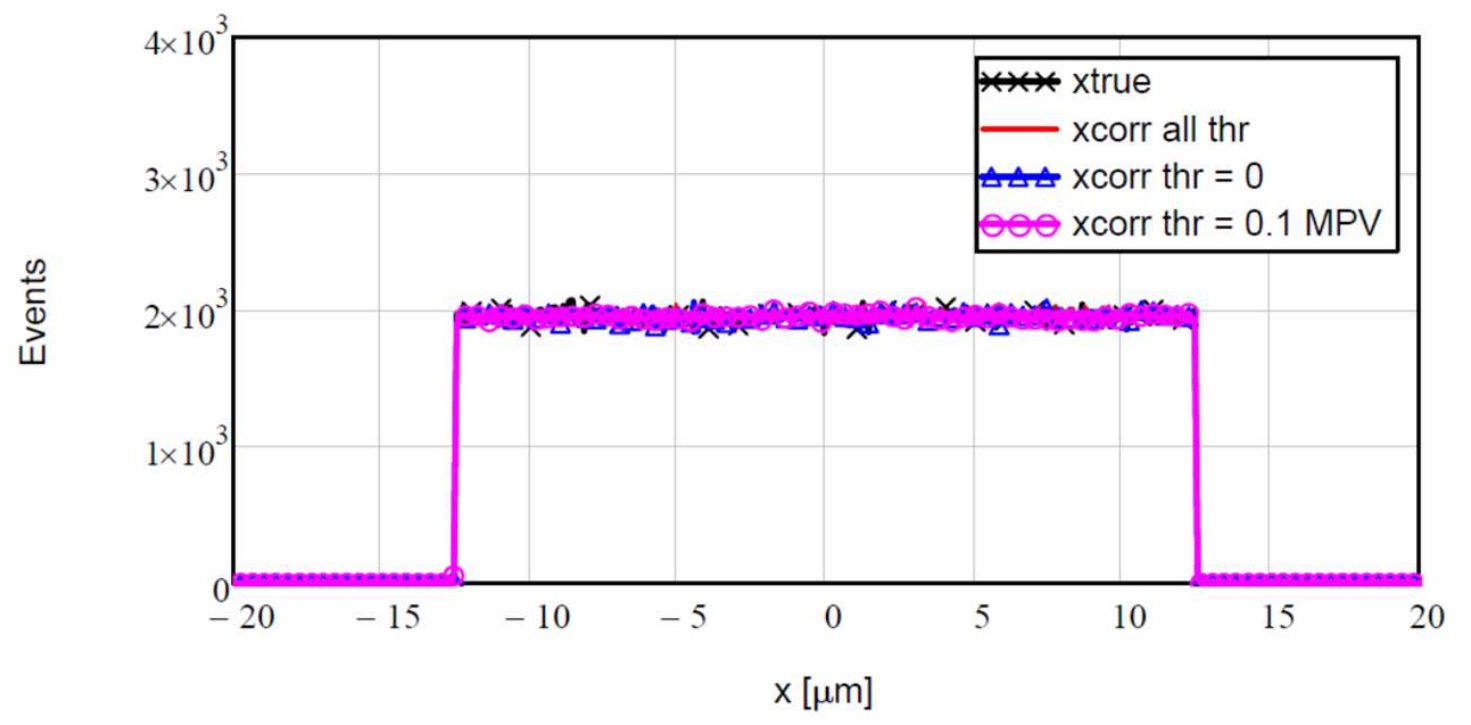}
    \caption{ }
    \label{fig:Fig_xcorth20}
   \end{subfigure}%
  \newline
  \centering
   \begin{subfigure}[a]{0.5\textwidth}
    \includegraphics[width=\textwidth]{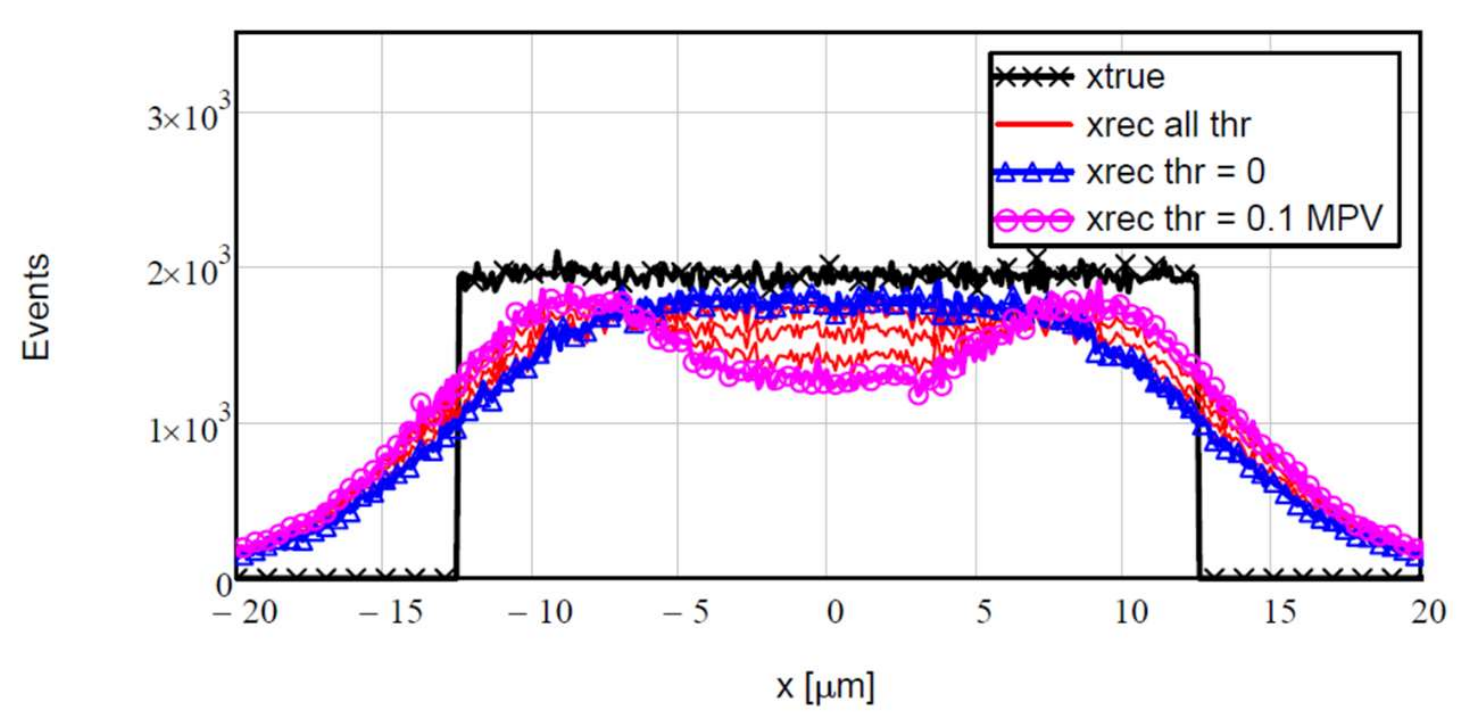}
    \caption{ }
    \label{fig:Fig_xrecth32}
   \end{subfigure}%
    ~
   \begin{subfigure}[a]{0.5\textwidth}
    \includegraphics[width=\textwidth]{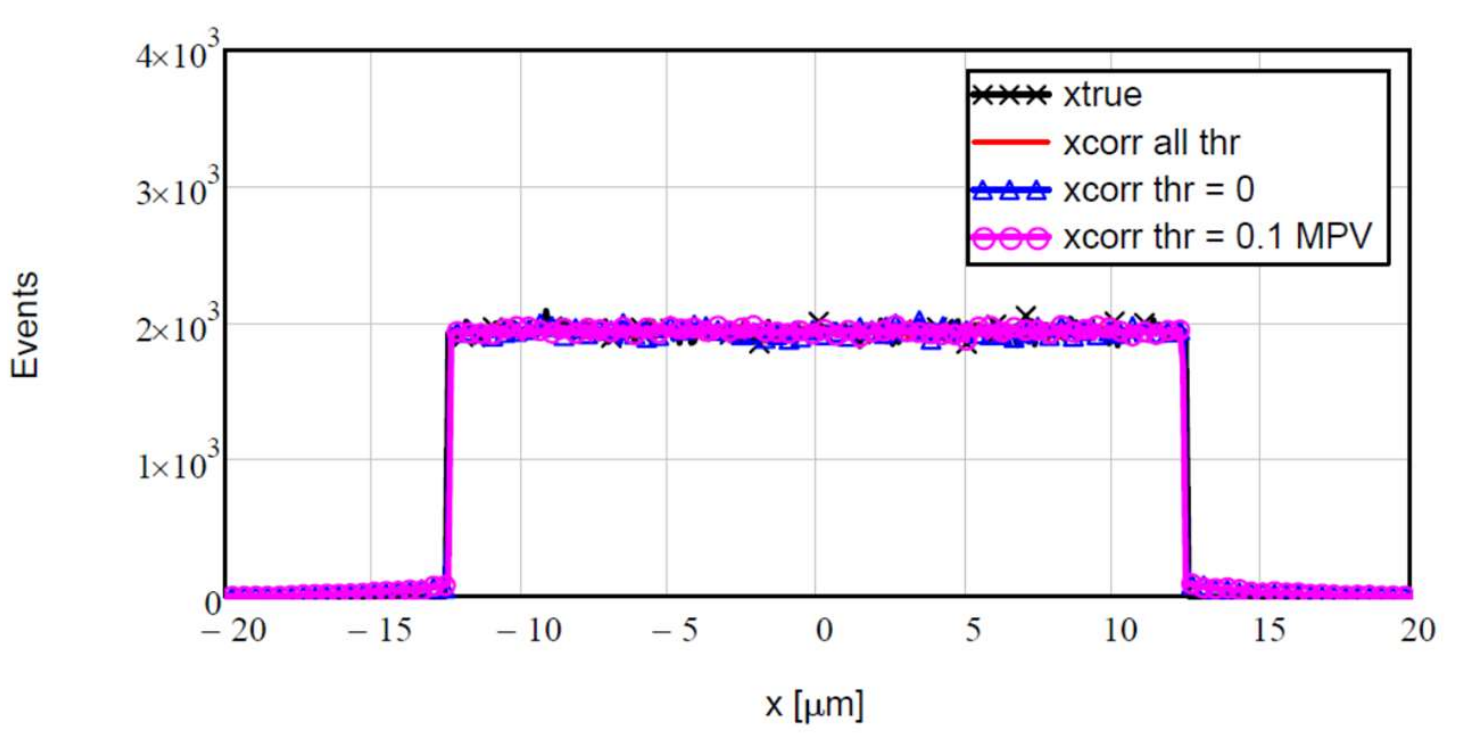}
    \caption{ }
    \label{fig:Fig_xcorth32}
   \end{subfigure}%

   \caption{Distributions $\mathrm{d}N/\mathrm{d}x_\mathit{rec}$, as a function of $\mathit{thr}$ for
   (a) $\theta = 0^{\,\circ}$,
   (c) $10^{\,\circ}$,
   (e) $20^{\,\circ}$,
   (g) $32^{\,\circ}$,
       and of $\mathrm{d}N/\mathrm{d}x_\mathit{corr}$ for
   (b) $\theta = 0^{\,\circ}$,
   (d) $10^{\,\circ}$,
   (f) $20^{\,\circ}$,
   (h) $32^{\,\circ}$.
   The thin red lines show the distributions for $\mathit{thr} = $ 2, 4, 6, and 8\,\% of the most probable value (MPV) of the charge distribution for $\theta = 0^{\,\circ}$, and thicker ones for $\mathit{thr} = 0$ and 10\,\%. }
  \label{fig:Fig_xth}
 \end{figure}

 In Fig.~\ref{fig:Fig_xth} the $x$~distribution of the simulated events reconstructed with the centre-of-gravity algorithm, $x_\mathit{rec}$, and with the method discussed in Sect.\,\ref{sect:Model}, $x_\mathit{corr}$, are compared to the flat $x_\mathit{true}$~distributions generated.
 The distributions for the $\mathit{thr} = 2$~\% to 8~\% values are shown as narrow lines, the ones for 0 and 10~\% with symbols.

 Figs.~\ref{fig:Fig_xrecth0} and \ref{fig:Fig_xcorth0} show the comparison for $\theta = 0^{\,\circ}$.
 For $cls = 1$, both $x_\mathit{rec}$ and $x_\mathit{corr}$ are reconstructed at $ x = 0$.
 The $x = 0$ bins are cut off in the histograms, so that the remaining events are visible on a linear scale.
 The regions adjacent to $x = 0$ are depleted of events, with a gap which increases with \emph{thr}, as already discussed in Sect.~\ref{sect:Model}.
 For $\mathit{cls} = 2 $ and $x_\mathit{rec}$ the number of events reconstructed at higher $|x|$~values are much smaller than the ones generated:
 They are reconstructed at $|x_\mathit{rec}| \ll  |x_\mathit{true}|$.
 As shown in Fig.~\ref{fig:Fig_xcorth0} the proposed method is able to correct this.

 The comparison for $\theta = 10^{\,\circ}$ are shown in Figs.~\ref{fig:Fig_xrecth10} and \ref{fig:Fig_xcorth10}.
 For $\theta = 10^{\,\circ}$ and $\mathit{thr} \geq 400$~e about 90~\% of the events have $\mathit{cls} = 2$ (Fig.~\ref{fig:Fig_clsth10}), and the number of events in the $x=0$~bin is reduced to about 10~\%.
 For the centre-of-gravity reconstruction, the $\mathit{cls} = 2 $ events are reconstructed correctly on average, however their $x$~distribution shows a complicated structure.
 The asymmetry for small $|x_\mathit{rec}|$ values is caused by the diffusion effect discussed in Sect.~\ref{sect:Qsurface}.
 The deviations at $|x_\mathit{rec}| \approx p_x / 2$ are mainly caused by electronics noise and the bias of the reconstruction method.
 The proposed method manages to reproduce the flat distribution of the generated events.
 Although it is expected to correct for the asymmetric bias in the central region, it can not correct for the effects of random electronics noise for $|x| \approx p_x /2$.
 Thus only a minor improvement of the position resolution can be expected.
 Similar observations are made for the $\theta = 20^{\,\circ}$ and $\theta = 32^{\,\circ}$ results.


 \begin{figure}[!ht]
   \centering
   \begin{subfigure}[a]{0.5\textwidth}
    \includegraphics[width=\textwidth]{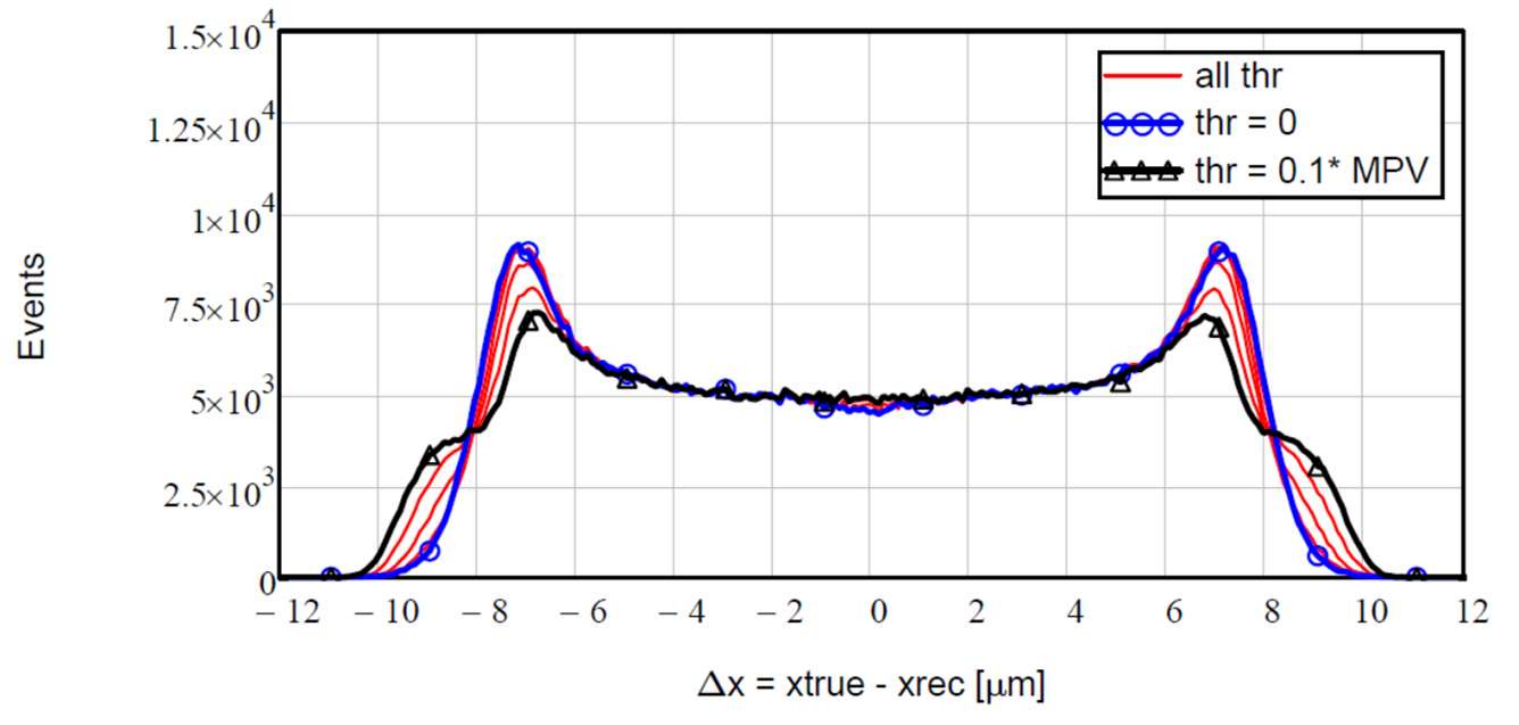}
    \caption{ }
    \label{fig:Fig_dxrecth0}
   \end{subfigure}%
    ~
   \begin{subfigure}[a]{0.5\textwidth}
    \includegraphics[width=\textwidth]{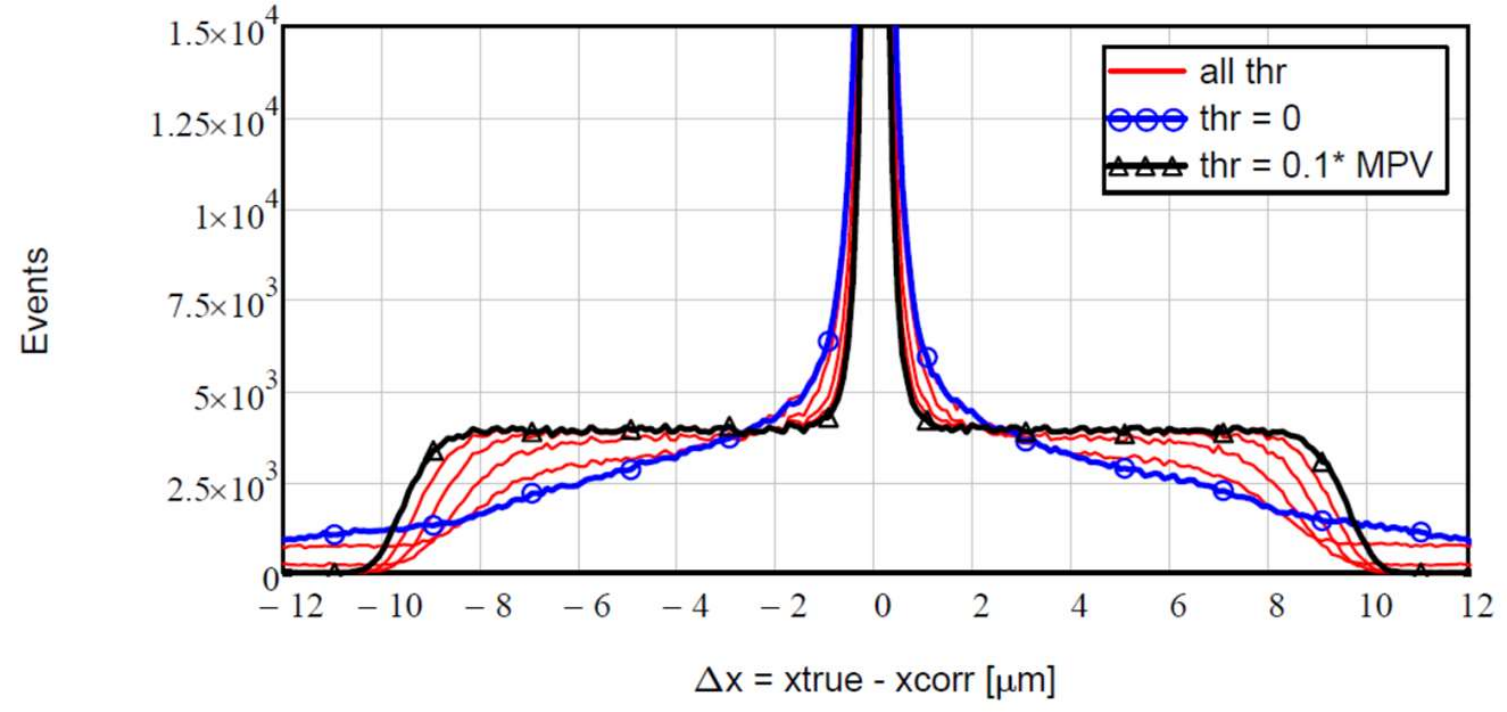}
    \caption{ }
    \label{fig:Fig_dxcorth0}
   \end{subfigure}%
  \newline
  \centering
   \begin{subfigure}[a]{0.5\textwidth}
    \includegraphics[width=\textwidth]{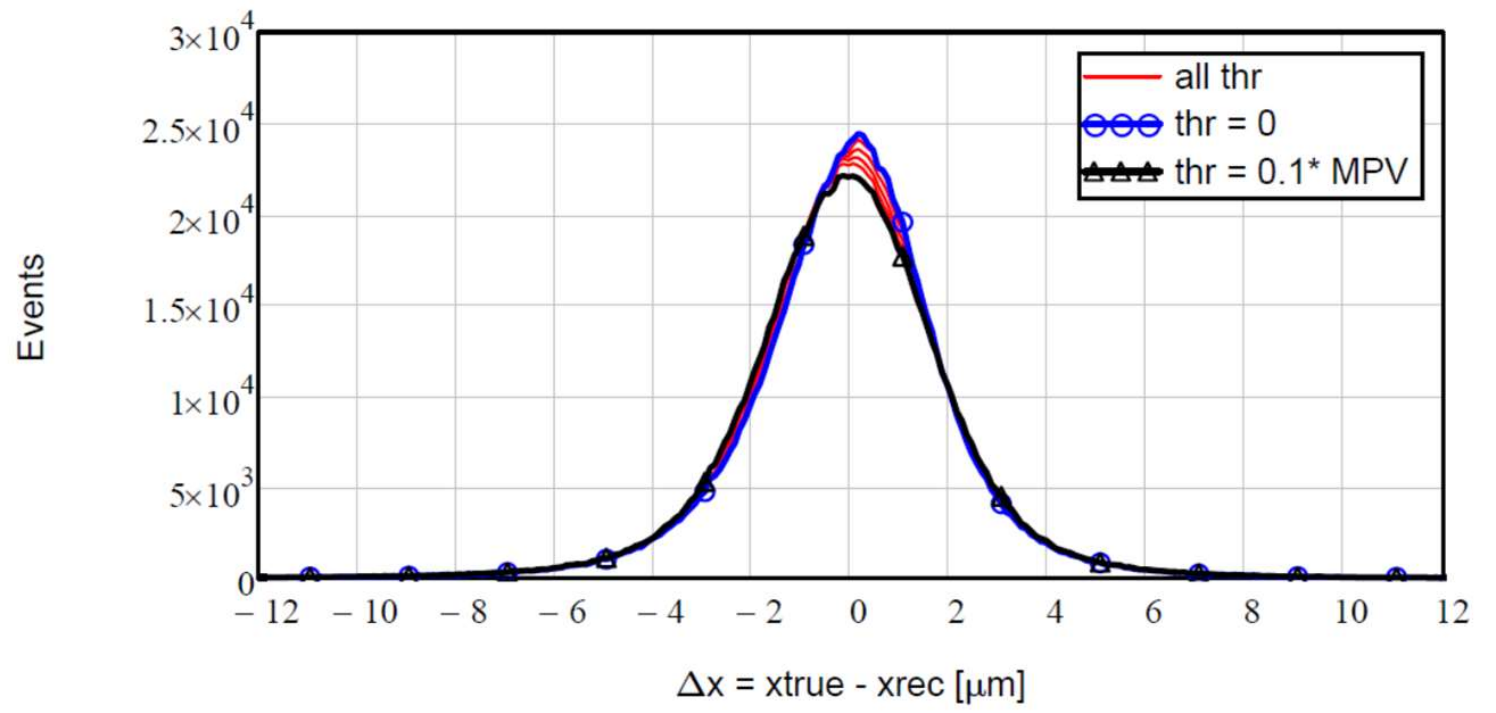}
    \caption{ }
    \label{fig:Fig_dxrecth10}
   \end{subfigure}%
    ~
   \begin{subfigure}[a]{0.5\textwidth}
    \includegraphics[width=\textwidth]{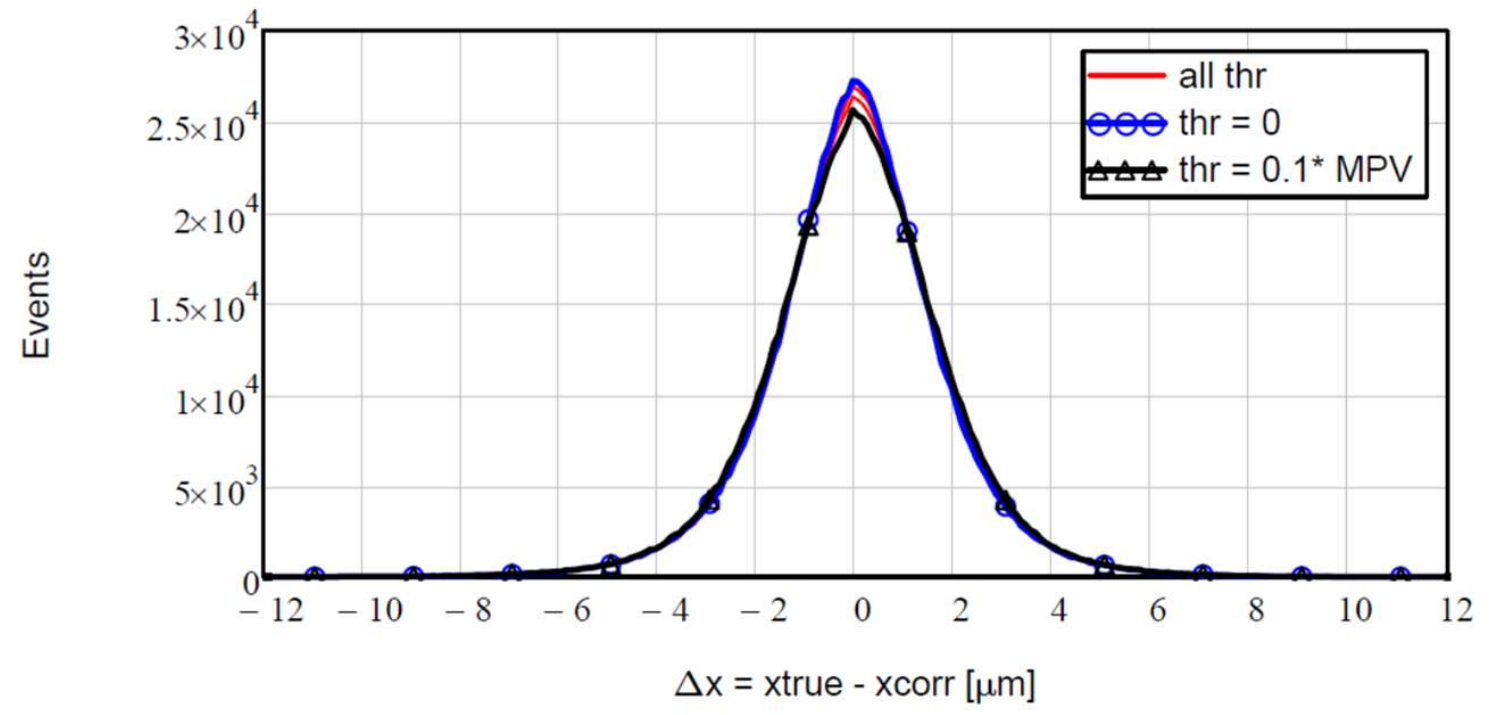}
    \caption{ }
    \label{fig:Fig_dxcorth10}
   \end{subfigure}%
  \newline
  \centering
   \begin{subfigure}[a]{0.5\textwidth}
    \includegraphics[width=\textwidth]{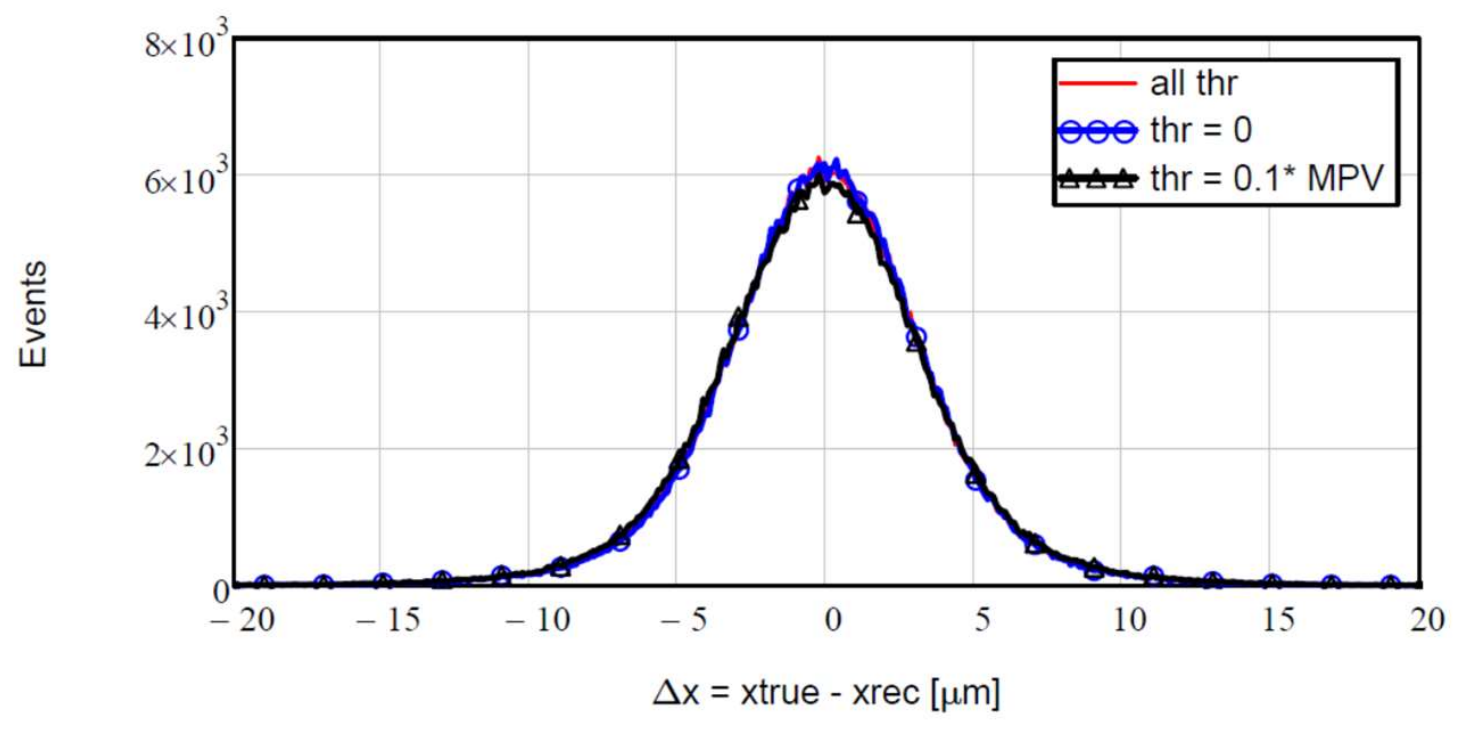}
    \caption{ }
    \label{fig:Fig_dxrecth20}
   \end{subfigure}%
    ~
   \begin{subfigure}[a]{0.5\textwidth}
    \includegraphics[width=\textwidth]{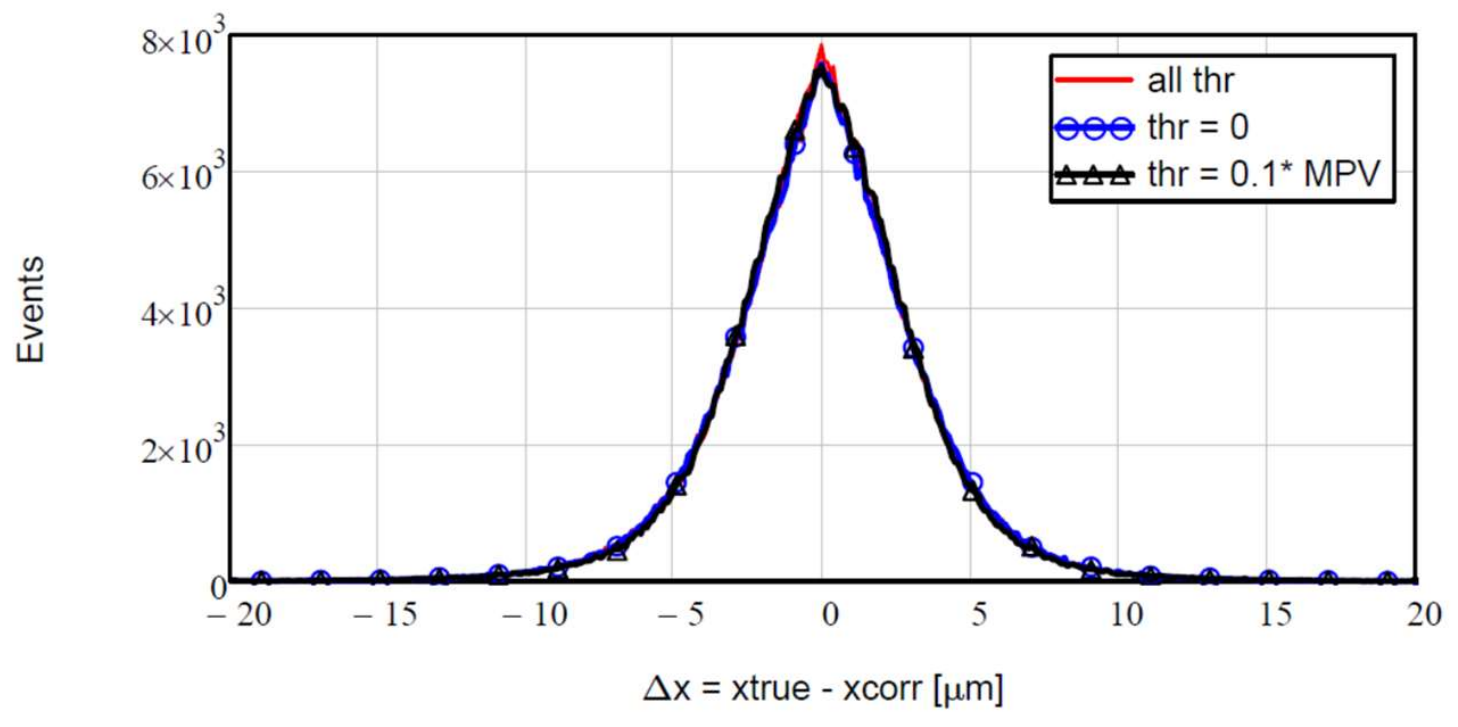}
    \caption{ }
    \label{fig:Fig_dxcorth20}
   \end{subfigure}%
  \newline
  \centering
   \begin{subfigure}[a]{0.5\textwidth}
    \includegraphics[width=\textwidth]{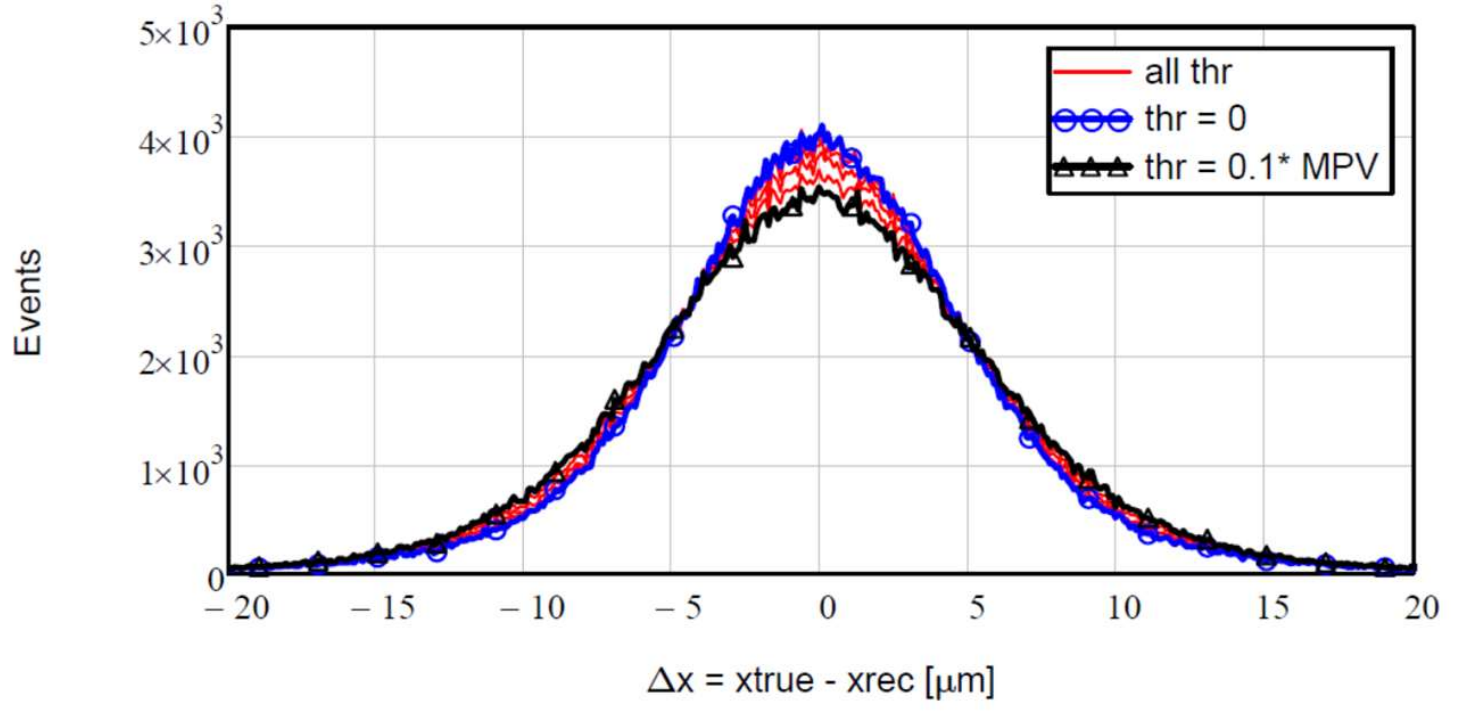}
    \caption{ }
    \label{fig:Fig_dxrecth32}
   \end{subfigure}%
    ~
   \begin{subfigure}[a]{0.5\textwidth}
    \includegraphics[width=\textwidth]{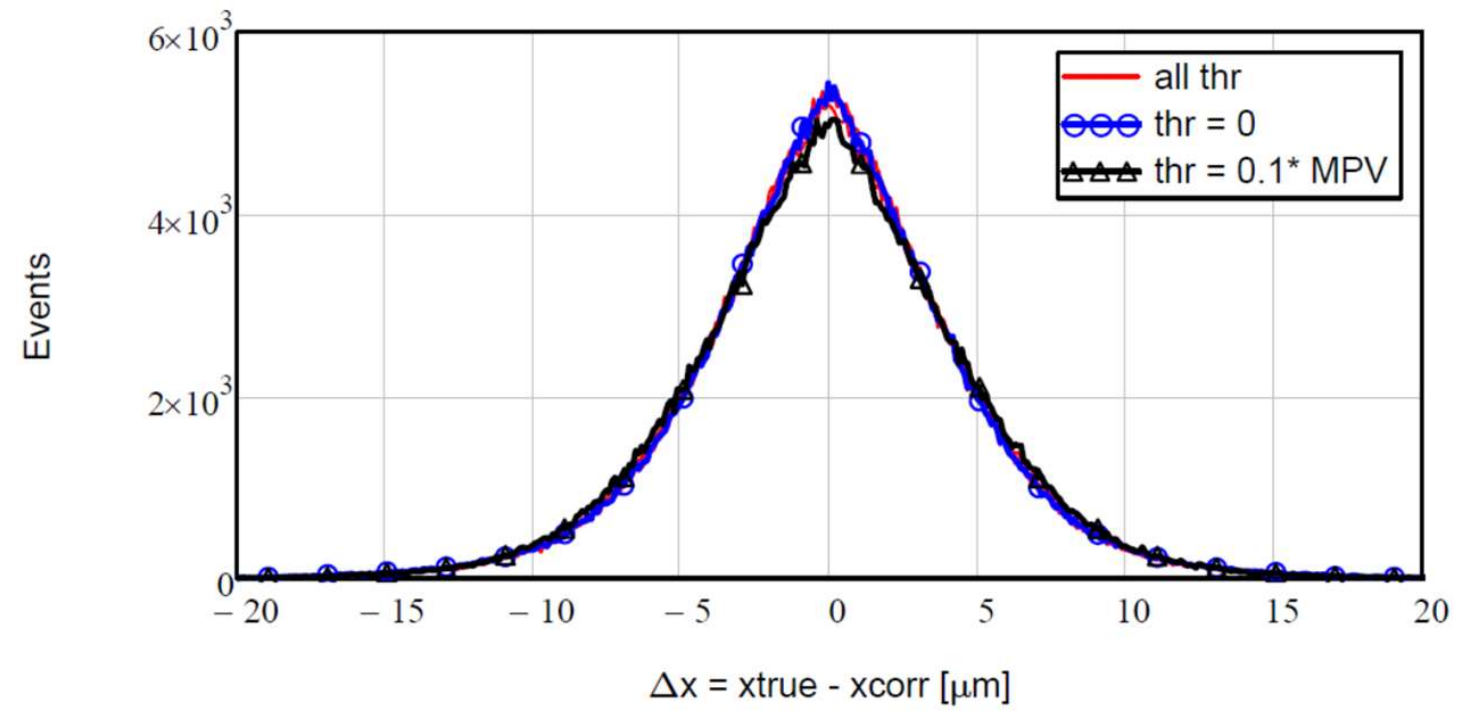}
    \caption{ }
    \label{fig:Fig_dxcorth32}
   \end{subfigure}%

    \caption{Distributions of the differences $\Delta x = x_\mathit{true} - x_\mathit{rec}$ as a function of $\mathit{thr}$ for
    (a) $\theta = 0^{\,\circ}$,
    (c) $10^{\,\circ}$,
    (e) $20^{\,\circ}$,
    (g) $32^{\,\circ}$,
       and of $\Delta x = x_\mathit{true} - x_\mathit{corr}$ for
    (b) $\theta = 0^{\,\circ}$,
    (d) $10^{\,\circ}$,
    (f) $20^{\,\circ}$,
    (h) $32^{\,\circ}$.
   The thin red lines show the distributions for $\mathit{thr} = $ 2, 4, 6, and 8\,\% of the most probable value (MPV) of the charge distribution for $\theta = 0^{\,\circ}$, and thicker ones for $\mathit{thr} = 0$ and 10\,\%. }
  \label{fig:Fig_dxth}
 \end{figure}

 Fig.\,\ref{fig:Fig_dxth} shows the position-response functions, $\mathrm{d}N/\mathrm{d} \Delta x$,
 for $\Delta x _\mathit{rec} = x _\mathit{true} - x _\mathit{rec}$ and
 for $\Delta x _\mathit{corr} = x _\mathit{true}- x _\mathit{corr}$ of the data shown in Fig.\,\ref{fig:Fig_xth}.
 In the literature $\Delta x $ is frequently called \emph{residual}.

\begin{figure}[!ht]
   \centering
   \begin{subfigure}[a]{0.5\textwidth}
    \includegraphics[width=\textwidth]{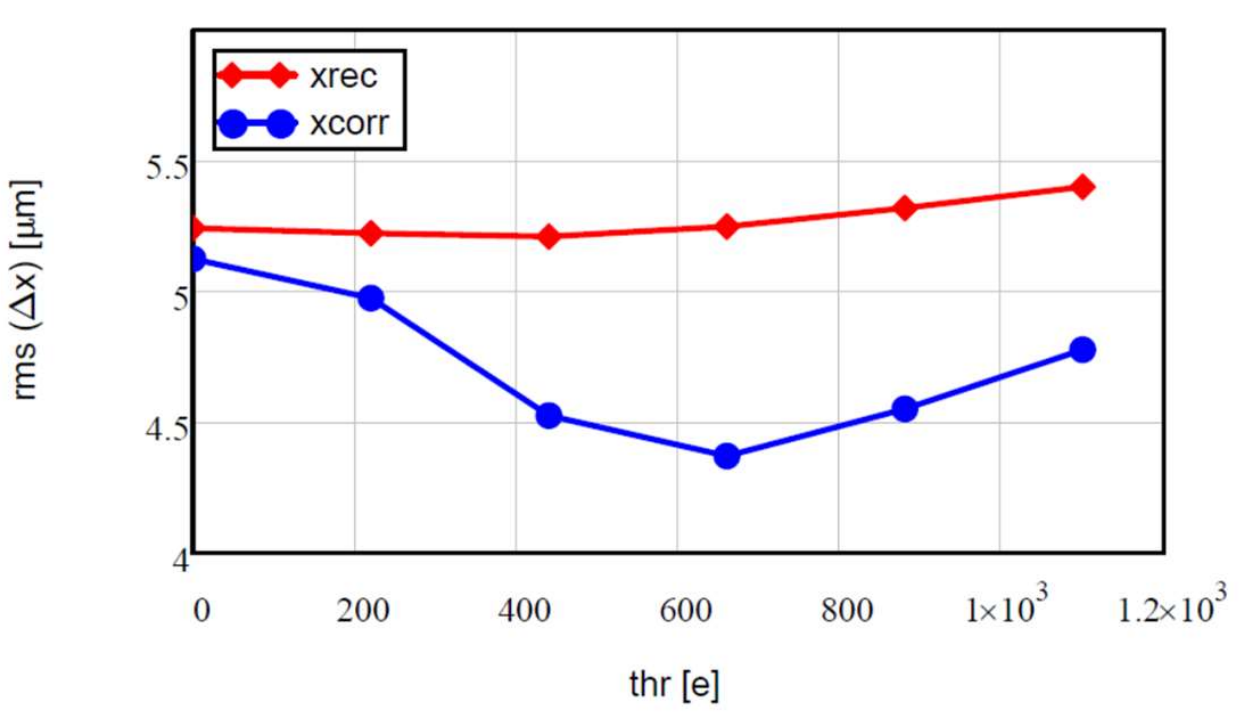}
    \caption{ }
    \label{fig:Fig_rmsdxth0}
   \end{subfigure}%
    ~
   \begin{subfigure}[a]{0.5\textwidth}
    \includegraphics[width=\textwidth]{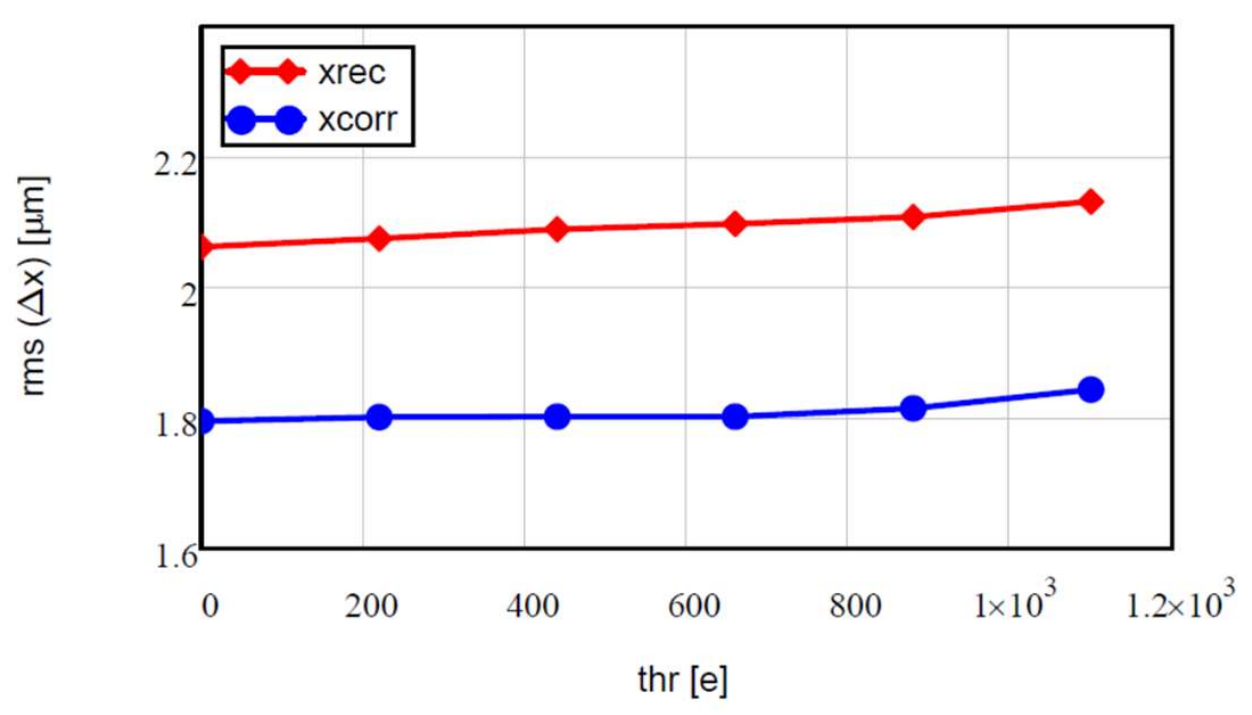}
    \caption{ }
    \label{fig:Fig_rmsdxth10}
   \end{subfigure}%
  \newline
       \begin{subfigure}[a]{0.5\textwidth}
    \includegraphics[width=\textwidth]{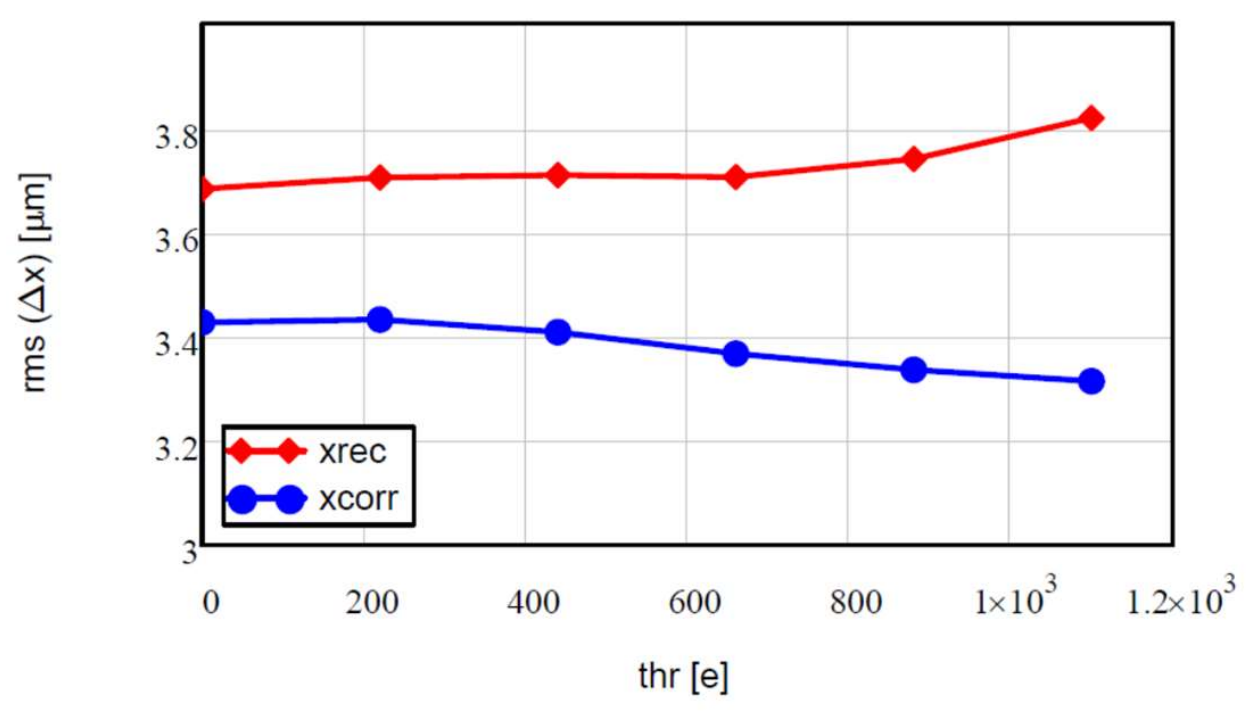}
    \caption{ }
    \label{fig:Fig_rmsdxth20}
   \end{subfigure}%
    ~
   \begin{subfigure}[a]{0.5\textwidth}
    \includegraphics[width=\textwidth]{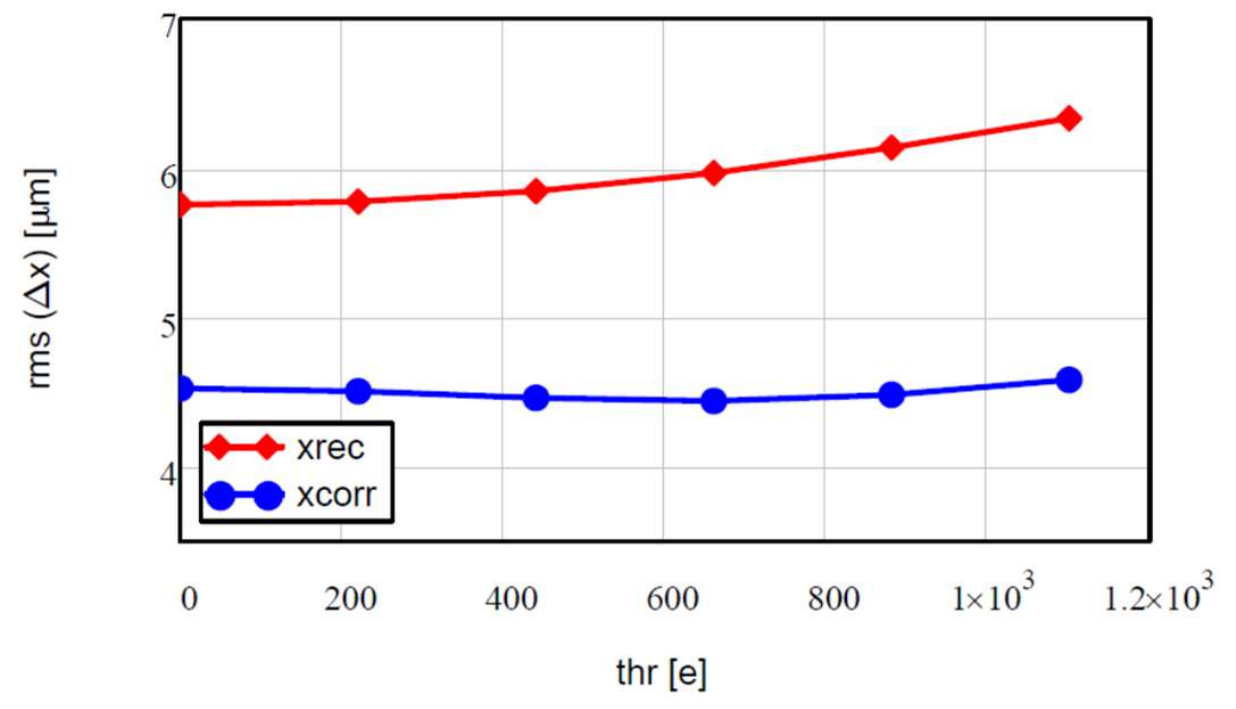}
    \caption{ }
    \label{fig:Fig_rmsdxth32}
   \end{subfigure}%
   \caption{\emph{rms} widths of the $\Delta x = x_\mathit{true} - x$~distributions for $x = x_\mathit{rec}$ and $x = x_\mathit{corr }$ as a function of $\mathit{thr}$ for
   (a) $\theta = 0^{\,\circ}$,
   (b) $10^{\,\circ}$,
   (c) $20^{\,\circ}$, and
   (d) $32^{\,\circ}$.
    }
  \label{fig:Fig_rmsdx}
 \end{figure}

 For $\theta = 0^{\,\circ}$ (Fig.~\ref{fig:Fig_dxrecth0}) the $\Delta x _\mathit{rec}$ distributions show peaks at $|\Delta x| \approx 7\,\upmu$m.
 They originate from $cls = 2$\,events, which are reconstructed at too small $|x|$ values.
 They are absent in the $\Delta x _\mathit{corr}$ distribution (Fig.~\ref{fig:Fig_dxcorth0}).
 The $\mathit{cls} = 2$\,events appear in a narrow peak at zero, and actually represent the events with the best position resolution.
 The flat part of the $\Delta x _\mathit{corr}$ distributions comes from the $\mathit{cls} = 1$\,events.
 For higher $\mathit{thr}$ values the fraction of $\mathit{cls} = 1$\,events increases and the flat part extends to higher $| \Delta x _\mathit{corr} |$\,values.
 For lower $\mathit{thr}$ values the fraction of the $\mathit{cls} = 2$\,events due to electronics noise increases, and higher $| \Delta x _\mathit{corr} |$\,values appear.
 Therefore there will be an optimal value of $\mathit{thr}$.
 This can be seen in Fig.~\ref{fig:Fig_rmsdxth0}, which shows the \emph{rms} of the $\Delta x$ distributions.
 At the optimal value, $\mathit{thr} \approx 660$~e, the average resolution is $\approx 4.4~\upmu$m for $x _\mathit{corr}$ compared to $\approx 5.2~\upmu$m for $x _\mathit{rec}$.

 For larger $\theta $ values the fraction of $\mathit{cls} = 1$ events decreases and the $\Delta x $ distributions for both $x _\mathit{rec}$ and $x _\mathit{corr}$ peak at $\Delta x = 0$.
 Fig.~\ref{fig:Fig_rmsdx} shows that for all $\theta $ values an improvement of the resolution of the proposed method is predicted.
 However for most angles the effect is small, and precise experimental data are required to see if such an improvement is actually observed.

 The main results of this sections are:
 For particles with normal incidence, $\theta = 0 ^{\,\circ}$, the proposed method is able to correct the strong bias of the centre-of-gravity algorithm for $cls > 1$\,events and significantly improves the overall position resolution.
 For larger angles $\theta $ the bias of the centre-of-gravity algorithm is small.
 The proposed method is able to correct this bias and predicts minor improvements of the position resolution.
 This however still has to be demonstrated with experimental data.
 The proposed method also allows determining the position-response function as a function of  reconstructed position. At $\theta = 0 ^{\,\circ}$ this function varies by as much as an order of magnitude, and therefore should be used in track fits.
 The position response functions have significant non-Gaussian tails and cannot be adequately described by their \emph{rms} values.

  \subsection{ Comparison of simulation results on cluster size and position resolution to test-beam data}
   \label{sect:ResAngles}

 In this section, results of the simulations are compared to test-beam results for silicon pixel sensors under study for the CMS Phase II Upgrade\,\cite{CMS:2017, Zoi:2021, Ebrahimi:2021, Feindt:2021, Schwandt:2018, Steinbrueck:2020}.
 The data for the comparison are taken from Refs.\,\cite{Zoi:2021, Ebrahimi:2021}.
 The measurements were performed in a 5.6\,GeV electron beam\,\cite{Diener:2019} at DESY, Hamburg.
 Three pixel sensors with 150\,$\upmu$m thickness and $25\,\upmu$m$\,\times 100\,\upmu$m pixel size were bump-bonded to PSI ROC4Sens chips\,\cite{Wiederkehr:2018, Rohe:2017} fabricated on 700\,$\upmu$m thick silicon wafers, and read out with 12 bit resolution.
 The sensors were operated at $U = 120$\,V and at $T = 20^{\,\circ}$C.
 The distance between the sensors was 2\,cm, and they could be rotated together so that the sensor planes remained parallel.
 The set-up is called "3-Master" (3M).
 Data for 25 angles, $\theta $, between beam axis and sensor normal from $0^{\,\circ}$ to $30^{\,\circ}$ were recorded.
 The rotation was around the $100\,\upmu$m pixel direction.
 In addition, at $\theta = 8.75^{\,\circ}$, data for beam energies between 1.2 and 6\,GeV were recorded in order to determine experimentally the influence of multiple scattering on the measured resolution.

\begin{figure}[!ht]
   \centering
   \begin{subfigure}[a]{0.5\textwidth}
    \includegraphics[width=\textwidth]{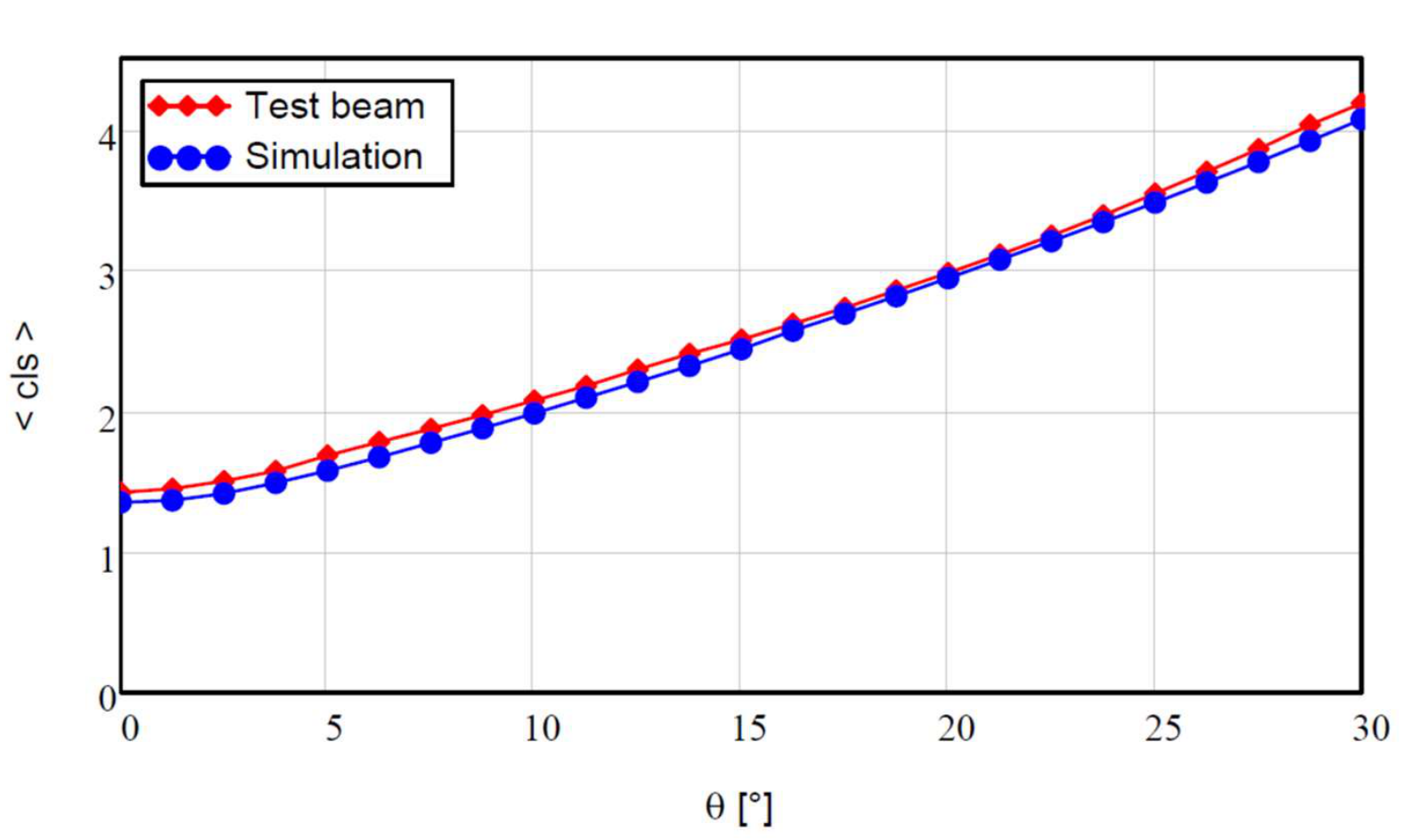}
    \caption{ }
    \label{fig:Fig_Cls_beam}
   \end{subfigure}%
    ~
   \begin{subfigure}[a]{0.5\textwidth}
    \includegraphics[width=\textwidth]{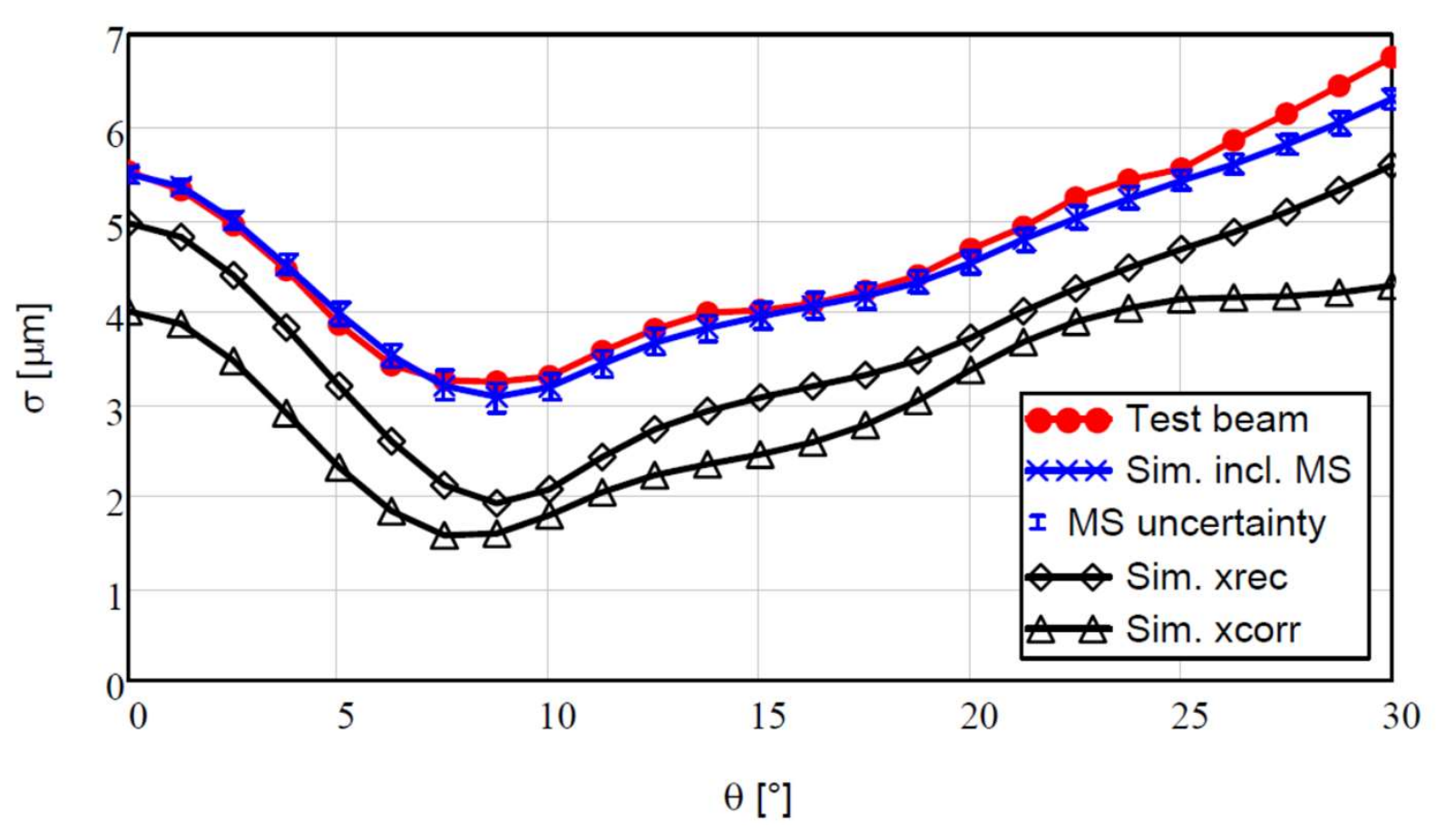}
    \caption{ }
    \label{fig:Fig_Sigth_beam}
   \end{subfigure}%
   \caption{Comparison of simulation with beam-test results for a sensor with $25\,\upmu$m pitch,  $150\,\upmu$m thickness at a bias voltage  $U = 120$\,V as a function of angle $\theta$.
   (a) Mean  cluster size, and
   (b) \emph{reduced rms} of the spatial resolution.
   Shown are
     the test-beam results, and for the simulation
     the reconstructed position using the centre-of-gravity method, $x_\mathit{rec}$,
     the method proposed in this paper, $x_\mathit{corr}$,
     and  $x_\mathit{rec}$ including the effect of multiple scattering in the test-beam set-up.  }
  \label{fig:Fig_Compbeam}
 \end{figure}

 The most probable signal from a 5.6\,GeV electron at normal incidence is about 11~000\,e (elementary charges).
 In the off-line analysis signals exceeding a threshold of approximately $ 660$\,e in contiguous pixels are grouped into clusters.
 Signals above threshold in pixels with the same $x$\,position (see Fig.\,\ref{fig:Fig_Sensor}) are added and counted as one when calculating the  cluster size, \emph{cls}.
 Fig.\,\ref{fig:Fig_Cls_beam} shows the measured $\langle \mathit{cls} \rangle $ as a function of $\theta $ (label \emph{Test beam}).

 The test-beam data are analysed in the following way:
 The $x$\,positions in the three sensors are reconstructed using the centre-of-gravity of the selected pixels, and the distribution of the difference of the position predicted from the outer sensors minus the position measured in the central sensor, $\Delta x_\mathrm{3M}$, is used for determining the position resolution:
 From the $\Delta x_\mathrm{3M}$\,distribution the \emph{reduced rms}, $ \delta _{\Delta x} $, is calculated iteratively by removing the events with $\Delta x_\mathrm{3M}$\,values exceeding $\pm \, 6$ times the \emph{reduced rms}.
 The measured single detector resolution, which is shown in Fig.\,\ref{fig:Fig_Sigth_beam} (label \emph{Test beam}), is defined as $\sigma _x = \delta _{\Delta x} / \sqrt{1.5}$~\cite{Zoi:2021, Ebrahimi:2021}.

 This procedure makes a number of assumptions:
 The three sensors have the same position resolution,
 the three position measurements are not correlated, and
 the shape of the combined response function of two sensors does not differ too much from the one of a single sensor.
 As discussed in Sect.\,\ref{sect:Model}, the centre-of-gravity reconstruction is biased, and the small distances between the detector planes and the small angular spread of the beam effectively result in correlations of the reconstructed positions in the three planes.
 These effects are most significant at small angles for events with  cluster size one in one or more sensors.
 In addition, only Gauss\,functions have the property that their convolutions are again Gauss\,functions.
 In Sect.\,\ref{sect:Model} it was shown that the response functions of single sensors can be very different from Gauss\,functions.
 Multiple scattering too, results in non-Gaussian tails.
 Nevertheless, it is expected that for angles $\theta \approx \mathrm{atan}(p/d) = 9.5^{\,\circ}$ and above, the position resolution can be determined reliably.

 In order to investigate the effects of multiple Coulomb scattering, data have been taken at the angle with the best resolution, $\theta _\mathit{best} = 8.75^{\,\circ}$, for energies, $E$, between 1.6\,GeV and 6\,GeV and $\sigma _x(E,\theta _\mathit{best} )$ determined.
 The straight line extrapolation of $\sigma_x ^2 (1/E^2)$ to $1/E^2 = 0$ gives $\sigma _\mathit{res}(\theta _\mathit{best}) = 2.4 \pm 0.1\,\upmu$m, the resolution corrected for multiple scattering, and the contribution from multiple scattering of $\sigma _\mathit{ms} (5.6\,\mathrm{GeV},\, \theta _\mathit{best} )= \sqrt{\sigma _x ^2 -\sigma _\mathit{res}^2} = (2.1 \pm 0.2)\,\upmu $m.
 The measured value of $\sigma _x(5.6\,\mathrm{GeV},\, \theta _\mathit{best}) = 3.2 \pm 0.2\,\upmu$m\,\cite{Zoi:2021, Ebrahimi:2021}.

 In the following, the analysis of the simulated data is described and the results compared to the experimental data.
  First, using the method described in Appendix~\ref{sect:Appendix_SpatialDistribution}, it has been verified that assuming a uniform spatial distribution is valid.
 For each of the 25 $\theta $\,values, $5 \times 10^5$ events uniformly distributed in the central electrode with pitch $p_x \times p_z$ were simulated as described in Sect.\,\ref{sect:Appendix_Signal}.
 For the number of pixels, $n_\mathit{px} = 3$ has been used for $\theta \leq 15^{\,\circ}$, $n_\mathit{px} = 5$ for larger angles, and $n_\mathit{pz} = 3$ for all angles.
 The \emph{reduced rms} for the simulated data, $\sigma _\mathit{xrec}$, obtained with similar cuts as for the experimental data, is shown as a function of $\theta $ in Fig.\,\ref{fig:Fig_Sigth_beam} (label Sim. xrec).
 The relation
 $ \delta _{\Delta x} ^2 = 1.5 \cdot \sigma _\mathit{xrec} ^2 + \sigma _\mathit{ms} ^2 \cdot \big( \cos (\theta _{\mathit{best}} ) / \cos(\theta) \big)^3 $
 is used for obtaining the \emph{reduced rms} of the simulated residual distributions including multiple scattering.
 The factor $\sqrt{1.5}$ relates the single sensor resolution with the 3-Master residual distribution.
 The power 3 takes into account that the angular spread of multiple scattering is approximately proportional to the square root of the path length, and that the residual of the experimental data is determined in the rotated coordinate system of the sensors, whereas for multiple scattering the coordinate system of the beam is relevant.
 Finally, $ \delta _{\Delta x} $ is divided by $\sqrt{1.5}$ for obtaining the simulated resolution including multiple scattering, which is compared in Fig.\,\ref{fig:Fig_Sigth_beam} to the measured resolution.

 Up to $\theta = 25^{\,\circ}$  the simulation describes the measurement within $0.2\,\upmu$m.
 At $30^{\,\circ}$ the difference is $0.4\,\upmu$m.
 This agreement demonstrates the quality of both data and simulation.
 The figure also compares the simulated resolution for $x_\mathit{rec}$, which uses the centre-of-gravity algorithm, to the one for $x_\mathit{corr}$, the method proposed in this paper.
 It is seen, that for all angles, however for small angles in particular, a significantly improved resolution is expected, as already discussed in Sect.\,\ref{sect:Threshold}.

\begin{figure}[!ht]
   \centering
   \begin{subfigure}[a]{0.5\textwidth}
    \includegraphics[width=\textwidth]{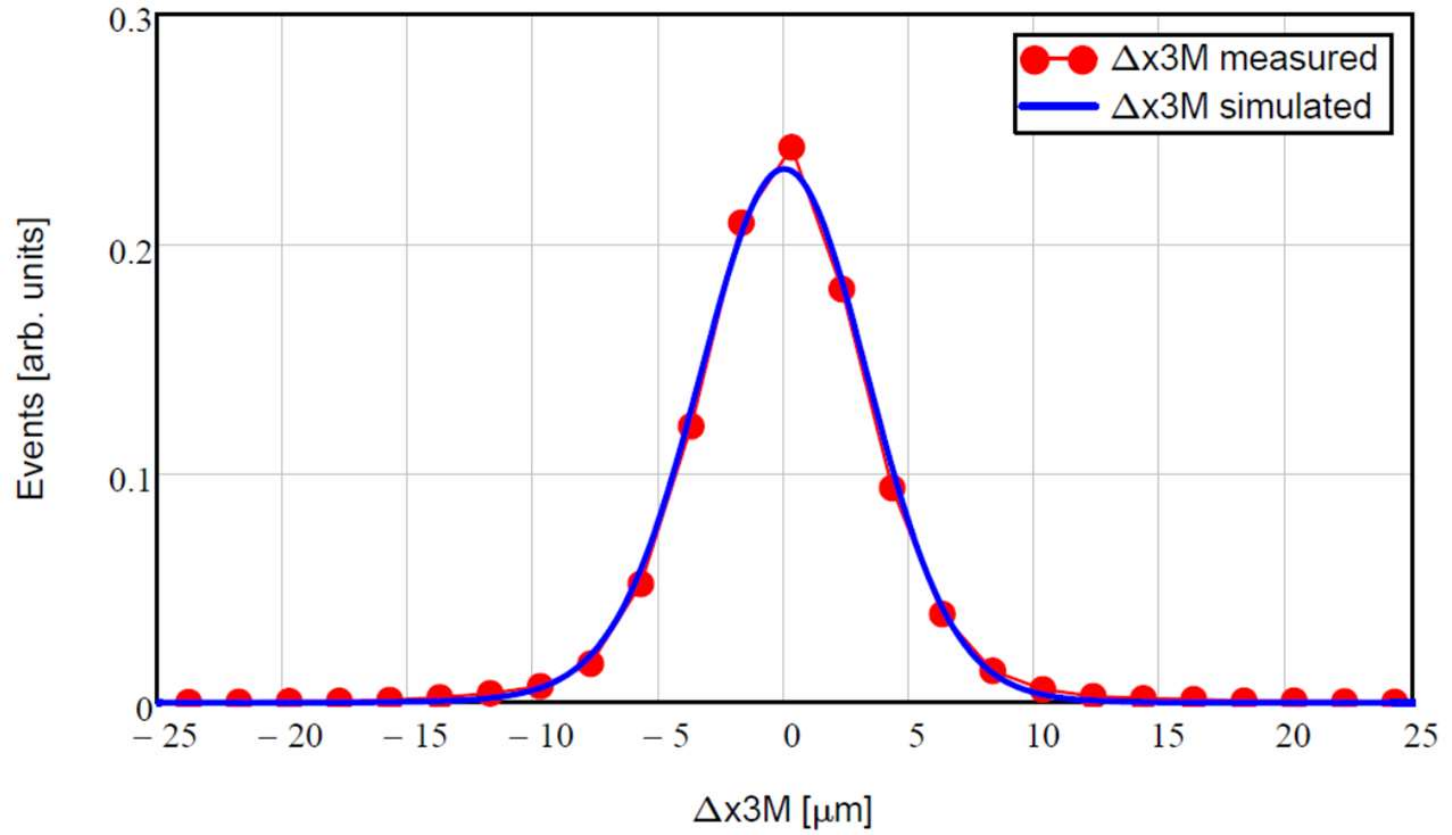}
    \caption{ }
    \label{fig:Fig_DX8p75-lin}
   \end{subfigure}%
    ~
   \begin{subfigure}[a]{0.5\textwidth}
    \includegraphics[width=\textwidth]{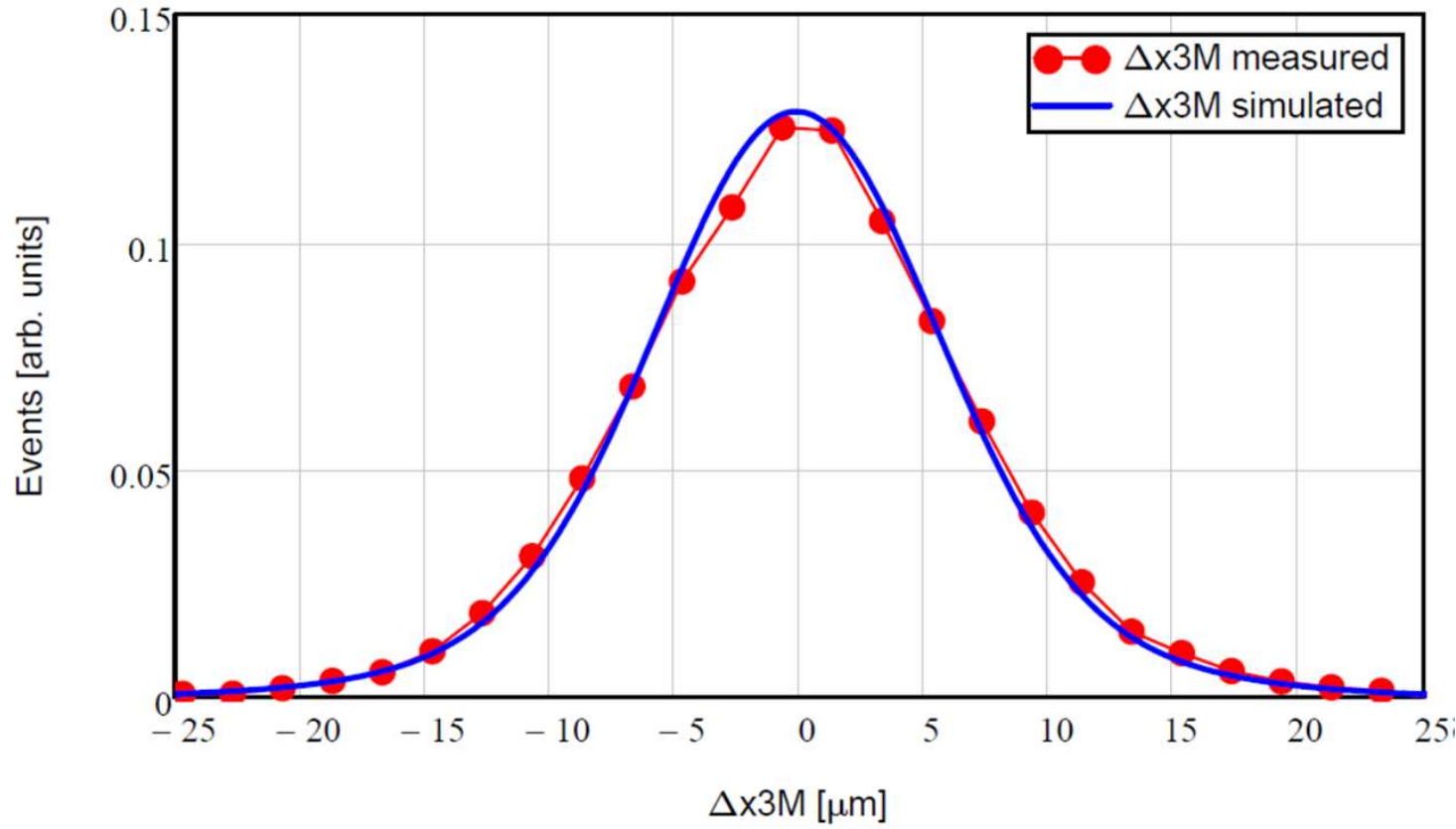}
    \caption{ }
    \label{fig:Fig_DX27p5-lin}
   \end{subfigure}%
\newline
   \begin{subfigure}[a]{0.5\textwidth}
    \includegraphics[width=\textwidth]{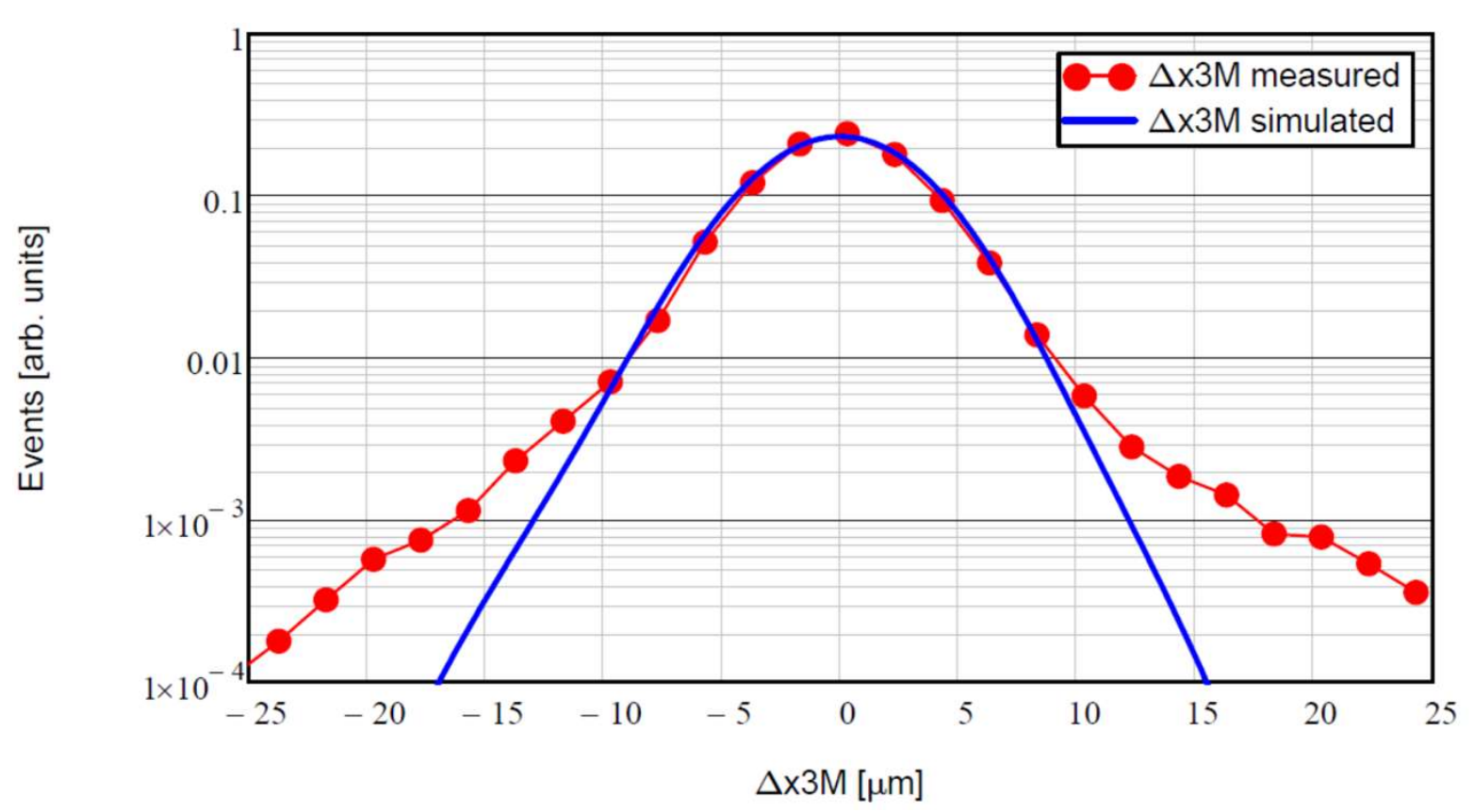}
    \caption{ }
    \label{fig:Fig_DX8p75-log}
   \end{subfigure}%
    ~
   \begin{subfigure}[a]{0.5\textwidth}
    \includegraphics[width=\textwidth]{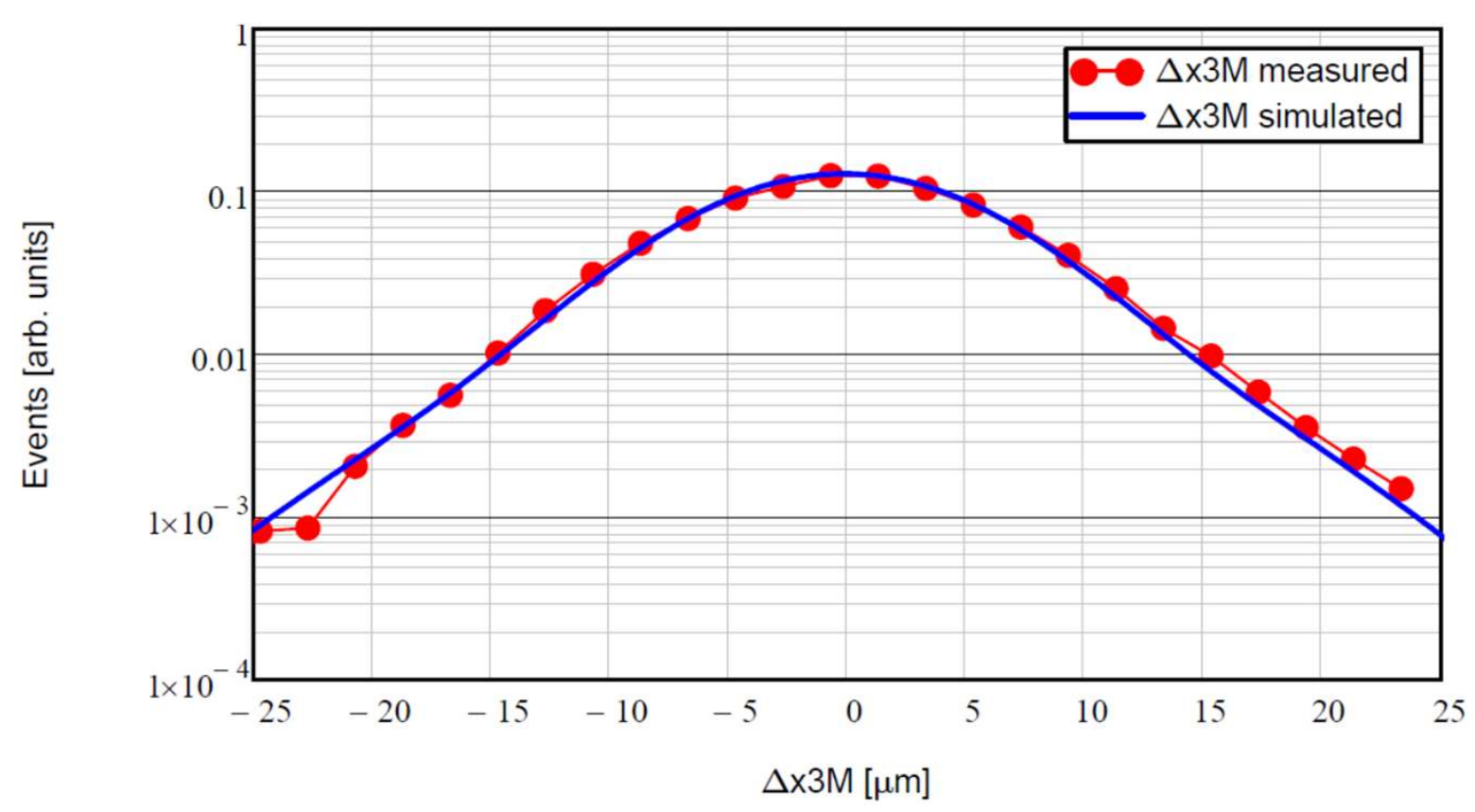}
    \caption{ }
    \label{fig:Fig_DX27p5-log}
   \end{subfigure}%
   \caption{Comparison of the measured to the simulated normalized $\Delta x_\mathrm{3M}$\,distributions.
    $\Delta x_\mathrm{3M}$ is the difference of the position predicted in the central sensor from the outer sensors minus the position determined in the central sensor, for the set-up with three parallel sensors, for
  (a) $\theta = 8.75 ^{\,\circ}$ in linear scale,
  (b) $\theta = 27.5 ^{\,\circ}$ in linear scale,
  (c) $\theta = 8.75 ^{\,\circ}$ in logarithmic scale, and
  (d) $\theta = 27.5 ^{\,\circ}$ in logarithmic scale.
   }
  \label{fig:DX}
 \end{figure}

 In Fig.\,\ref{fig:DX} measured and simulated normalized $\Delta x_\mathrm{3M}$\,distributions for the data described above are compared for the two angles, $\theta = 8.75 ^{\,\circ}$ and $\theta = 27.5 ^{\,\circ}$.
 $\Delta x_\mathrm{3M}$ is the difference of the position predicted in the central sensor from the outer sensors minus the position determined in the central sensor for the 3M set-up.
 For the simulated data, the $\Delta x_\mathrm{3M}$\,distributions are obtained by scaling  the $\Delta x$\,values of the simulated response functions $\mathrm{d}N / \mathrm{d}\Delta x$ (examples are shown in Fig.\,\ref{fig:Fig_dxth}) by the factor $\sqrt{1.5}$, and then convolving the scaled distribution with a Gauss function with $\sigma _\mathit{ms}(\theta) $, which accounts for multiple Coulomb scattering.
 Fig.\,\ref{fig:DX} shows that overall the simulations describe the measured data quite accurately.
 The deviations for larger $|\Delta x_\mathrm{3M}|$ seen in Fig.\,\ref{fig:Fig_DX8p75-log} are ascribed to simplifying assumptions in the analysis of the data from three detector planes and to the use of a Gauss function for multiple scattering, which ignores large-angle scattering events.

 As discussed in Sect.\,\ref{sect:Model}, at small angles $\theta $ the reconstructed positions in the three sensors are correlated because of the bias of the centre-of-gravity method.
 As a result the $\Delta x _\mathrm{3M}$\,distribution shows narrow peaks \cite{Zoi:2021}, mainly but not only from events with  cluster size one in one, or more than one sensor.
 The simulation of the $\Delta x _\mathrm{3M}$\,distribution at small angles requires the knowledge of the relative alignment of the three sensors with an accuracy of $ \lesssim 1\,\upmu$m.

 The main results of the comparison of simulation results with test-beam data for angles $\theta $ between $0^{\,\circ}$ and $30^{\,\circ} $ presented in this section are:
 The average  cluster sizes agree to better than 0.1, and
 the \emph{reduced rms} of the measured sensor resolution, which includes the effects of multiple scattering, agrees within $0.2 \, \upmu$m up to $25^{\,\circ} $, and the central parts of the measured $\Delta x_\mathrm{3M}$\,distributions are well described.

  \section{Summary and conclusions}
   \label{sect:Conclusions}

 The topic of the paper is the position reconstruction of signals from segmented detectors.
 It is shown that the frequently used centre-of-gravity method can result in a strong bias of the reconstructed position.
 The bias is particularly strong if the width of the signal distribution at the position of the read-out electrodes is less than their size.
 For events with signals in more than one read-out electrode, a method is presented which corrects the bias of the centre-of-gravity method.
 With the help of simulated events a significant improvement of the position resolution is demonstrated for charged particles traversing with different angles a pixel sensor with a pixel pitch of $25\,\upmu$m~$\times 100\,\upmu$m.
 In addition to improving the position resolution, the method provides an estimate of the position-response function for every reconstructed event.
 As the response function depends strongly on the position of the particle relative to the electrodes, it should be used for track fits.
  The proposed method does not achieve a higher precision than previously used methods, but it is more easily implemented and does not require additional information beyond the signals recorded in the individual detector elements.
  It can also be applied to other reconstruction methods than the centre-of-gravity algorithm, and for the position reconstruction in other types of segmented detectors, like calorimeters.

 As a next step, the method should be applied to experimental data for silicon sensors without and with radiation damage in order to confirm the improvement predicted from the simulated events.
 In addition, the proposed method can be applied to other position reconstruction methods, like the \emph{head-tail}
 algorithm.
 Also the use of the method for other detectors is strongly encouraged.

 As a by-product, a simulation program for silicon pixel sensors has been developed.
 With the program about 500~k~events per minute can be simulated for silicon pixel sensors with 100\,\% charge collection.
 It has been verified that the generated events provide an accurate description of test-beam data.
 Missing in the program are the effects of energetic $\delta $\,electrons and the Lorentz drift in magnetic fields, which however can be implemented.
 Last but not least, the program can be extended to the simulation of radiation-damaged silicon sensors with losses of charge carriers during their drift in the sensor.
 This however, requires the knowledge of the electric field, the position-dependent trapping probabilities separately for electrons and holes, and the weighting field for radiation-damaged sensors.

 \begin{appendices}
   \section{Appendices}

   \label{sect:Appendix}

  \subsection{Signal simulation}
   \label{sect:Appendix_Signal}

 This Appendix gives details of the program presented in Sect.~\ref{sect:Simulation} for simulating the signals produced by energetic charged particles in a silicon pixel sensor with 100\,\% charge collection.
 The aim is to have a flexible program which is fairly realistic and which can generate 500\,k\,events in about 1 minute on a standard PC.
 Fig.\,\ref{fig:Fig_Sensor} of Sect.~\ref{sect:Simulation} displays the cross section and the top view of the simulated pixel sensor, defines the coordinate system and shows a track traversing the sensor in the $x$ - $y$~plane with the angle $\theta $ relative to the $y$~axis.
 Table\,\ref{tab:InputSim} lists the parameters used in the simulation.

\begin{table} [!ht]
  \centering
   \begin{tabular}{c|c|c}
     Symbol & Value & Description \\
   \hline  \hline
     $d$ & $150 \, \upmu$m & Sensor thickness\\
     $n_\mathit{px}$   & 3, 5, 7 & Number of pixels in $x$~direction\\
     $n_\mathit{pz}$   & 3 & Number of pixels in $z$~direction\\
     $p_x$ &  25 $\upmu $m & Pixel pitch in $x$~direction\\
     $p_z$ & 100 $\upmu $m & Pixel pitch in $z$~direction\\
     $n_y$ & 15, 30 & Number of $y$-steps for energy loss\\
     $\Delta x_\mathit{min}$, $\Delta z_\mathit{min}$ & $2\,\upmu$m & see text \\
     $N_d$ & $4.5 \times 10^{12}\,\mathrm{cm}^{-3}$ & $p$-doping density\\
   \hline
     $U$ & 120, 150 V & Bias voltage\\
     $T$ & 20$^{\, \circ} $C & Temperature\\
     $x_\mathit{true}$ & $-p_x/2 \leq  x_\mathit{true} < p_x/2$ & $x$ position of track at $y = 0$\\
     $z_\mathit{true}$ & $-p_z/2 \leq  z_\mathit{true} < p_z/2$ & $z$ position of track at $y = 0$\\\
     $\theta $ & $0 ^{\, \circ}$ to $30 ^{\, \circ}$ & Particle angle relative to $y$~axis \\
   \hline
     $\mathit{MPV}$ & 11~000\,e & Most probable charge for $\theta = 0 ^{\, \circ}$ \\
     $\Delta E_\mathit{max} $ & 2.0 & Maximal charge~/~MPV \\
   \end{tabular}
    \caption{Input parameters used in the simulation program.
    \label{tab:InputSim}}
  \end{table}

 For a pad sensor of thickness $d$, $p$-doping density, $N_d$, the $n^+p$~junction at $y = d/2$, and bias voltage $U$, the absolute value of the electric field above full depletion is approximately
 \begin{equation}\label{equ:Ey}
   E_y(y) = U/d + y \cdot q_0 \cdot N_d / \varepsilon _\mathit{Si},
 \end{equation}
 with the elementary charge $q_0$,  the dielectric constant of silicon, $\varepsilon _\mathit{Si} $\footnote{The value 11.7 has been assumed for the  relative dielectric constant of silicon.},
 and the coordinate system shown in Fig.\,\ref{fig:Fig_Sensor}.
 This only approximates the electric field of a sensor with segmented electrodes.
 Close to the electrodes the electric field is more complicated and depends on a number of parameters:
 the width of the pixel implants,
 the doping of the bulk and of the $p^+$ implants required to electrically separate the electrodes,
 the boundary conditions on the SiO$_2$\,surface in-between the electrodes,
 and the dark current.
 The charge sharing between electrodes for charges generated at $x$\,values close to their boundaries is influenced by this field distribution.
 A study of the charge collection close to the sensor surface is reported in Ref.\,\cite{Poehlsen:2013}, with the conclusion that even in non-radiation damaged sensors the electric field close to the electrodes is only poorly known as important information required for TCAD simulations is lacking.

 Next, the simulation of the $e$-$h$\,pairs and the charges $Q$ induced in the electrodes by a particle traversing the sensor at position ($x_\mathit{true}$, 0, $z_\mathit{true}$) at an angle $\theta $ to the $y$~axis is presented.
 In order to take into account the local energy-loss fluctuations, the sensor depth is divided into $n_y$ segments of depth $\Delta y = d/n_y$.
 For every $\Delta y$\,interval a random number for the energy loss $\Delta E$ is generated for the particle-path length $\Delta y /\cos (\theta)$ as described in Ref.\,\cite{Bichsel:1988} and Appendix\,\ref{sect:Appendix_Eloss}.

 An electron cloud with the distribution $\delta (x) \cdot \delta (z)$ at $y = y_0$, diffuses to a 2-D~Gaussian charge distribution when drifting to $y = d/2$ with the variance
 \begin{equation}\label{equ:sigma}
   \sigma ^2 (y_0) = \frac{2\, k_B \, T}{q_0} \int _{y_0} ^{d/2}\frac{\mathrm{d}y} {E_y (y)},
 \end{equation}
 with the Boltzmann constant $k_B$, and the absolute temperature $T$.
 It is remarkable that $\sigma (y_0)$ does not depend on the mobility, which significantly simplifies the simulation.
 This follows from
 \begin{equation*}
   \sigma ^2(\tau)=\frac{2\,k_B \,T }{q_0} \, \mu_e \cdot \tau \Rightarrow
   \frac{\mathrm{d} \sigma ^2 }{\mathrm{d} \tau} = \frac{2\,k_B \,T}{q_0}\, \mu _e \Rightarrow
   \frac{\mathrm{d} \sigma ^2 }{\mathrm{d} y} = \frac{\mathrm{d} \sigma ^2 }{\mathrm{d} \tau} / \frac{\mathrm{d} y }{\mathrm{d} \tau} = \frac{2\,k_B \,T}{q_0 \, E_y (y) },
 \end{equation*}
 where the symbol $\tau $ is used for time.
 For $E = 0$, which however does not occur in a real sensor, the expression of Eq.\,\ref{equ:sigma} is not defined, and $\sigma ^2 (\tau) = 2 \cdot k_B \cdot T \cdot \mu _{e/h} \cdot \tau / q_0$ has to be used.

 Inclined tracks in the $x$ - $y$~plane traversing a distance $\Delta y$ cover an $x$\,range of width $\Delta x = \Delta y \cdot |\tan (\theta)|$, which is taken into account by convolving the Gauss distributions with a box distribution of width $\Delta x' = \max (\Delta x, \Delta x _\mathit{min} )$.
 In $z$, the Gauss distribution is convolved with a box distribution of width $\Delta z_\mathit{min}$.
 The convolved Gauss distribution of a charge cloud drifting from ($0,y_0,0)$ to $y = d/2$ is denoted $G^\star (x, y = d/2, z; y_0)$.
 The parameters $\Delta x _\mathit{min} $ and $\Delta z _\mathit{min} $, which are only relevant for the charges produced  close to the electrode boundaries, are introduced to describe the transverse width of the initial charge carrier distribution and the effects of charge sharing due to the electric field in the region of the electrode boundaries.
 The choice $\Delta x _\mathit{min} = \Delta z _\mathit{min} = 2\,\upmu$m is quite arbitrary, but the exact value hardly influences the results.
 The charge induced in a given electrode is obtained from

 \begin{equation}\label{ewqu:Qi}
   Q_i = \sum _{iy =  1} ^{n_y}  \int _\mathit{pixel} G^\star (x = x _\mathit{true} + y_{iy }\cdot \tan(\theta), y = d/2, z = z _\mathit{true}; y_{iy} )~\mathrm{d}x ~ \mathrm{d}z .
 \end{equation}
 To speed up the simulation, the $n_y$ cumulative distributions of the different $G^\star $ functions are calculated only once for a given angle $\theta$.

 The effect of fluctuations in the charge cloud during its drift has been estimated using multinomial distributions, which describe the fluctuations of the signals in individual electrodes for a given total signal.
 For the $150\,\upmu$m thick sensors investigated, the number of charge carriers generated by energetic particles ($ \mathrm{MPV} \approx 11~000$\,e) is sufficiently large compared to the \emph{rms} noise of 250\,e, so that this effect can be neglected.
 However, this is not the case for thin detectors with low electronics noise.

 Energetic $\delta $\,electrons which travel a finite distance and thus change the shape of the generated charge distribution, are so far not implemented in the simulation.
 Therefore, the results of the simulation become unreliable for charge values exceeding $\approx 2$\,times the most probable one.

 In a further analysis step, the electronics noise is simulated by adding a Gaussian-distributed random number to every $Q_i$, threshold cuts are made, the  cluster size is calculated and the $x$ and $z$~positions reconstructed as discussed in the main part of the paper.



  \subsection{Energy loss simulation}
   \label{sect:Appendix_Eloss}

 This Appendix describes the method used to generate random numbers for the energy loss, $\Delta E$, of energetic charged particles traversing a distance $t$ in silicon.
 This is achieved by the following steps, most of which are described in detail in Ref.\,\cite{Bichsel:1988}.
  \begin{enumerate}
    \item The mean number of energy-loss interactions of a particle traversing the distance $t$ is $\mu = t/t_1$, where $t_1$ is the mean spatial distance between energy-loss events.
    \item The $\Delta E$ distribution for $n$ energy-loss events, $\mathrm{d}P^{(n)}(\Delta E) /\mathrm{d}\Delta E $, is obtained by the $n-1$-fold convolution of the single-event $\Delta E$ distribution $\mathrm{d}P^{(1)}(\Delta E) / \mathrm{d}\Delta E $. For $n = 0$, $\mathrm{d}P^{(0)}(\Delta E) / \mathrm{d}\Delta E  = \delta (\Delta E )$.
    \item The $\Delta E$ distribution for the distance $t$ is calculated as the sum over the $\mathrm{d}P^{(n)}(\Delta E)/\mathrm{d}\Delta E $ distributions weighted with the Poisson distribution with mean $\mu $: $\mathrm{d}P(\Delta E)/\mathrm{d}\Delta E = \sum _{n \, = \, 0} ^\infty \big( (\mu ^{n} / n\,!) \cdot e^{- \mu} \cdot (\mathrm{d}P^{(n)}(\Delta E)/\mathrm{d}\Delta E ) \big) $.
    \item The inverse of the normalised cumulative distribution of $\mathrm{d}P{(\Delta E)}/\mathrm{d}\Delta E $ is used to generate random numbers for the energy loss $\Delta E$ of particles passing the distance $t$ in silicon.
  \end{enumerate}

\begin{figure}[!ht]
   \centering
   \begin{subfigure}[a]{0.5\textwidth}
    \includegraphics[width=\textwidth]{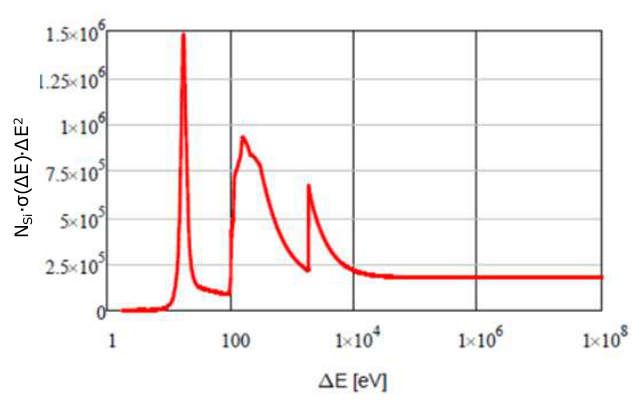}
    \caption{ }
    \label{fig:Fig_dE1}
   \end{subfigure}%
    ~
   \begin{subfigure}[a]{0.5\textwidth}
    \includegraphics[width=\textwidth]{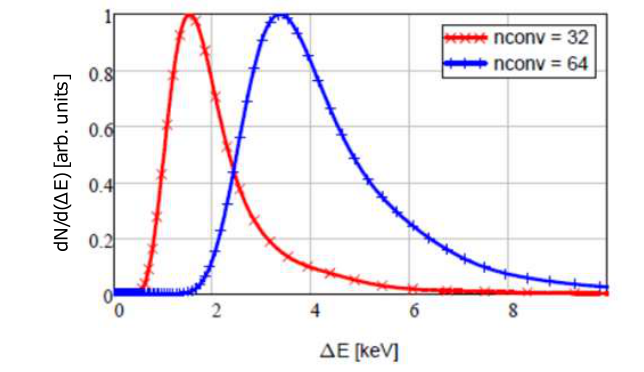}
    \caption{ }
    \label{fig:Fig_dEconv}
   \end{subfigure}%
   \caption{ (a) Differential cross section for a single energy-loss event, $\mathrm{d} \sigma ^{(1)} (\Delta E ) / \mathrm{d} \Delta E $ of 5\,GeV/c pions in silicon multiplied by $N_{Si}$,  the number of Si atoms per unit volume, and $\Delta E ^2$.
   (b) $\Delta E$ distribution for 32 (64) energy-loss events obtained by convolving $\mathrm{d} \sigma ^{(1)} (\Delta E ) / \mathrm{d} \Delta E $ 32 (64) times.
   The curves are normalised to 1 at their maxima.
   The mean distance between energy-loss events calculated using the data shown in (a) is $0.2602\,\upmu$m, and 32 (64) convolutions correspond to a path length of 8.33 (16.65)\,$\upmu$m.
   }
  \label{fig:Fig_Conv}
 \end{figure}

 A program for calculating the $\Delta E$\,distribution for a single energy-loss event based on Ref.\,\cite{Bichsel:1988} is available at Ref.\,\cite{Bichsel:code}, where also
 $N_{Si} \cdot \Delta E ^2 \cdot \mathrm{d} \sigma ^{(1)} (\Delta E ) / \mathrm{d} \Delta E $
 for 5\,GeV/c pions for the $\Delta E$ range from 1.796\,eV to 174.2\,MeV in 1699 logarithmic $\Delta E$\,bins can be accessed.
 Fig.\,\ref{fig:Fig_dE1} shows these data.
 $N_{Si} = 4.99 \times 10^{22}$\,cm$^{-3}$ is the number of Si atoms per cm$^{-3}$.
 The differential cross section multiplied by $\Delta E^2$ is shown because of the $1/\Delta E^2$\,dependence of the Rutherford cross section.
 The structures below 10\,keV reflect the atomic structure of Si and the constant at higher $\Delta E$ the Rutherford formula.
 $N_{Si} \cdot \mathrm{d} \sigma ^{(1)} (\Delta E ) / \mathrm{d} \Delta E ) \cdot \mathrm{d} \Delta E$ is the inverse of the mean distance of events with an energy loss between $\Delta E$ and $\Delta E + \mathrm{d}\Delta E$, and its integral is $1/t_1$, the inverse of the mean distance between energy-loss events;
For the data of Fig.\,\ref{fig:Fig_dE1}, $t_1 = 0.2602\,\upmu$m.

 The convolved spectrum $h(f, g)$ of two data sets $(t_i, f_i)$ and $(t_i, g_i)$ with bin widths $\Delta  t_i$ is obtained by
 \begin{equation}\label{equ:Convolution}
   h_j = \sum _{i \, = \, 0} ^{j} \big( \Delta t _i \cdot f_i \cdot g(t = t_j - t_i) \big),
 \end{equation}
 where for $g(t)$ a linear interpolation between the $g _i$\,values is used.
 The differential cross sections for $n$ energy-loss events are

   \begin{equation}\label{equ:ConSig}
    \frac{\mathrm{d} \sigma ^{(n)} (\Delta E) }{\mathrm{d} \Delta E} = \left\{
           \begin{array}{cc}
    \vspace{1mm}
        \delta (\Delta E) & \hbox{\rm{for}\,\, $n = 0 $} \\
        h \big(\mathrm{d}\sigma ^{(1)}/\mathrm{d} \Delta E,\,\mathrm{d} \sigma ^{(n-1)} / \mathrm{d} \Delta E \big)
        & \hbox{\rm{for}\,\, $ n \geq 1 $.} \\
           \end{array}
         \right.
  \end{equation}

 For $n = 0$, no energy is lost,
 for $n = 1$, the energy-loss cross section is the single-event energy-loss cross section, and
 for $n > 1$, the cross section is calculated iteratively using the convolution of Eq.\,\ref{equ:Convolution}.

 Next, the mean number of energy-loss events $\mu = \Delta t / t_1$ for the path length $\Delta t = \Delta y / \cos( \theta )$ is calculated, and the probability density $\mathrm{d} P / \mathrm{d} \Delta E$ is obtained from the Poisson-weighted sum
 \begin{equation}\label{eq:PoissonSum}
   \frac{\mathrm{d} P }{\mathrm{d} \Delta E} = \sum _{n \, = \, 0} ^\infty
   \frac{\mu ^n \cdot e^{- \mu}}{n\,!} \cdot \frac{\mathrm{d}P^{(n)}} {\mathrm{d}\Delta E}.
   \Big( \frac{\mu ^n \cdot e^{- \mu}}{n\,!} \cdot \frac{\mathrm{d}P^{(n)}} {\mathrm{d}\Delta E} \Big).
 \end{equation}
 The functions $\mathrm{d} P ^{(n)} / \mathrm{d} \Delta E$ are the normalised cross sections $\mathrm{d} \sigma ^{(n)} / \mathrm{d} \Delta E$, and $\mathrm{d} P / \mathrm{d} \Delta E$ the energy-loss probability density for the path length $\Delta t$.

 For generating $\mathrm{d} P / \mathrm{d} \Delta E$-distributed random numbers $\Delta E _i$, the standard method using the probability distribution $P(\Delta E) = \int _0 ^{\Delta E}(\mathrm{d} P / \mathrm{d} \xi \cdot \mathrm{d} \xi )$ is followed.
 A random number, $r_i$, which is uniformly distributed between 0 and 1 is generated and $\Delta E_i = P^{(-1)} (r_i)$, where $P^{(-1)}$ is the inverse of $P (\Delta E) $.

\begin{figure}[!ht]
   \centering
   \begin{subfigure}[a]{0.5\textwidth}
    \includegraphics[width=\textwidth]{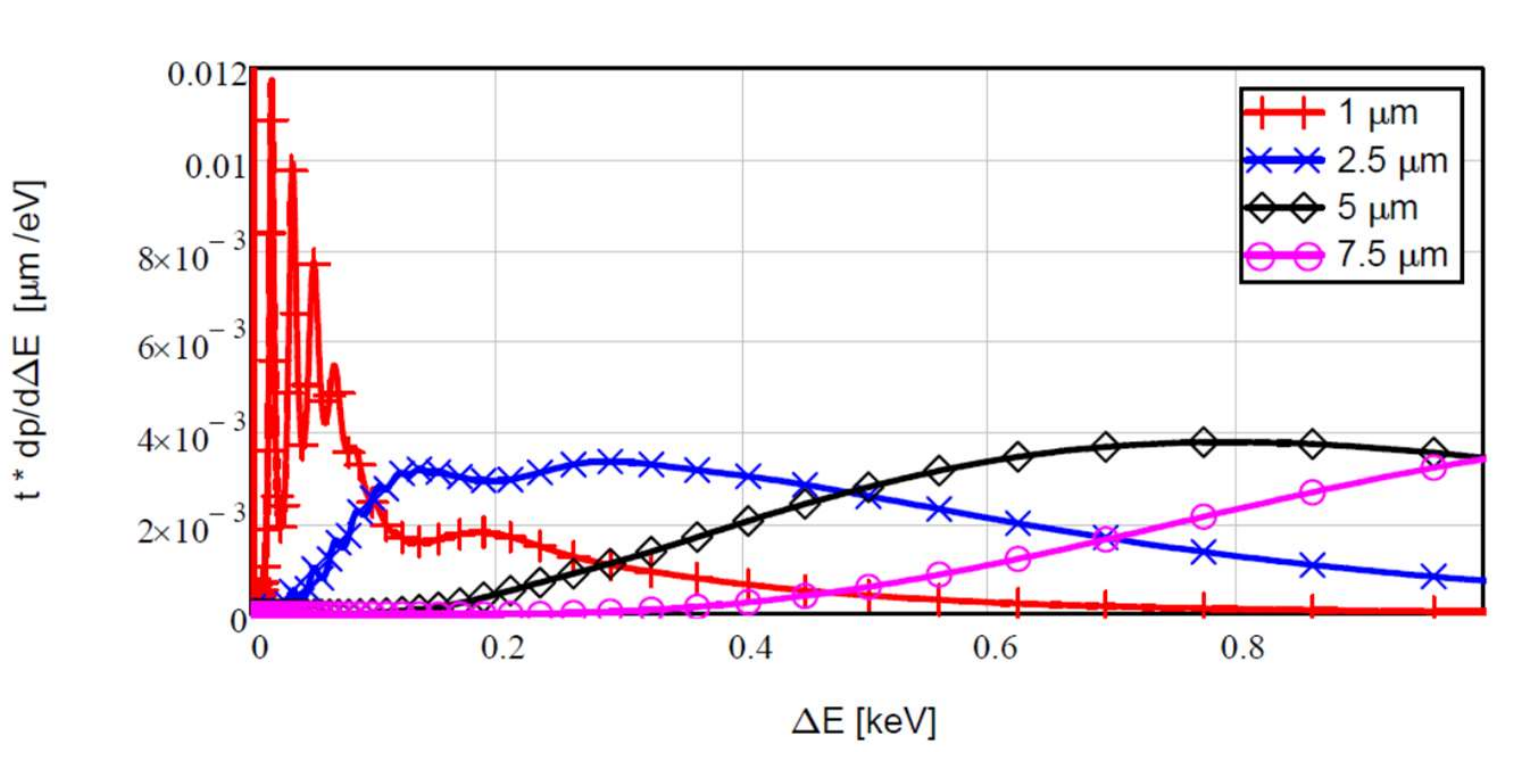}
    \caption{ }
    \label{fig:Fig_MClowt}
   \end{subfigure}%
    ~
   \begin{subfigure}[a]{0.5\textwidth}
    \includegraphics[width=\textwidth]{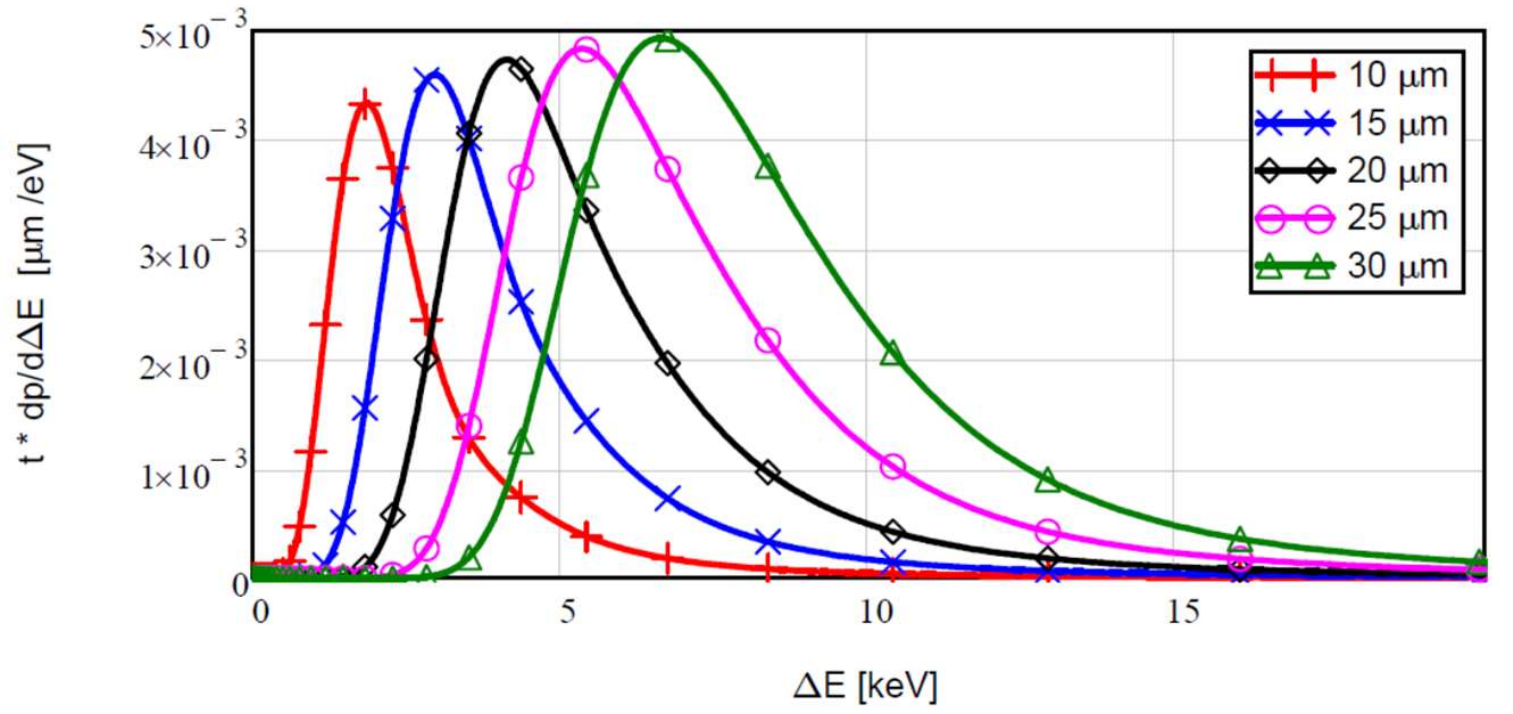}
    \caption{ }
    \label{fig:Fig_MChight}
   \end{subfigure}%
   \caption{Product of the Si-path length $t$ and the probability density function $\mathrm{d} P / \mathrm{d} \Delta E$ for different $t$ for $10^6$ Monte Carlo events each.
  (a) $t$\, range $1\,\upmu$m to $7.5\,\upmu$m, and
  (b) $t$\, range $10\,\upmu$m to $30\,\upmu$m.
   }
  \label{fig:Fig_MC}
 \end{figure}

 Fig.\,\ref{fig:Fig_MC} shows for $10^6$ generated events the product $t \cdot \mathrm{d} P / \mathrm{d} \Delta E$ for several Si-path lengths, $t$.
 As the mean $\langle \Delta E \rangle \propto t$,  $\mathrm{d} P / \mathrm{d} \Delta E$ is multiplied with $t$ for visibility reasons.
 For $t = 1\,\upmu$m, which corresponds to an average of 4 energy-loss events, the structures caused by the single energy-loss cross sections are visible, and there is also a significant number of $\Delta E = 0$ events.
 For $t \gtrsim 5\,\upmu$m the structure disappears, the fraction of $\Delta E = 0$ events is negligible and the probability density distribution approaches a \emph{Landau curve}.
 One also notices, that at the most probably value $t \cdot \mathrm{d} P / \mathrm{d} \Delta E$ increases with $t$, which is caused by the decrease of the relative width with $t$.
 For the simulation of the detector response in Sect.\,\ref{sect:Resolution} the $t$\,range is $15\,\upmu$m to $20\,\upmu$m.

  \subsection{Spatial distribution of tracks over the electrode pitch}
   \label{sect:Appendix_SpatialDistribution}

 For the simulations and the correction method a uniform $\mathrm{d}^2N/(\mathrm{d}x_\mathit{true} \mathrm{d}z_\mathit{true})$ distribution is assumed.
 In addition, it is stated that if this is not the case, the cumulative distribution will deviate from a straight line, however, the method sketched in Fig.\,\ref{fig:Fig_Nx-true} can still be applied.
 In this Appendix a method for determining d$N$/d$x_\mathit{true}$ and d$N$/d$z_\mathit{true}$  from the measured spatial distribution of the particles over the sensor, is described.

 The measured 1D-spatial distribution is given by a histogram with $n$ bins $(\mathit{Nb}_i,\mathit{xb}_i)$, where $\mathit{Nb}_i$  is the number of events in the bin centred at position $\mathit{xb}_i$;  $b$ stands for \emph{beam}, as such measurements are typically performed in test beams.
 For simplicity the $\mathit{xb}_i$ are assumed to be uniformly spaced, with spacing $\Delta \mathit{xb}$.
 Next, it is assumed that the event numbers per $\Delta \mathit{xb}$ inside the individual bins can be obtained by a linear interpolation, which gives for bins $i = 2$ to $n$--1:
 \begin{equation}\label{equ:Nbxi}
  \mathit{Nb}\big(\xi ^{(i)}\big) \approx \mathit{Nb}_i  + \frac{\mathit{Nb}_{i+1} - \mathit{Nb}_{i - 1}}{2 \cdot \Delta \mathit{xb}} \cdot \xi^{(i)} \hspace{5mm} \mathrm{for} \hspace{5mm} - \Delta \mathit{xb}/2 \leq \xi^{(i)} < \Delta \mathit{xb}/2,
 \end{equation}
 where $\xi^{(i)}$ is the distance from the bin centre.
 Ignoring a minor corrections for the edge bins,
 the sum over all bins is
\begin{equation}\label{equ:SumNbxi}
 \mathit{Nb}(\xi) \approx \sum _{i = 1} ^{n}\mathit{Nb}(\xi ^{(i)}) =
 \mathit{N}_\mathit{tot} + \frac{\mathit{Nb}_n -\mathit{Nb}_1}{2 \cdot \Delta \mathit{xb}} \cdot \xi,
\end{equation}
 with the total number of events $N_\mathit{tot}$. From this follows the mean probability density distribution for the electrodes of pitch $p$
\begin{equation}\label{equ:ProbBeam}
  \mathrm{d}P(x)/ \mathrm{d}x \approx \frac{1}{p} \cdot \Big( 1 + \frac{Nb_n - Nb_1}{ 2 \cdot N_\mathit{tot} \cdot \Delta \mathit{xb}} \cdot x \Big)
  \hspace{5mm} \mathrm{for} \hspace{5mm} -p/2 \leq x \leq p/2.
\end{equation}
 For the data discussed in Sect.~\ref{sect:ResAngles}, typical values of the $x$~slope are a few $10^{-3}$\,cm$^{-1}$, which justifies the assumption of a uniform d$N$/d$x_\mathit{true}$~distribution for the simulation and a linear cumulative distribution for the correction of $x_\mathit{rec}$.
 It also implies that the linear interpolation used in Eq.~\ref{equ:Nbxi} is adequate.
 \end{appendices}

  \section*{Acknowledgements}
   \label{sect:Acknowledgement}

 We acknowledge the support of A.~Ebrahimi and F.~Feindt by the BMBF, the German Federal Ministry of Education and Research, funding code 05H19GUCC9,
 and of I.~Zoi by the Emmy-Noether program (HI 1952/1-1) of the DFG, the German Research Foundation.
 We thank D.~Dannheim for valuable comments to a preliminary version of the paper.

  \section*{References}
   \label{sect:Bibliography}

\end{document}